\documentclass[12pt,a4paper]{article}
\usepackage[utf8]{inputenc}
\usepackage[margin=1.8cm]{geometry}
\usepackage{amsmath,amsthm,mathtools,amssymb,lscape,mathabx}
\usepackage{thmtools,thm-restate,multicol}
\usepackage{mathrsfs}
\usepackage{proba}
\usepackage{booktabs}
\usepackage{threeparttable,subfigure,subfloat,float}
\usepackage{stackengine}
\usepackage{enumerate}
\usepackage{natbib}
\usepackage{graphicx,setspace,multirow}
\usepackage{dsfont}
\usepackage{caption}
\usepackage{xcolor}
\usepackage{chngcntr}
\usepackage{apptools}
\usepackage{xr}
\AtAppendix{\counterwithin{lemma}{section}}

\usepackage{hyperref}
\hypersetup{
    unicode=false,          
    pdftoolbar=true,        
    pdfmenubar=true,        
    pdffitwindow=false,     
    pdftitle={},    
    pdfauthor={},     
    pdfsubject={},   
    pdfcreator={},   
    pdfproducer={},  
    pdfkeywords={}, 
    pdfnewwindow=true,      
    colorlinks=false,       
    linkcolor=red,          
    citecolor=green,        
    filecolor=magenta,      
    urlcolor=cyan           
}

\newtheorem{Def}{Definition}
\newtheorem{assumption}{Assumption}
\newtheorem{Pro}{Proposition}

\newtheorem{Rem}{Remark}

\numberwithin{equation}{section}

\def\m{\mathcal}
\def\b{\boldsymbol}

\def\E{\mathbb{E}}
\def\R{\mathbb{R}}

\def\1{\mathds{1}}
\def\S{\mathcal{S}}

\DeclareSymbolFont{sfoperators}{OT1}{cmss}{m}{n}
\DeclareSymbolFontAlphabet{\mathsf}{sfoperators}

\makeatletter
\def\operator@font{\mathgroup\symsfoperators}
\makeatother

\begin{document}

\title{Do We Exploit all Information for Counterfactual Analysis? Benefits of Factor Models and Idiosyncratic Correction}

\vspace{-0.5cm}
\author{{\textbf{Jianqing Fan}}\\
\small{Department of Operations Research and Financial Engineering} \\
\small{Princeton University}
\and
{\textbf{Ricardo Masini}} \\
\small{Center for Statistics and Machine Learning, (CSML),  Princeton University}\\
\small{Sao Paulo School of Economics (EESP), Getulio Vargas Foundation}
\and
{\textbf{Marcelo C. Medeiros}} \\
\small{Department of Economics}\\
\small{Pontifical Catholic University of Rio de Janeiro (PUC-Rio)}\\
\vspace{-0.8cm}}

\maketitle

\vspace{-0.9cm}

\begin{abstract}
\noindent
Optimal pricing, i.e., determining the price level that maximizes profit or revenue of a given product, is a vital task for the retail industry.  To select such a quantity, one needs first to estimate the price elasticity from the product demand. Regression methods usually fail to recover such elasticities due to confounding effects and price endogeneity. Therefore, randomized experiments are typically required. However, elasticities can be highly heterogeneous depending on the location of stores, for example. As the randomization frequently occurs at the municipal level, standard difference-in-differences methods may also fail. Possible solutions are based on methodologies to measure the effects of treatments on a single (or just a few) treated unit(s) based on counterfactuals constructed from artificial controls. For example, for each city in the treatment group, a counterfactual may be constructed from the untreated locations. In this paper, we apply a novel high-dimensional statistical method to measure the effects of price changes on daily sales from a major retailer in Brazil. The proposed methodology combines principal components (factors) and sparse regressions, resulting in a method called Factor-Adjusted Regularized Method for Treatment evaluation (\texttt{FarmTreat}). The data consist of daily sales and prices of five different products over more than 400 municipalities. The products considered belong to the \emph{sweet and candies} category and experiments have been conducted over the years of 2016 and 2017. Our results confirm the hypothesis of a high degree of heterogeneity yielding very different pricing strategies over distinct municipalities.



\noindent
\textbf{Acknowledgments}: We wish to thank an associate editor and three anonymous referees for very insightful comments. Fan's research was supported by NSF grants DMS-1712591,  DMS-2052926, DMS-2053832 and ONR grant N00014-19-1-2120.
Masini's and Medeiros' research was partially supported by  CNPq and CAPES. We are also in debt with Thiago Milagres for helping us with the dataset and all the team from the D-LAB@PUC-Rio for providing a superb research environment.

\end{abstract}

\doublespacing

\newpage

\setlength{\abovedisplayskip}{1.5pt}
\setlength{\belowdisplayskip}{1.5pt}

\section{Introduction}
The evaluation of treatment effects on a single (or just a few) treated unit(s) based on counterfactuals (i.e., the unobservable outcome had there been no intervention) constructed from artificial controls has become a popular practice in applied statistics since the proposal of the synthetic control (SC) method by \citet*{aAjG2003} and \citet*{aAaDjH2010}. Usually, these artificial (synthetic) controls are built from a panel of untreated peers observed over time, before and after the intervention.

The majority of methods based on artificial controls relies on the estimation of a statistical model between the treated unit(s) and a potentially large set of explanatory variables coming from the peers and measured before the intervention. The construction of counterfactuals poses a number of technical and empirical challenges. Usually, the dimension of the counterfactual model to be estimated is large compared to the available number of observations and some sort of restrictions must be imposed. Furthermore, the target variables of interest are non-stationary. Finally, conducting inference on the counterfactual dynamics is not straightforward. Although the original work by \citet*{aAjG2003} is able to handle some of these challenges, a number of extensions has been proposed; see \citet*{nDgI2016}, \citet*{sAgI2017}, or \citet*{aA2021} for recent discussions. Motivated by an application to the retail industry where the optimal prices of different products have to be determined, we develop a new methodology to construct counterfactuals which nests several other methods and efficiently explores all the available information. Our proposed method is well-suited for both stationary and non-stationary data as well as for high- or low-dimensional settings.

\subsection{Heterogeneous Elasticities and Optimal Prices}\label{S:App}
The determination of the optimal price of products is of great importance in the retail industry. By optimal price we mean the one that either maximizes profit or revenue. To determine such quantity we need first to estimate the price elasticity from the demand side. This is not an effortless task as standard regression methods usually fail to recover the parameter of interest due to confounding effects and the well-known endogeneity of prices.

Our novel dataset consists of daily prices and quantities sold of five different products for a major retailer in Brazil, aggregated at the municipal level. The company has more than 1,400 stores distributed in over 400 municipalities, covering all the states of the country.\footnote{Due to a confidentiality agreement, we are not allowed to disclosure either the name of the products or the name of the retail chain.} The chosen products differ in terms of magnitude of sales and in importance as a share of the company's total revenue. The overarching goal is to compute optimal prices at a municipal level via counterfactual analysis. Our method determines the effects in sales due to price changes and provides demand elasticities estimates which will be further used to compute the optimal prices.

To determine the optimal price of each of the products, a randomized controlled experiment has been carried out. More specifically, for each product, the price was changed in a group of municipalities (treatment group), while in another group, the prices were kept fixed at the original level (control group). The magnitudes of price changes across products range from 5\% up to 20\%. Furthermore, for three out of five products, the prices were increased, and for the other two products there was a price decrease. The selection of the treatment and control groups was carried out according to the socioeconomic and demographic characteristics of each municipality as well as to the distribution of stores in each city. Nevertheless, it is important to emphasize three facts. First, we used no information about the quantities sold of the product in each municipality, which is our output variable, in the randomization process. This way, we avoid any selection bias and can maintain valid the assumption that the intervention of interest is independent of the outcomes. Second, although according to municipality characteristics, we keep a homogeneous balance between groups, the parallel trend hypothesis is violated, and there is strong heterogeneity with respect to the quantities sold and consumer behavior in each city, even after controlling for observables. This implies that price elasticities are quite heterogeneous and optimal prices can be remarkably different among municipalities. Finally, there are a clear seasonal pattern in the data as well as common factors affecting the dynamics of sales across different cities.

Our results confirm the heterogeneous patterns in the intervention effects, yielding different elasticities and optimal prices across municipalities. In addition, the impacts also differ across products. Overall, the effects of price changes are statistical significant in more than 20\% of the municipalities in the treatment group and the optimal prices in terms of profit maximization are usually below the actual ones. Therefore, we recommend that the optimal policy in terms of profit maximization is to change the prices in the cities where the effects were statistically significant. Further experiments may be necessary to evaluate the effects of price changes in the cities were it was not possible to find statistically significant results.

\subsection{Methodological Innovations}

Driven by the empirical application discussed in the previous subsection, this paper proposes a methodology that includes both principal component regression (factors) and sparse linear regression for estimating counterfactuals for better evaluation of the effects on the sales of a set of products after price changes. It does not impose either sparsity or approximate sparsity in the mapping between the peers and the treated by using the information from hidden but estimable idiosyncratic components. Furthermore, we show that when the number of post-intervention observations is fixed, tests like the ones proposed in \citet*{rMmM2019} or \citet*{vCkWyZ2020a}, can be applied. Finally, we also consider a high-dimensional test to answer the question whether the use of idiosyncratic component actually leads to better estimation of the treatment effect. Our framework can be applied to much broader context in prediction and estimation and hence we leave more abstract and general theoretical developments to a different paper \citep{jFrMmM2021}.

The proposed method consists of four steps, called {\sl FarmTreat}. In the first one, the effects of exogenous (to the intervention of interest) variables are removed, for example, heterogeneous deterministic (nonlinear) trends, seasonality and other calendar effects, and/or known outliers. In the second step, a factor model is estimated based on the residuals of the first-step model. The idea is to uncover a common component driving the dynamics of the treated unit and the peers. This second step is key when relaxing the sparsity assumption. To explore potential remaining relation among units, in the third step a LASSO regression model is established among the residuals of the factor model, which are called the idiosyncratic components in the factor model. Sparsity is only imposed in this third step and it is less restrictive than the sparsity assumption in the second step. Note that all these three steps are carried out in the pre-intervention period. Finally,  in the forth step, the model is projected for the post-intervention period under the assumption that the peers do not suffer the intervention. Inspired by \citet*{jFyKkW2020}, we call the methodology developed here \texttt{FarmTreat}, the factor-adjusted regularized method for treatment evaluation.

The procedure described above is well suited either for stationary data or in the case of deterministic nonlinear and heterogeneous trends. In case of unit-roots, the procedure should be carried out in first-differences under the assumption that factors follow an integrated process (with or without drift). In this case, our result follows from Section 7 in \citet*{jBsN2008}. After the final step, the levels of both the target variable and the counterfactual can be recovered and the inference conducted.

We show that the estimator of the instantaneous treatment is unbiased. This result enables the use of residual ressampling procedures, as the ones in \citet*{rMmM2019} or \citet*{vCkWyZ2020a}, to test hypotheses about the treatment effect without relying on any asymptotic result for the post-intervention period. The testing procedures proposed in \citet*{rMmM2019} or \citet*{vCkWyZ2020a} are similar with the crucial difference that the first paper considers models estimated just with the pre-intervention sample, while the second paper advocates the use of the full sample to estimate the models. As shown by the authors and confirmed in our simulations, using the full data yields much better size properties in small samples.

We believe our results are of general importance for the following reasons. First and most importantly, the sparsity or approximate sparsity assumptions on the regression coefficients do not seem reasonable in applications where the cross-dependence among all units in the panel are high. In addition, due to the cross-dependence, the conditions needed for the consistency of LASSO or other high-dimensional regularization methods are violated \citep*{jFyKkW2020}. Second, first filtering for trends, seasonal effects and/or outliers seem reasonable in order to highlight the potential intervention effects by removing uninformative terms. Finally, modeling remaining cross-dependence among the treated unit and a sparse set of peers are also important to gather all relevant information about the correlation structure about the units.

Under the hypothesis that the treatment is exogenous which is standard in the synthetic control literature, we have an unbiased estimator for the treatment effect on the treated unit for each period after the intervention.  In the case the treatment is exogenous with respect only to the peers, we can identify the effects of a specific intervention on the treated unit, i.e., the time of a single intervention is fully known. This might be the quantity of interest in several macroeconomic applications as, for instance, the effects of Brexit on the United Kingdom economy fixing the date of the event.

We conduct a simulation study to evaluate the finite-sample properties of the estimators and inferential procedures discussed in the paper. We show that the proposed method works reasonably well even in very small samples. Furthermore, as a case study, we estimate the impact of price changes on product sales by using a novel dataset from a major retail chain in Brazil with more than 1,400 stores in the country. We show how the methods discussed in the paper can be used to estimate heterogeneous demand price elasticities, which can be further used to determine optimal prices for a wide class of products. In addition, we demonstrate that the idiosyncratic components do provide useful information for better estimation of elasticities.

\subsection{Comparison to the Literature}

Several papers in the literature extend the original SC method and derive estimators for counterfactuals when only a single unit is treated. We start by comparing with \citet*{cCrMmM2018}. Differently from this paper, we do neither impose sparsity nor our results are based on pre- and post-intervention asymptotics. We just require the pre-intervention sample to diverge in order to prove our results. Furthermore, by combing a factor structure with sparse regression we relax the (weak) sparsity assumption on the relation between the treated unit and its peers. In addition, we allow for heterogeneous trends which may not be bounded as in the case of the aforementioned paper; for a similar setup to \citet*{cCrMmM2018}, see \citet*{kLdB2017}. Masini and Medeiros (2019,2020)\nocite{rMmM2019,rMmM2020} consider a synthetic control extension when the data are nonstationary, with possibly unit-roots. However, the former paper imposes weak-sparsity on the relation between the treated unit and the peers and the later only handles the low-dimensional case. The low-dimensional non-stationary case is discussed in many other papers. See, for example, \citet*{cHhsCskW2012}, \citet*{mOyP2015}, \citet*{zDlZ2015}, and \citet*{kL2020}, among many others.

Compared to Differences-in-Differences (DiD) estimators, the advantages of the many estimators based on the SC method are three folds. First, we do not need the number of treated units to grow. In fact, the workhorse situation is when there is a single treated unit. The second, and most important difference, is that our methodology has been developed for situations where the $n-1$ untreated units may differ substantially from the treated unit and cannot form a control group, even after conditioning on a set of observables. For instance, in the application in this paper, the dynamics of sales in a specific treated municipality cannot be perfectly matched by any other city exclusively. On the other hand, there may exist a set of cities where the combined sales are close enough to ones of the treated unit in the absence of the treatment. Another typical example in the literature if to explain the gross domestic product (GDP) of a specific region by a linear combination of GDP from several untreated regions; see \citet*{aAjG2003}. Finally, SC methods and their extensions are usually consistent even without the parallel trends hypothesis.

More recently, \citet*{lGtM2016} generalize DiD estimators by estimating a correctly specified linear panel model with strictly exogenous regressors and interactive fixed effects represented as a number of common factors with heterogeneous loadings. Their theoretical results rely on double asymptotics when both $T$ (sample size) and $n$ (number of peers) go to infinity. The authors allow the common confounding factors to have nonlinear deterministic trends, which is a generalization of the linear parallel trend hypothesis assumed when DiD estimation is considered. Our method differs from \citet*{lGtM2016} in a very important way as we consider cross-dependence among the idiosyncratic units after the common factors have been accounted for.

Finally, we should compare our results with Chernozhukov and Wüthrich and Zhu (2020a,b)\nocite{vCkWyZ2020a,vCkWyZ2020b}. \citet*{vCkWyZ2020a} propose a general conformal inference method to test hypotheses on the counterfactuals which can be applied to our model setup as discussed above. When the sample size is small we strongly recommend the use of the approach described in \citet*{vCkWyZ2020a} to conduct inference on the intervention effects. \citet*{vCkWyZ2020b} propose a very nice generalization of \citet*{cCrMmM2018} with a new inference method to test hypotheses on intervention effects under high dimensionality and potential nonstationarity. However, their approach differs from ours in three aspects. First, and more importantly, their results are based on both pre- and post-intervention samples diverging. Second, their inferential procedure is designed to test hypothesis only on the average effect. Our procedure can be applied to a wide class of hypothesis tests. Finally, they impose that exactly the same (stochastic) trend is shared among all variables in the model. This is a more restricted framework than the one considered here.

\subsection{Organization of the Paper}

The rest of the paper is organized as follows. We give an overview of the proposed method and the application in Section \ref{S:overview}. We present the setup and assumptions in Section \ref{S:Setup} and state the key theoretical result in  Section \ref{S:Theory}. Inferential procedures are presented in Section \ref{S:Inference}. We present the results of a simulation experiment in Section \ref{S:Simulations}. Section 4 is devoted to provide guidance to practitioners and  a discussion of the empirical application can be found in Section \ref{S:Applications}.  Section \ref{S:Conclusion} concludes the paper. Finally, the proof of our theoretical result and additional empirical results are relegated to the Supplementary Material.

\section{Methodology}\label{S:overview}

The dataset is a realization of $\{Z_{it}, \b W_{it}: 1\leq i\leq n, 1\leq t\leq T\}$, in which $Z_{it}$ is the variable of interest and $\b W_{it}$ describes potential covariates, including seasonal terms and/or deterministic (nonlinear and heterogeneous) trends, for example. Suppose we are interested in estimating the effects on the variable $Z_{1t}$ of the first unit after an intervention that occurred at $T_0+1$. We estimate a counterfactual based on the peers $\b Z_{-1t}:=(Z_{2t},\ldots, Z_{nt})'$ that are assumed to be unaffected by the intervention. We allow the dimension of $\b{Z}_{-1t}$ to grow with the sample size $T$, i.e. $n:=n_T$. We also assume that $\b{W}_{it}$ are not affected by the intervention. Our key idea is to use both information in the latent factors and idiosyncratic components and we name the methodology as \texttt{FarmTreat}.

The procedure is thus summarized by the following steps:
\begin{enumerate}
\item
For each unit $i=1,\ldots,n$, run the regression:
\[
Z_{it}=\b\gamma_i'\b{W}_{it}+R_{it},\quad t=1,\ldots,T^*,
\]
and compute $\widehat{R}_{it}:=Z_{it} - \widehat{\b\gamma}_i'\b{W}_{it}$, where $T^*=T_0$ for $i=1$ and $T^*=T$, otherwise. This step removes heterogeneity due to $\b{W}_{it}$. As mentioned before, $\b{W}_{it}$ may include an intercept, any observable factors, dummies to handle seasonality and outliers, and determinist (polynomial) trends, for example. In the case of our particular application, $Z_{it}$ represents the daily quantity of a product sold per store in a municipality $i$ and $\b{W}_{it}$ includes a constant, six dummy variables for the days of the week and a linear trend.

\item
Write $\b{R}_t:=(R_{1t},\ldots,R_{nt})'$, which is the cross-sectional data $\b{Z}_t:=(Z_{1t}, \cdots, \b{Z}_{nt}')'$ after the heterogeneity adjustments. Fit the factor model
\[
\b{R}_t=\b\Lambda\b{F}_t+\b{U}_t,
\]
where $\b{F}_t$ is an $r$-dimensional vector of unobserved factors, and $\b\Lambda$ is an unknown $n\times r$ loading matrix and $\b{U}_t$ is an $n$-dimensional idiosyncratic component. The second step consists of using the panel data $\{\widehat {\b R}_t\}_{t=1}^T$ to learn the common factors $\b{F}_t$ and factor loading matrix $\b\Lambda$ and compute the estimated idiosyncratic components by
\[
\widehat{\b U}_t=\widehat{\b R}_t - \widehat{\b\Lambda}\widehat{\b F}_t,
\]
where $\widehat{\b U}_t=\left(\widehat{U}_{1t},\ldots,\widehat{U}_{nt}\right)'$.  There is a large literature on high-dimensional factor analysis; see Chapter 10 of the book by \cite{FLZZ20} for details. One important point is that we should not use data after $T_0$ for the treated unit. There are many possibilities to handle this issue that are discussed in Section \ref{S:Practice}.
\item
The third step is to use the idiosyncratic component to further augment the prediction on the treatment unit. It consists of first testing for the null of no remaining cross-sectional dependence (optional). If the null is rejected, fit the model in the pre-intervention period
\[
\widehat{U}_{1t}=\b\theta_1'\widehat{\b{U}}_{-1t} +V_{t},\quad t=1,\ldots,T_0,
\]
by using LASSO, where $\widehat{\b{U}}_{-1t}=\left(\widehat{U}_{2t},\ldots,\widehat{U}_{nt}\right)'$. Namely, compute
\begin{equation}\label{eq:LASSO}
\widehat{\b\theta}_1=\arg\min\left[\sum_{t=1}^{T_0}\left(\widehat{U}_{1t}-\b\theta_1'\widehat{\b{U}}_{-1t}\right)^2+\xi\|\b\theta_1\|_1\right].
\end{equation}

This step uses cross-sectional regression of the idiosyncratic components to estimate the effects in the treated unit. It is approximately the same as using $\widehat{\b{F}}_t$ and $\widehat{\b{U}}_{-1t}$ to predict $\widehat{R}_{1t}$ with the sparse regression coefficients for $\widehat{\b{U}}_{-1t}$, due to the orthogonality between $\{\widehat{\b{F}}_t\}_{t=1}^T$ and $\{\widehat{\b{U}}_{t}\}_{t=1}^T$.
The model includes sparse linear model on $\b{R}_t$ as a specific example (see \eqref{eq-adj} below) and the required model selection conditions are more easily met due to the factor adjustments. It also encompasses the principal component regression (PCR) in which $\widehat{\b\theta}_1= 0$, namely, using no cross-sectional prediction.

\item
Finally, the intervention effect $\delta_t$ is estimated for $t>T_0$ as
\begin{equation}\label{eq:treat}
\widehat{\delta}_t = Z_{1t} - \left(\widehat{\b\gamma}_1'\b{W}_{1t} + \widehat{\b\lambda}_1'\widehat{\b F}_t + \widehat{\b\theta}_1'\widehat{\b{U}}_{-1t}\right).
\end{equation}
where $ \widehat{\b\lambda}_1$ is the estimated loading of unit $1$, the first row of $\widehat {\b \Lambda}$. During the post treatment period, the realized factors $\widehat{\b F}$ are learned without using $R_{1,t}$.
\item
Use the estimator \eqref{eq:treat} to test for null hypothesis of no intervention effect in the form described by \eqref{E:H0}.
\end{enumerate}

The innovations of our approach in estimating counterfactuals are multi-folds. For simplicity, let us suppose that we have no $\b W_{it}$ component, so that $\b R_t = \b Z_t$.  First of all, the proposed procedure explores both the common factors and the dependence among idiosyncratic components. This not only makes use of more information, but also makes the newly transformed predictors less correlated. The latter makes the variable selection much easier and prediction more accurate. Note that factor regression (principal component regression) to estimate counterfactuals is a special case when $\b \theta_1 = 0$. Clearly, the method explores the sparsity of $\b \theta_1$ to improve the performance and also includes the case of sparse regression on $\b Z_{-1t}$ to estimate counterfactuals as in \citet*{rMmM2019}, where counterfactuals are estimated as
\begin{equation} \label{eq2.3}
 Z_{1t} = \b \theta_1 ' \b Z_{-1t} + \epsilon_t, \quad t = 1, \cdots, T_0.
\end{equation}
However, the variables $\b Z_{-1t}$ are highly correlated in high dimensions as they are driven by common factors, which makes variable selection procedures inconsistent and prediction ineffective. Instead, \citet*{jFyKkW2020} introduces the idea of lifting, called factor adjustments. Using the factor model in step 2, we can write the linear regression model \eqref{eq2.3} as
\begin{equation}\label{eq-adj}
  Z_{1t}  =  \b \theta_1 ' \b \Lambda_{-1} \b F_t +  \b \theta_1 ' \b U_{-1t} + \epsilon_t,
\end{equation}
where $\b \Lambda_{-1}$ and $\b U_{-1t}$ are defined as $\b \Lambda$ and $\b U_{t}$ without the first row. When we take $\b \lambda_1 = \b \theta_1'\b \Lambda_{-1}$, this reduces to use sparse regression to estimate the counterfactuals, but now use more powerful \texttt{FarmSelect} of  \citet*{jFyKkW2020} to fit the sparse regression. Again, \texttt{FarmSelect} imposes the condition $\b \theta_1'\b \Lambda_{-1}$ as the regression coefficients of $\b F_t$. Our method does not require this constraint. This flexibility allows us to apply our new approach even when the sparse linear model does not hold.

Finally, we also consider a test for the contribution of the idiosyncratic components by testing the null hypothesis that $\b\theta_1=\b 0$. Note that this is a high-dimensional hypothesis test, which is equivalent to testing the uncorrelatedness between the idiosyncratic component $U_{1t}$ for the treated unit and those from the untreated units $\b U_{-1t}$ in the pre-intervention period.

\section{Assumptions and Theoretical Result}\label{S:Setup}

\subsection{Assumptions}\label{S:Assumptions}

Suppose we have $n$ units (municipalities, firms, etc.) indexed by $i=1,\dots,n$. For every time period $t=1,\ldots,T$, we observe a realization of a real valued random vector $\b{Z}_t:=(Z_{1t},\ldots,Z_{nt})'$.\footnote{We consider a scalar variable for each unit for the sake of simplicity, and the results in the paper can be easily extended to the multivariate case.} We assume that an intervention took place at $T_0+1$, where $1<T_0<T$. Let $\mathcal{D}_t\in\{0,1\}$ be a binary variable flagging the periods where the intervention for unit 1 was in place. Therefore, following Rubin's potential outcome framework, we can express $Z_{it}$ as
\begin{equation*}
Z_{it}= \mathcal{D}_t Z_{it}^{(1)} + (1-\mathcal{D}_t) Z_{it}^{(0)},
\end{equation*}
where $Z_{it}^{(1)}$ denote the potential outcome when the unit $i$ is exposed to the intervention and $Z_{it}^{(0)}$ is the potential outcome of unit $i$ when it is not exposed to the intervention.

We are ultimately concerned with testing the hypothesis on the potential effects of the intervention in the unit of interest, i.e., the treatment effect on the treated. Without loss of generality, we set unit 1 to be the one of interest. The null hypothesis to be tested is:
\begin{equation}\label{E:H0}
\mathcal{H}_0:\b{g}(\delta_{T_0+1},\ldots,\delta_{T})=\b{0},
\end{equation}
where $\delta_t:=Z_{1t}^{(1)}-Z_{1t}^{(0)},\quad \forall t>T_0$, and $\b{g}(\cdot)$ is a vector-valued continuous function. The general null hypothesis \eqref{E:H0} can be specialized to many cases of interest, as for example:
\[
\mathcal{H}_0:\frac{1}{T-T_0}\sum_{t=T_0+1}^T\delta_t=0
\quad \textnormal{or} \quad
    \mathcal{H}_0:\delta_t=0,\,\forall t>T_0.
\]

It is evident that for each unit $i=1,\ldots,n$ and at each period $t=1,\dots,T$, we observe either $Z_{it}^{(0)}$ or $Z_{it}^{(1)}$. In particular, $Z_{1t}^{(0)}$ is not observed from $t=T_0+1$ onwards. For this reason, we henceforth call it the \emph{counterfactual} -- i.e., what  $Z_{1t}$ would have been like had there been no intervention (potential outcome).

The counterfactual is constructed by considering a model in the absence of an intervention:
\begin{equation}\label{E:main_model}
{Z}_{1t}^{(0)} =\m{M}\left(\b{Z}_{-1t}^{(0)};\b{\theta}\right) + V_t,\quad t=1,\ldots,T,
\end{equation}
where $\b{Z}_{-1t}^{(0)}:=(Z_{2t}^{(0)},\ldots, Z_{nt}^{(0)})'$ be the collection of all control variables (all variables in the untreated units).\footnote{We could also have included lags of the variables and/or exogenous regressors into $\b{Z}_{-1t}$, but again, to keep the argument simple, we have considered only contemporaneous variables; see \citet*{cCrMmM2018} for more general specifications.}, $\m{M}:\m{Z}\times\b\Theta\rightarrow\mathbb{R}$, $\m{Z}\subseteq\R^{n-1}$, is a known measurable mapping up to a vector of parameters indexed by $\b\theta\in\b\Theta$ and $\b\Theta$ is a parameter space. A linear specification (including a constant) for the model $\m{M}(\b Z_{0t};\b\theta)$ is the most common choice among counterfactual models for the pre-intervention period.  \texttt{FarmTreat} uses a more sophisticated model.

Roughly speaking, in order to recover the effects of the intervention, we need to impose that the peers are unaffected by the intervention in the unit of interest. Otherwise our counterfactual model would be invalid. Specifically we consider the following key assumption
\begin{assumption}[\textbf{Intervention Independence}]\label{Ass:ind}
$\b Z_t^{(0)}$ is independent of $\m{D}_s$ for all $1\leq s,t\leq T$.
\end{assumption}

\begin{Rem}
Assumption \ref{Ass:ind} identifies the treatment effect on the treated. If only $ \b Z_{-1t}^{(0)}$ is independent of $\m{D}_s$ for all $1\leq s,t\leq T$, we can recover the effect of the intervention on the treated unit given that $T_0$ is deterministic and known. This later case is typical in papers on SC.
\end{Rem}

The main idea is to estimate (\ref{E:main_model}) using just the pre-intervention sample, $t=1,\ldots,T_0$, since under Assumption \ref{Ass:ind}, $\b Z_{t}^{(0)}=(\b Z_{t}^{(0)}|\m{D}_t = 0) = (\b Z_{t}|\m{D}_t=0)$ for all $t$.  Consequently, the estimated counterfactual for the post-intervention period, $t=T_0+1,\dots,T$, becomes
$\widehat{Z}_{1t}^{(0)} :=\m{M}(\b Z_{-1t};\widehat{\b\theta}_{T_0})$. Under some sort of stationary assumption on $\b Z_{t}$, in the context of a linear model, \citet*{cHhsCskW2012} and \citet*{cCrMmM2018}, show that $\widehat{\delta}_t:=Z_{1t} - \widehat{Z}_{1t}^{(0)}$ is an unbiased estimator for $\delta_t$ as the pre-intervention sample size grows to infinity in the low and high dimensional sparse case respectively.

We model the units in the absence of the intervention as follows.

\begin{assumption}[\textbf{DGP}]\label{Ass:DGP} The process $\{Z_{it}^{(0)}:1\leq i\leq n , t\geq 1\}$ is generated by
\begin{equation}\label{E:DGP}
Z_{it}^{(0)} =\b\gamma_{i}' \b W_{it}+\b\lambda_i'\b F_t +U_{it}
\end{equation}
where $\b\gamma_i\in\R^{k}$ is the vector of coefficients of the $k$-dimensional observable random vector $\b W_{it}$ of attributes of unit $i$,  $\b F_t$ is a $r$-dimensional vector of common factors and $\b\lambda_i$ its respective vector of loads for unit $i$; and $U_{it}$ is a zero mean idiosyncratic shock. Finally, we assume that $\b W_{it}$, $\b F_t$ and $U_{it}$ are mutually uncorrelated.
\end{assumption}

The reason to include $\b W_{it}$ is to accommodate an intercept, heterogeneous deterministic trends, seasonal dummies or any other exogenous (possibly random) characteristic of unit $i$ that the practitioner judges to be helpful in the construction of the counterfactual. As mentioned before we include an intercept, dummies to account for the effects of different days of the week and a linear trend. Other possibilities could be dummies of nation-wide promotions and/or holidays, for example.

In case of stochastic heterogeneous trends, we let the factors follow a random walk with (or without) drift: $\b{F}_t=\b\mu+\b{F}_{t-1}+\b\eta_t$, where $\{\b\eta_t\}$ is a second-order stationary vector process. When this is the case, the methodology must be applied in first-differences and levels should be reconstructed in the end. Therefore, our approach can be directly applied even if the units have heterogeneous and stochastic trends.

Our counterfactual model is the sample version of the projection of $Z_{1t}^{(0)}$ onto the space spanned by $(\b W_{1t}, \b F_t, \b U_{-1,t})'$. Under Assumption \ref{Ass:DGP} the counterfactual can be taken as
\begin{equation}\label{E:counterfactual_model}
Z_{1t}^{(0)} = \b\gamma_{1}' \b W_{1t}+\b\lambda_1'\b F_t +\b \theta_1'\b U_{-1t} + V_t,
\end{equation}
where $\b\theta_1$ is the coefficient of the linear regression of $U_{1t}$ onto $\b U_{-1t}$ and $V_t$ the respective projection error.

\subsection{Theoretical Guarantees}\label{S:Theory}

In order to state our result in a precise manner we consider the technical assumption below. First, let $\b W_{S,it}$ denotes the sub-vector of $\b W_{it}$ after the exclusion of all deterministic (non-random) components (constant, dummies, trends, etc). We state the assumption for the case where the unobserved factors are stationary. For non-stationary (unit-root) factors, the only difference is to state the conditions for the first-differences of the variables involved: $\Delta \b W_{S,it}$, $\Delta \b{F}_t$, and $\Delta \b{U}_t$. In this case, our results can be derived following \citet*{jBsN2008}, Section 7. Note that, as mentioned before, if the interest lies on the intervention effects on the levels of the series, after the final step, the levels of both the target variable and the counterfactual can be recovered and the inferential procedures can be applied unaltered.

\begin{assumption}[\textbf{Regularity Conditions}]\label{Ass:Moments}
There is a constant $0<C<\infty$ such that:
\begin{enumerate}[(a)]
\item
The covariance matrix of $\b W_{S,it}$ is non-singular;
\item
$\E|\b W_{S,it}|^p\leq C $ and $\E |U_{it}|^{p+\epsilon}\leq C $ for some $p\geq 6$ and $\epsilon>0$ for $i=1,\ldots,n$, $t=1,\ldots,T$;
\item The process $\{(\b W_{S,t}',\b F_t',\b U_t')',t\in \Z\}$ is weakly stationary with strong mixing coefficient $\alpha$ satisfying $\alpha(m)\leq \exp(-2cm)$ for some $c>0$ and  for all $m\in\Z$;
\item $\|\b\theta_1\|_\infty \leq C$;
\item $\kappa_0:=\kappa\left[\E(\b U_t \b U_t'),\S_0,3\right]\geq C^{-1}$ where $\kappa[\cdot]$  is the compatibility condition defined in \eqref{E:GIF} in the Supplementary Material and $\S_0:=\{i:\theta_{1,i}\neq 0\}$.
\end{enumerate}
\end{assumption}

Condition (a) is necessary for the parameters $\b\gamma_i$, $i=1,\ldots,n$, to be well defined. Conditions (b) and (c) taken together allow the law of large numbers for strong mixing processes to be applied to appropriately scaled sums. In particular, $(b)$ bounds the $p$-th plus moment uniformly. However, if $U_{it}$ has exponential tails as contemplated in Assumption 3 in \citet*{jFrMmM2021}, we could state a stronger result in terms of the allowed number of non-zero coefficients as a fraction of the sample size. The mixing rate in condition (c) can be weaken to polynomial rate at the expense of an interplay between (c) and the conditions appearing in Proposition \ref{C:Main}.

Finally, conditions $(d)$ and $(e)$ in Assumption \ref{Ass:Moments} are regularity conditions on the high-dimensional linear model to be estimated by the LASSO in step 3. Condition (e) ensures the (restricted) strong convexity of the objective function, which is necessary for consistently estimate $\b\theta_1$ when $n>T$. In effect, it uniformly lower bounds the minimum restricted $\ell_1$-eigenvalue of the covariance matrix of $\b U_t$. For simplicity, the bounds appearing in (d) and (e) are assumed to hold uniformly. However, both conditions could be somewhat relaxed to allow $\|\b\theta_1\|_\infty$ to grow slowly and/or $\kappa_0$ decreases slowly to 0 as $n$ diverges. Once again, at the expense of having both terms included in the conditions of Proposition \ref{C:Main}.

\begin{Pro}\label{C:Main} Under Assumptions \ref{Ass:ind}--\ref{Ass:Moments}, assume further that:
\begin{enumerate}[(a)]
    \item There is a bounded sequence $\eta:=\eta_{n,T}$ such that $\|\widehat{\b U}- \b U\|_{\max} =O_P(\eta)$; and
    \item $|\S_0| = O\left(\left\{\eta\left[(nT)^{1/p}+ \eta\right] + \frac{n^{4/p}}{\sqrt{T}}\right\}^{-1}\right).$
\end{enumerate}
If the penalty parameter $\xi$ in \eqref{eq:LASSO} is set to be at the order of $
\frac{n^{2/p}}{\sqrt{T}} +\eta T^{1/p}$ then, as $T_0\to\infty$, $\|\widehat{\b\theta}_1-\b\theta_1\|_1= O_P\left(\xi|\m{S}_0|\right)$,
and for every $t>T_0$:
\[
\widehat{\delta}_t-\delta_t = V_t  + O_P\left\{|\S_0|\left[\eta(nT)^{1/p} + \frac{n^{3/p}}{\sqrt{T}}\right]\right\},
\]
where $V_t$ is the stochastic component not explainable by untreated units defined by \eqref{E:counterfactual_model}
\end{Pro}

\begin{Rem} Conditions (a) and (b) are high level assumptions that translate into a restriction on the estimation rate in steps 1 and 2 of the proposed methodology, which in turn puts an upper bound on the number of non-zero coefficients in $\b\theta_1$ (sparsity) in order for the estimation error to be negligible. The rate $\eta$ can be explicitly obtained in terms of $n$ and $T$ by imposing conditions on projection matrix of $\b W_{it}$ and the factor model. For the former, we need uniform consistencies of both the factor and the loadings estimators that take into account the projection error in the previous step. In a more general setup, \citet*{jFrMmM2021} state conditions under which $\eta =\frac{n^{6/p}}{T^{1/2-6/p}}+\frac{T^{1/p}}{\sqrt{n}}$.
\end{Rem}

Proposition \ref{C:Main} is key for our inference procedure discussed in Section \ref{S:Inference}. For instance, it can be used to argue that $\widehat{\delta}_t-\delta_t = V_t +o_p(1)$ provided that $ |\S_0|\left[\eta(nT)^{1/p} + \frac{n^{3/p}}{\sqrt{T}}\right] = o(1)$. Since $V_t$ is zero mean by construction,  as $T_0\to\infty$, $\widehat \delta_t$ is an unbiased estimator for $\delta_t$ for every  post-intervention period. Furthermore, as described below, we can estimate the quantiles of $V_t$ using the pre-intervention residuals to conduct a valid inference on $\delta_t$.

\subsection{Testing for Intervention Effect}\label{S:Inference}

We test the null of no intervention effects based on estimators $\{\widehat{\delta}_t\}_{t>T_0}$ and the results of Masini and Medeiros (2019,2020)\nocite{rMmM2020}\nocite{rMmM2019} and \citet*{vCkWyZ2020a}. Let $T_2:=T-T_0$ be the number of observations after the intervention and define a generic continuous mapping $\b\phi:\R^{T_2}\rightarrow\R^b$ whose argument is the $T_2$-dimensional vector $(\widehat{\delta}_{T_0+1}-\delta_{T_0+1},\ldots, \widehat{\delta}_{T}-\delta_T)'$ with given treatment effects
$\delta_{T_0+1}, \cdots, \delta_T$.

We are interested in the distribution of $\b{\widehat{\phi}}:=\b\phi(\widehat{\delta}_{T_0+1}-\delta_{T_0+1},\dots, \widehat{\delta}_{T}-\delta_T)$ under the null \eqref{E:H0}, where $\b \phi$ is a given statistic. The typical situation is the one where the pre-intervention period is much longer than the post intervention period, $T_0\gg T_2$. Frequently, it could be well the case that $T_2=1$. However, $V_t$ does not vanish as in most cases there is a single treated unit. Nevertheless, under strict stationarity of the process $\{V_t\}$ and unbiasedness of the treatment effect estimator,  it is possible to resample the pre-intervention residuals following either the procedure described in Masini and Medeiros (2019,2020)\nocite{rMmM2020}\nocite{rMmM2019} or the one in \citet*{vCkWyZ2020a} to compute the sample quantile of the statistic of interest. As pointed out earlier, the main difference between the two approaches is that the former estimate the counterfactual model using only pre-treatment observations while the later considers the estimation using the full sample.

Under the asymptotic limit taken on the pre-invention period $(T_0\to\infty)$, by Proposition~\ref{C:Main}, we have that $\b{\widehat{\phi}}-\b\phi_0=o_P(1)$, where $\b\phi_0:=\b\phi(V_{T_0+1},\dots,V_T)$.  Thus, the distribution of $\b{\widehat{\phi}}$ can be estimated by that of $\b\phi_0$. Consider the construction of $\b{\widehat{\phi}}$ using only blocks of size $T_2$ of consecutive observations from the pre-intervention sample. There are $T_0-T_2+1$ such blocks denoted by $\b{\widehat{\phi}}_j:=\b\phi(\widehat{V}_j,\dots,\widehat{V}_{j+T_2-1}),\,j=1,\ldots,T_0-T_2+1$, where $\widehat{V}_t:=Z_{1t} - \left(\widehat{\b\gamma}_1'\b{W}_{1t} + \widehat{\b\lambda}_1'\widehat{\b F}_t + \widehat{\b\theta}_1'\widehat{\b{U}}_{-1t}\right)$ for the pre-intervention period. The estimators $\widehat{\b\gamma}_1$, $\widehat{\b\lambda}_1$, $\widehat{\b F}_t$, $ \widehat{\b\theta}_1'$, and $\widehat{\b{U}}_{-1t}$ use either the pre-intervention or the full sample depending on the inferential approach chosen by the practitioner.

For each $j$, we have that  $\b{\widehat{\phi}}_j-\b\phi_j=o_P(1)$ where $\b\phi_j:=\b\phi(V_j,\dots,V_{j+T_2-1})$ and $\b\phi_j$ is equal in distribution to $\b\phi_0$ for all $j$. Hence, we propose to estimate the distribution $\m{Q}_T(\b x):=\P(\b{\widehat{\phi}}\leq \b x)$ by its empirical distribution
\[
\widehat{\m{Q}}_T(\b x):=\frac{1}{T_0-T_2+1}\sum\limits_{j=1}^{T_0-T_2+1}\1(\b{\widehat{\phi}}_j\leq \b x),
\]
where, for a pair of vectors $\b a, \b b\in\R^d$, we say that $\b a\leq \b b\iff a_i\leq  b_i,\forall i$. See Masini and Medeiros (2019,2020)\nocite{rMmM2020}\nocite{rMmM2019} and \citet*{vCkWyZ2020a} for further details.

\subsection{Testing for Idiosyncratic Contributions}
The question of statistical and practical interest is if the idiosyncratic component contributes the estimation of the treatment effect.  To answer this question, write \eqref{E:DGP} as:
\[
\b Z_t =\b\Gamma \b W_t + \b \Lambda \b F_t + \b U_t,\qquad t\in\{1,\ldots,T\},
\]
where $\b Z_t:=(Z_{1t},\ldots, Z_{nt})'$, $\b U_t:=(U_{1t},\ldots, U_{nt})'$, and $\b W_t:=(\b W_{1t}',\ldots, \b W_{nt}')'$. The $(n\times nk)$ block diagonal matrix $\b \Gamma$ has blocks given by $(\b\gamma_1',\dots \b\gamma_n')$. Finally, $\b\Lambda:=(\b\lambda_1,\dots,\b\lambda_n)'$.

Let  $\b\Pi :=(\pi_{ij})_{1\leq i,j\leq n}$ denote the $(n\times n)$ covariance matrix of $\b U_t$. Our method exploits the sparsity of the off-diagonal elements of $\b\Pi$. In particular, we are interested in testing whether $\b U_{-1t}$ has linear prediction power on the treated unit $U_{1t}$.  This amounts to the following high-dimensional hypothesis test: $\mathcal{H}_{0}: \pi_{1j} = 0,\; \forall\; 2\leq j\leq n$.

In order to conduct the test we propose the following test statistic $S:=\|\b Q\|_\infty$, where $\b Q:= \frac{1}{\sqrt{T_0}}\sum_{t=1}^{T_0} \b D_t$, $\b D_t:=\widehat{U}_{1t}\widehat{\b U}_{-1t}$, and  $\widehat{U}_{it} := \hat R_{it}- \widehat{\b\lambda_i}'\widehat{\b F}_t$. Also let $c^*(\tau)$ be the $\tau$-quantile of the Gaussian bootstrap $S^*:=\|\b Q^*\|_\infty$, where $\b Q^*|\b Z, \b W\sim \m{N}(\b 0,\widehat{\b\Upsilon})$. For a given symmetric kernel $k(\cdot)$ with $k(0)=1$ and bandwidth $h>0$ (determining the number of lags), we have that
\[
\widehat{\b\Upsilon}:=\sum_{|\ell|<T_0}k(\ell/h) \widehat{\b M}_\ell\quad\textnormal{with}\quad \widehat{\b M}_\ell :=\tfrac{1}{T_0}\sum_{t = \ell +1}^{T_0} \b D_t\b D_{t-\ell}'
\]
is the estimator of the long-run covariance matrix $\b \Upsilon:=\V\widetilde{\b Q}$,  where $\widetilde{\b Q}:=\tfrac{1}{\sqrt{T_0}}\sum_{t=1}^{T_0} U_{1t}\b U_{-1t}$. Notice that $\widehat{\b\Upsilon}$ is just the Newey-West estimator if $k(\cdot)$ is chosen to be the triangular kernel. More generally, the choice of kernels can be made in class of kernels described in \citet*{andrews91}. The validity of such a method has been proved in \cite{jFrMmM2021} under a more general setting. In particular, the authors show under some regularity conditions
\[\sup_{\tau\in(0,1)}|\P(S\leq c^*(\tau)) -\tau|=o(1)\quad \text{under $\m{H}_0$}.\]

\section{Guide to Practice}\label{S:Practice}

In this section we provide practical guidance to the implementation of the \texttt{FarmTreat} method.

The first step involves the definition of the variables in $\b W_{it}$. This is, of course, application dependent. Nevertheless, typical candidates are deterministic functions of time, i.e, $f(t)$, in order to capture trends, an intercept to remove the mean, seasonal dummies or other calendar effects, or any other dummies to remove potential outliers. Unit-root tests on the variable of interest may also be important in order to decide whether first-differences of the data should be taken or not.

The second step is the estimation of $\b \Lambda$ and the sequence of factors $\{\b F_t,\,t\in\Z\}$ for the full sample, before and after the intervention. Therefore, we cannot just rely on pre-intervention period to estimate the factors. On the other hand, if we use all the observations from the treated unit, we will bias our estimation under the alternative of nonzero treatment effects. Therefore, there are two possible ways to estimate the factors and the factor loadings:
\begin{enumerate}
\item
A simple approach is to estimate the factors and factor loadings without the treated unit. In order to estimate the loadings $\widehat {\b \lambda}_1$ of the first unit, we then regress $R_{1t}$ on the estimated factors. This is the approach adopted in both simulations and in the empirical application.
\item
The imputed approach is to use the imputation $\widehat {\b \lambda}_1' \widehat{\b F}_t$ for the post intervention period of the treated unit 1 and  then apply the whole data to reestimate the factor and factor loadings. $\widehat {\b \lambda}_1$ and $\widehat{\b F}_t$ are estimated with just the pre-intervention period.
\item
Note that $\E(\b R_t)=\b 0$ by definition. Hence, we can replace the post-intervention observations of $R_{1t}$ by 0 in order to carry the factor analysis. As the number of post-intervention observations is expected to be quite small, this replacement will have negligible effects. It is important to notice, however, that we do this just to estimate the factors.
\end{enumerate}

To determine the number of factors we advocate the use of the eigenvalue ratio test \citep*{sAaH2013}. Other possibility is the use of one of the information criteria discussed in \citet*{jBsN2002}.

After the estimation of the common factor structure, we can test for remaining cross-dependence using the test described in Section \ref{S:Inference}. In the case of rejection of the null of no remaining dependence, the last step consists of a LASSO regression. This step of testing is optional for evaluating the treatment effect, as the sparsity of LASSO includes no effect as a specific example.  Nevertheless, it is an interesting statistical problem whether the idiosyncratic component contributes to the prediciton power. For selecting the penalty parameter in LASSO, we recommend the use of an information criterion, such as the BIC as in \citet*{rMmM2019}.

The final step is to test the null hypothesis concerning the intervention effects. When the pre-intervention sample is small, we follow \citet*{vCkWyZ2020a} and estimate the models under the null. Note that in this case we should re-estimate the model using the full sample.

\section{Simulations}\label{S:Simulations}

In this section we report simulations results to study the finite sample behavior of the method proposed in this paper. We consider the following data generating process:
\begin{equation}\label{E:simul}
\begin{split}
Z_{it}&=\delta_{it}+\b\gamma_i'\b{W}_t+R_{it},\qquad R_{it}=\b\lambda_{i}'\b{F}_{t}+U_{it},\\
\b{F}_{t}&=0.8 \b{F}_{t-1} + \b{V}_{t},\qquad
U_{it}=
\begin{cases}
\b\beta'\b{U}_{-1t} + \varepsilon_{it},&\textnormal{if}\,i=1,\\
\varepsilon_{it},&\textnormal{otherwise,}
\end{cases}
\end{split}
\end{equation}
where $\{\varepsilon_{it}\}$ is a sequence of independent and normally distributed zero-mean random variables with variance equal to $0.25$ if $i=1$ and $\b\beta\neq\b{0}$ or variance equal to $1$ if $i>1$ or $\b\beta=\b{0}$. $\b{V}_{t}$ is a sequence of independent and normally distributed zero-mean random vectors taking values on $\R^2$ such that $\E(\b{V}_t\b{V}_t')=0.25\times\b{I}$, and $\E(\varepsilon_{it}\b{V}_s)=\b{0}$, for all $i,t,$ and $s$. $\b{W}_{it}$ consists of a constant, a linear trend, and two independent Gaussian random variables with mean and variance equal to $1$. The parameters are set as follows: $\b\gamma_i$ is $(p+2$)-dimensional vector where, for each replication, the first entry is randomly pick from a Gaussian random variable with zero mean and variance 1; the second term is randomly selected from an Uniform distribution between -5 and 5; and the last two elements are Gaussian distributed with mean 0.5 and variance 1. For each replication, the elements of $\b\lambda_i$, $i>1$, are drawn independently from a normal distribution with mean two and unit variance and, for $i=1$, the elements of $\b\lambda_i$ are drawn from a normal distribution with mean -6 and variance 0.04. The first two elements of $\b\beta$ are either set to 0.5 and the rest is set to zero or we set all the elements equal to zero. We consider the following sample sizes: $T_0=50,75,100,150,250,500$ and $1000$; and $T_2=1$. For each sample size, $n$ is set as $n=\{T,2T,3T\}$. The number of factors is set to two. For size simulations, $\delta_{it}=0$ for all $i$ and $t$. For power simulations, $\delta_{it}=2$ for $i=1$ and $t=T_0+1$.

Tables \ref{T:simul_est_1} and \ref{T:simul_est_2} show descriptive statistics for the counterfactual estimation. The table depicts the mean, the median and the mean squared error (MSE) for $\delta_{T_0+1}$ under the null and alternative hypotheses, respectively. Three cases are considered. In the first one, the factor structure is neglected and a sparse LASSO regression of the first unit against the remaining ones is estimated. This is the ArCo methodology put forward by \citet*{cCrMmM2018}. The second one is equivalent to the approach of \citet*{lGtM2016}, where a pure factor model is considered. Finally, the \texttt{FarmTreat} approach is considered, which encompasses the previous two methods as a specific example. We also report, between brackets, the same statistics when the full sample is used to estimate the counterfactual model as advocated by \citet*{vCkWyZ2020a}.

Tables \ref{T:simul_est_1} and \ref{T:simul_est_2} show descriptive statistics for the counterfactual estimation. The table depicts the mean, the median and the mean squared error (MSE) for $\delta_{T_0+1}$ under the null and alternative hypotheses, respectively. Three cases are considered. In the first one, the factor structure is neglected and a sparse LASSO regression of the first unit against the remaining ones is estimated. This is the \texttt{ArCo} methodology put forward by \citet*{cCrMmM2018}. The second one is equivalent to the approach of \citet*{lGtM2016},
where a pure factor model is considered; we call this method the Principal Component Regression (\texttt{PCR}).
Finally, the \texttt{FarmTreat} approach is considered, which encompasses the previous two methods as a specific example. We also report, between brackets, the same statistics when the full sample is used to estimate the counterfactual model as advocated by \citet*{vCkWyZ2020a}.

From the inspection of the results in the tables, it is clear that the biases for estimating of the treatment effect are small and MSEs decrease as the sample size increase, as expected. Furthermore, the \texttt{ArCo} delivers very robust estimates, but the MSE can be substantially reduced by the \texttt{FarmTreat} methodology. Therefore, there is strong evidence supporting methodology derived in this paper, which is consistency with our theoretical results. Second, as already shown in the simulations in \citet*{cCrMmM2018}, the performance of the pure factor model is poor in terms of MSE.  This is particularly the case when $n$ or $T$ is small, since the factors are not well estimated. When this happens, the prediction power of the idiosyncratic components comes to rescue (comparing the performance with \texttt{FarmSelect}). This demonstrates convincingly the need of using the idiosyncratic component to augment the prediction. When comparing with the results when the full sample is used to estimate the model, two facts emerge from the tables. First, when the null hypothesis is true, the gains of using the full sample are undebatable. However, when the null is false, using the full sample is a bad idea, specially when $T_0$ is small. This is somewhat expected as in the later case we are including observations affected by the intervention in the estimation sample.

Table \ref{T:simul_size} presents the empirical size of the ressampling test when there is a single observation after the intervention and the counterfactual is estimated according to the methods described above. It is clear that size distortions are high when $T_0$ is small. The size converges to the nominal one as the sample increases. On the other hand, using the full sample to estimate the models correct the distortions
and are strongly recommended in the case of small samples.

Table \ref{T:simul_power} shows the empirical power. The ressampling approach delivers high power, specially when ArCo and \texttt{FarmTreat} methodologies are considered. On the other hand, the test looses a lot of power when the full sample is considered. This is expected as the estimator of the treatment effect will be biased, specially in small samples.

\begin{figure}
\caption{MSE Ratio}
\begin{spacing}{}
\scriptsize
The figure reports the ratio of the mean squared errors (MSE) of the \texttt{FarmTreat} methodology and Principal Component Regression (PCR)
when there is no cross-dependence among idiosyncratic components.
\vspace{0.3cm}
\end{spacing}
\centering
\includegraphics[width=\linewidth]{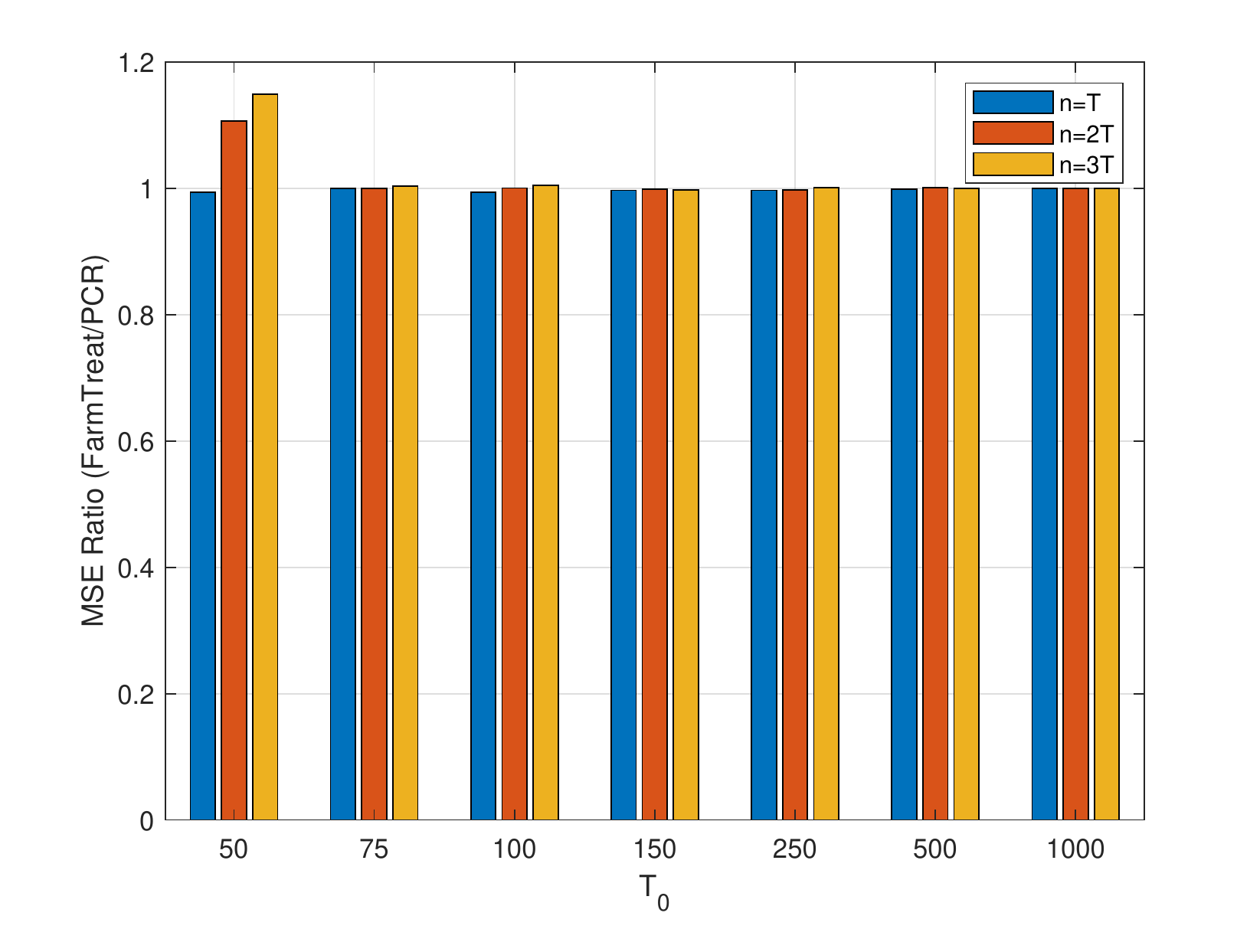}
\label{F:ratio}
\end{figure}

Figure \ref{F:ratio}  compares the MSEs of \texttt{PCR} and \texttt{FarmTreat} when DGP has no idiosyncratic contribution, i.e., $\b\beta=\b{0}$.   This case favors to \texttt{PCR}.  As we can see, \texttt{FarmTreat} achieves comparable results to \texttt{PCR}, indicating that the methodology is quite robust.

\section{Price Elasticity of Demand}\label{S:Applications}

\subsection{Data Description}

As described in Section \ref{S:App} the goal is to determine the optimal price of products for a large retail chain in Brazil. The optimal prices should be computed for each city. Our dataset consists of the daily prices and quantities sold of five different products, aggregated at the municipal level. The company's more than 1400 stores differ substantially across and within municipalities, ranging from small convenience stores with a limited selection of products up to very large ones, selling everything from sweets to home appliances and clothes. The stores can be street stores or can be located in shopping malls. The products sold are divided into several departments. Here, we consider products from the \emph{Sweets and Candies} unit. The chosen products differ in terms of magnitude of sales, price range, and in importance as a share of the company's revenue. For example, the median daily sales per store over the available period and across municipalities vary from 0 (Product V) to 35 units (Product II).

Our sample consists of about 50\% of the municipalities where there are stores. As the number and size of stores differ across municipalities, we divide the daily sales at each city by the number of stores in that particular location. To determine the optimal price of each of the products (in terms of profit or revenue maximization) and avoid confounding effects, a randomized controlled experiment has been carried out. For each product, the price was changed in a set of municipalities (treatment group), while in another group, the prices were kept fixed at the original level (control group). Note that the randomization is carried out at the city-level not at the store-level. With the application of our methodology optimal prices can be computed for each city in the treatment group as well as other levels of aggregation. In order to determine the prices for the locations in the control group, the experiment can be repeated by inverting the groups in a second batch of experiments. Here, we will report the results concerning the first group of randomized experiments.

The selection of the treatment and control groups was carried out according to the socioeconomic and demographic characteristics as well as to the distribution of stores in each city. The following variables were used: human development index, employment, GDP per capita, population, female population, literate population, average household income (total), household income (urban areas), number of stores, and number of convenience stores. Details about the method can be found in the Supplementary Material. As mentioned in the Introduction, we used no information about the quantities sold of the product in each municipality to create the treatment and control groups. Therefore, we avoid any selection bias, maintaining valid the assumption that the intervention of interest is independent of the outcomes.

It is important to highlight that although the experiment is randomized, traditional differences-in-differences estimators cannot be considered as the goal is to estimate the price elasticities at the municipal level which is exactly the same level of the randomization. Nevertheless, we can rely on differences-in-differences to estimate the intervention effects at the country level.

\subsection{Results}

In this section we report the results of the experiment described in the previous subsection.
Table \ref{T:experiments} describes each one of the experiments carried out for each product. The table shows the sample date, the period of the experiment (usually two weeks), the type of the experiment (if the price was increased or decreased), the magnitude of the price change, and the number of municipalities in the treatment ($n_1$) and control groups ($n_0$). $n$ is the total number of municipalities considered. $n$, $n_0$, and $n_1$ vary according to the product, but we omit the product identification to simplify notation.

For each day $t$, $q_{it}^{(j)}$ represents, for municipality $i$, the quantities sold of product $j$, where $i=1,\ldots,n$, $t=1,\ldots,T$, and $j=1,\ldots,5$. For convenience of notation assume that $i\in\{1,\ldots,n_0\}$ represents cities in the control group and $i\in\{n_0+1,\ldots,n\}$ indexes the municipalities in the treatment group. Finally, define $\widetilde{q}_{it}^{(j)}=q_{it}^{(j)}/n^s_{i}$, where $n^s_{i}$ is the number of stores at location $i$. The analysis is carried on for $\widetilde{q}_{it}^{(j)}$.

\begin{figure}
\caption{Data for Product I.}
\begin{spacing}{}
\scriptsize
Panel (a) reports the daily sales divided by the number of stores aggregated for all cities as well as for the treatment and control groups.
The plot also indicates the date of the intervention.
Panels (b) and (c) display the distribution of the average sales per store over
time across municipalities in the treatment and control groups, respectively.
Panels (d) and (e) present fan plots of sales across municipalities in the treatment and control groups for each given time point.
The black curves represent the cross-sectional mean over time and the vertical green line indicates the date of intervention.
\vspace{0.2cm}
\end{spacing}
\centering
\includegraphics[width=\linewidth]{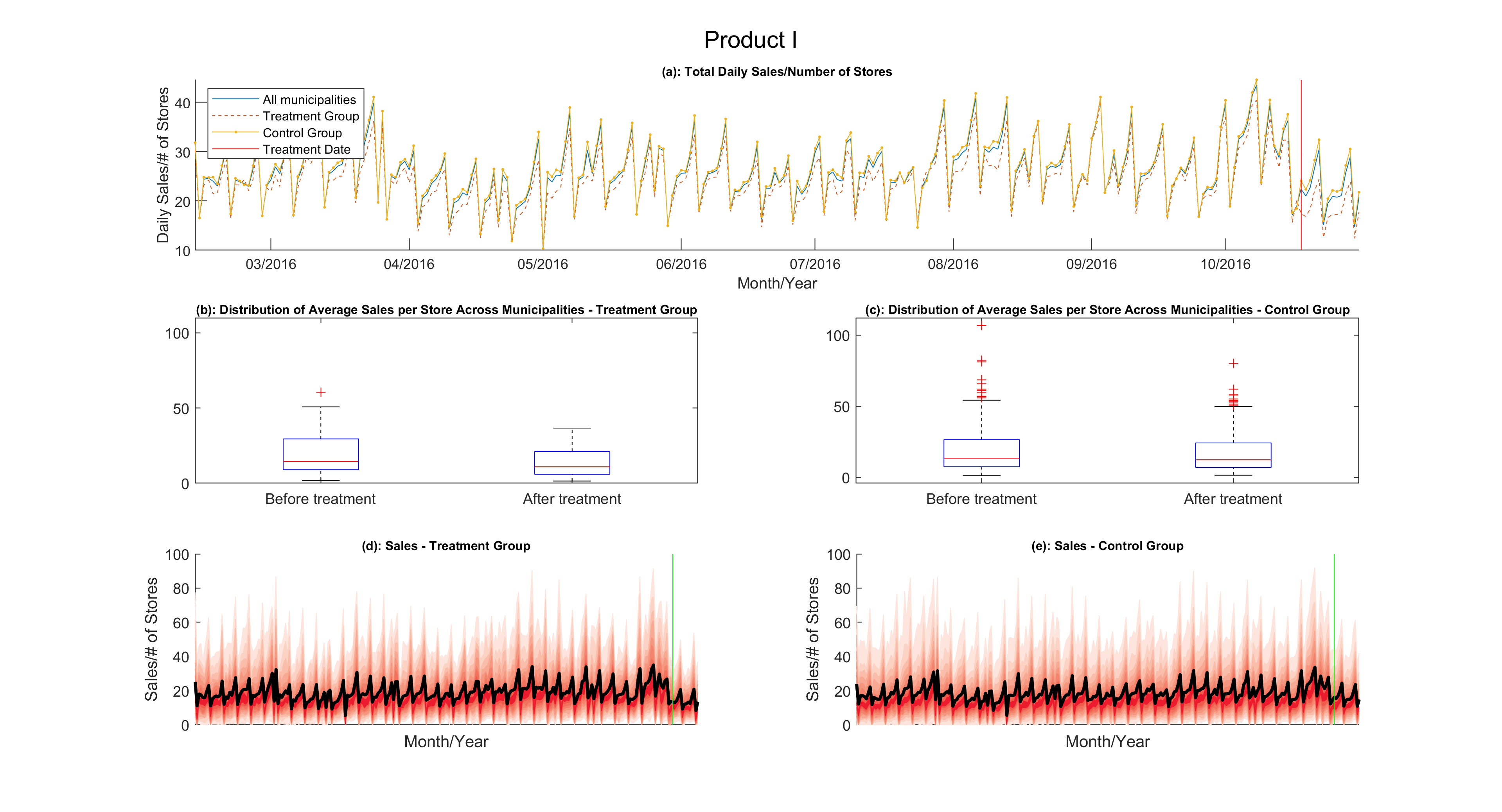}
\label{F:lasa1}
\end{figure}

Figure \ref{F:lasa1} shows the data for the first product considered in the application. The data for the remaining products are displayed in Figures \ref{F:lasa2}--\ref{F:lasa5} in the supplementary material. Panel (a) in the figures reports the daily sales at each group of municipalities (all, treatment, and control) divided by the number of stores in each group. More specifically, the plot shows the daily evolution of $q_{\textnormal{all},t\cdot}^{(j)}=\frac{1}{s}\sum_{i=1}^nq_{it}^{(j)}$, $q_{\textnormal{control},t\cdot}^{(j)}=\frac{1}{s_0}\sum_{i=1}^{n_0}q_{it}^{(j)}$, and $q_{\textnormal{treatment},t\cdot}^{(j)}=\frac{1}{s_1}\sum_{i=n_0+1}^nq_{it}^{(j)}$. The plot shows the data before and after price changes and the intervention date is represented by the horizontal line.
Panels (b) and (c) display the distribution across municipalities of the time averages of $\widetilde{q}_{it}^{(j)}$, before and after the intervention and for the treatment and control groups, respectively.
Panels (d) and (e) present fan plots for the evolution of $\widetilde{q}_{it}^{(j)}$. The black curves there represent the cross-sectional means over time.

Several facts emerge from the plots. First, the dynamics of sales change depending of the product and the sample considered. Second, there is a weekly seasonal pattern in the data which is common to all products. The big spikes for Products II and IV, observed in Panel (a), are related to major promotions. We selected these particular products and sample to illustrate that our methodology is robust to outliers. One point that deserves attention is that promotions took place in both control and treatment groups and, therefore, do not have any harmful implication to our methodology. Eyeballing the graphs in Panel (a) of Figures \ref{F:lasa1} and \ref{F:lasa5}, we observe a substantial drop in sales before the start of the experiment and happened in both control in treatment groups. This experiment clearly shows the benefits of our method in comparison, for instance, with the before-and-after (BA) estimator. BA estimator does not take into account common trends or global shocks that affects both treatment and control groups. Finally, a point to highlight concerning Products IV and V is the fact the daily sales are quite small as compared to the other three products. For instance, the average daily sales per store is less than one unit for Product V. One of the reasons for the drop in sales for Product V just before the intervention is a large drop in the number of available units in some of the municipalities. For example, in about 3\% of the municipalities, both in the treatment and control groups, there were not a single unit of the product available to be sold. As we are going to see later, this will have an impact on the results obtained for this specific product. Finally, by observing Panels (d) and (e), we notice a significant heterogeneity across municipalities.

The models are estimated at the municipal level. For each product and each municipality, we run a first-stage regression of $\widetilde{q}_{it}^{(j)}$ on seven dummies for the days of the week, a linear deterministic trend and the number of stores that are open at municipality $i$ on day $t$. For the municipalities in the control group the above regression is estimated with the full sample. For the municipalities in the treatment group we use data only up to time $T_0$. The second step consists of estimating factors for the first-stage residuals. We select the number of factors, $k$, by the eigenvalue ratio test. In the third step, we run a LASSO regression of each idiosyncratic component of treated units on the idiosyncratic terms of the control group. As described in Section \ref{S:Practice}, the penalty parameter is determined by the BIC. Finally, we compute the counterfactual for each municipality $i=1,\ldots,n_1$ for $t=T_0+1,\ldots,T$: $\widehat{\widetilde{q}}_{it}^{(j)}$. We also compute the instantaneous and average intervention impact as $\widehat{\delta}_{it}^{(j)}=\widetilde{q}_{it}^{(j)}-\widehat{\widetilde{q}}_{it}^{(j)}$ and $\widehat{\Delta}_i^{(j)}=\frac{1}{T-T_0}\sum_{t=T_0+1}^T\widehat{\delta}_{it}^{(j)}$, respectively. We test the null hypothesis of  intervention effect, $\mathcal{H}_0:\delta^{(j)}_{it}=0\,\forall t\geq T_0$, with the ressampling procedure with either $\phi(\widehat \delta_{T_0+1},\ldots, \widehat  \delta_T)=\sum_{t=T_0+1}^T \widehat  \delta_t^2$
or $\phi(\widehat \delta_{T_0+1},\ldots, \widehat \delta_T)=\sum_{t=T_0+1}^T|\widehat  \delta_t|$. We also test the for daily effects.

Under the hypothesis of linear demand function, price elasticities $\epsilon_{ij}$ for each municipality $i$ and product $j$ can be recovered as
$\widehat{\epsilon}_{ij}=\frac{\widehat\beta_{ij}p_{ij,T_0-1}}{\overline{Q}_{ij}}$, where $\widehat{\beta}_{ij}=\frac{\widehat{\Delta}_{ij}}{N_i\Delta_{p_j}}$, $\widehat{\Delta}_{ij}$ is the estimated average effect for municipality $i$ and product $j$, $N_i$ is the number of stores, $\Delta_{p_j}$ is the price change, $p_{ij,T_0-1}$ is the price before the intervention and $\overline{Q}_{ij}$ is the average counterfactual quantity sold. Finally, optimal prices for profit maximization can be determined by:
\[
p_{ij}^*=\frac{(1-\mathsf{Taxes}_{ij})(\overline{Q}_{ij}-\widehat\beta_{ij}p_{ij,T_0-1})-\widehat\beta_{ij}\times\mathsf{Costs}_{ij}}{-2\widehat\beta_{ij}(1-\mathsf{Taxes}_{ij})},
\]
where $\mathsf{Taxes}_{ij}$ and $\mathsf{Costs}_{ij}$ are the municipality-product-specific tax and costs,respectively.

Table \ref{T:estimation} reports, for each product, the minimum, the 5\%-, 25\%-, 50\%-, 75\%-, and 95\%-quantiles, maximum, average, and standard deviation for several statistics. We consider the distribution over the all treated municipalities. In Panel (a) in the table we report the results for the R-squared of the pre-intervention model. Panel (b) displays the average intervention effect over the experiment period. Panels (c) and (d) depict the results for the $p$-values of the ressampling test described in Section \ref{S:Inference} for the null hypothesis of no intervention effect with the square or absolute value statistic, respectively. Panel (e) presents the results for the $p$-values of the null hypothesis of no idiosyncratic contribution. Table \ref{T:price} presents, for each product, the same descriptive statistics for the estimated elasticities and the percentage difference between the estimated optimal price and the current price. Contrary to what we show in Table \ref{T:estimation}, in Table \ref{T:price} we report only results with respect to the municipalities where the estimated average effects have the correct sign (positive when there is a price reduction and negative when there is a price increase) and are statistically significant at the 10\% level. The last column in the table shows the fraction of municipalities where the above criterium is satisfied.

Additional results are displayed in Figure \ref{F:lasa_res1} and Figures \ref{F:lasa_res2}--\ref{F:lasa_res5} in the Supplementary Material. For each product, Panel (a) in the figures displays a fan plot of the $p$-values of the ressampling test for the null hypothesis $\mathcal{H}_{0}: \delta_t = 0$ for each given $t$ after the treatment, using the test statistic $\phi(\widehat \delta_t) = |\widehat \delta_t|$, which is the same as using the test statistic $\widehat \delta_t^2$. The black curve represents the cross-sectional median across time $t$. Panel (b) shows an example for one municipality. The panel shows the actual and counterfactual sales per store for the post-treatment period. 95\% confidence intervals for the counterfactual path are also displayed.

Several facts emerge from the results. First, the average R-squared are quite high for Products II and IV and moderate for Products I and III. This fact provides some evidence that, on average, the estimated models are able to properly describe the dynamics of the sales. For Products II and IV this finding is even more pronounced as in the worst case, the R-squared are 0.4669 and 0.4028, respectively. On the other hand, the model for Product V yields very low R-squared. A potential reason for the poor fit is the fact that sales per store of Product I are very small and there are some municipalities that displays no sales in some days.

The second finding is related to the estimation of the average intervention effect ($\Delta$). As expected, the estimated mean effects ($\widehat\Delta$) have the correct sign for Products I--IV, on average. For Product I, $\Delta$ has the correct sign (negative) for 90\% of the municipalities and is statistically significant at the 10\% level in about 37\% of all treated cities. Among the cities with $\widehat\Delta>0$, in only one we find statistical significance at the 10\% level. For Product II, $\widehat\Delta$ has the correct sign for 89\% of the treated municipalities and the results are significant in about 33\% of all the treated cities. For Product III, the numbers are similar. However, in none of the cities where $\Delta$ has been estimated with the opposite sign, the effects are significant. For Product IV, the estimated average treatment effect has the correct sign in 59\% of the cities. Fortunately, in the 41 cases where the estimates have the wrong sign, the results are significant in only three of them. For Product V the estimates have the correct sign in only 35\% of the cities. However, in only four cities the results with the wrong sign are statistically significant at the 10\% level. The reasons for poor results concerning Product V are possibly twofold. First, as mentioned before, the sales per store are quite small and the in-sample fit is poor. Second, there was a stock problem around the time of the experiment. Figure \ref{F:inventory} in the Supplementary Material displays the evolution of the distribution of available product units across municipalities. From the inspection of Panel (a) in the figure it is clear that the distribution changes around the experiment dates, pointing to large decrease in stocks for Product V.

It is worth comparing the results in Panel (b) of Table \ref{T:estimation} with the ones if we use the before-and-after (BA) estimator to compute the average treatment effects. The BA estimator for each municipality is just average sales over the period after the intervention minus the average sales over the days before the intervention. The results are reported in Table \ref{T:estimation_ba} in the Supplementary Material. As expected from our previous discussion, the BA over estimates the effects of the price changes, specially for Products I and II, and yields estimates with the wrong sign for Product IV. For Product V, the BA estimates are even more negative that the ones from the \texttt{farmTreat} methodology.

A final fact from the inspection of Panel (e) is that the contribution of the idiosyncratic terms to construct the counterfactual is statistically relevant in several cases.

Now we turn attention to Table \ref{T:price}. If we focus only on the cities with estimated average effects that have the correct sign and where the such effects are statistically significant, we estimate very high elasticities on average; see Panel (a) in Table \ref{T:price}. From Panel (b), we note that on average prices must be decreased.

\section{Conclusions}\label{S:Conclusion}

In this paper we proposed a new method to estimate the effects of interventions when there is potentially only one (or just a few) treated units. The outputs of interest are observed over time for both the treated and untreated units, forming a panel of time series data. The untreated units are called peers and a counterfactual to the output of interest in the absence of intervention is constructed by writing a model relation the unit of interest to the peers. The novelty of this paper concerns how this model is constructed. We combine factor models with sparse regression on the idiosyncratic components. This model includes both the principal component regression and sparse regression on the original measurements as  specific cases. The main advantage of our proposal is that we avoid the usual assumption of (approximate) sparsity and make model selection consistency conditions easier to be satisfied. A formal test is also proposed to prove the case for using the idiosyncratic components.

In terms of practical application we show how our methodology can be used to compute optimal prices for products from the retail industry in Brazil. Our results indicate optimal prices substantially lower than the current prices adopted by the company.

\bibliography{ref}

\begin{figure}
\caption{Results for Product I.}
\begin{spacing}{}
\scriptsize
Panel (a) displays a fan plot, across $n_1$ municipalities in the treatment group, of the $p$-values of the re-sampling test
for the null $\mathscr{H}_{0}: \delta_{t} = 0$ at each time $t$ after the treatment. The black curve represents the median $p$-value
across municipalities over $t$. Panel (b) shows an example for one municipality.
The panel depicts the actual and counterfactual sales per store for the post-treatment period.
95\% confidence intervals for the counterfactual path is also displayed.
\vspace{0.3cm}
\end{spacing}
\centering
\includegraphics[width=\linewidth]{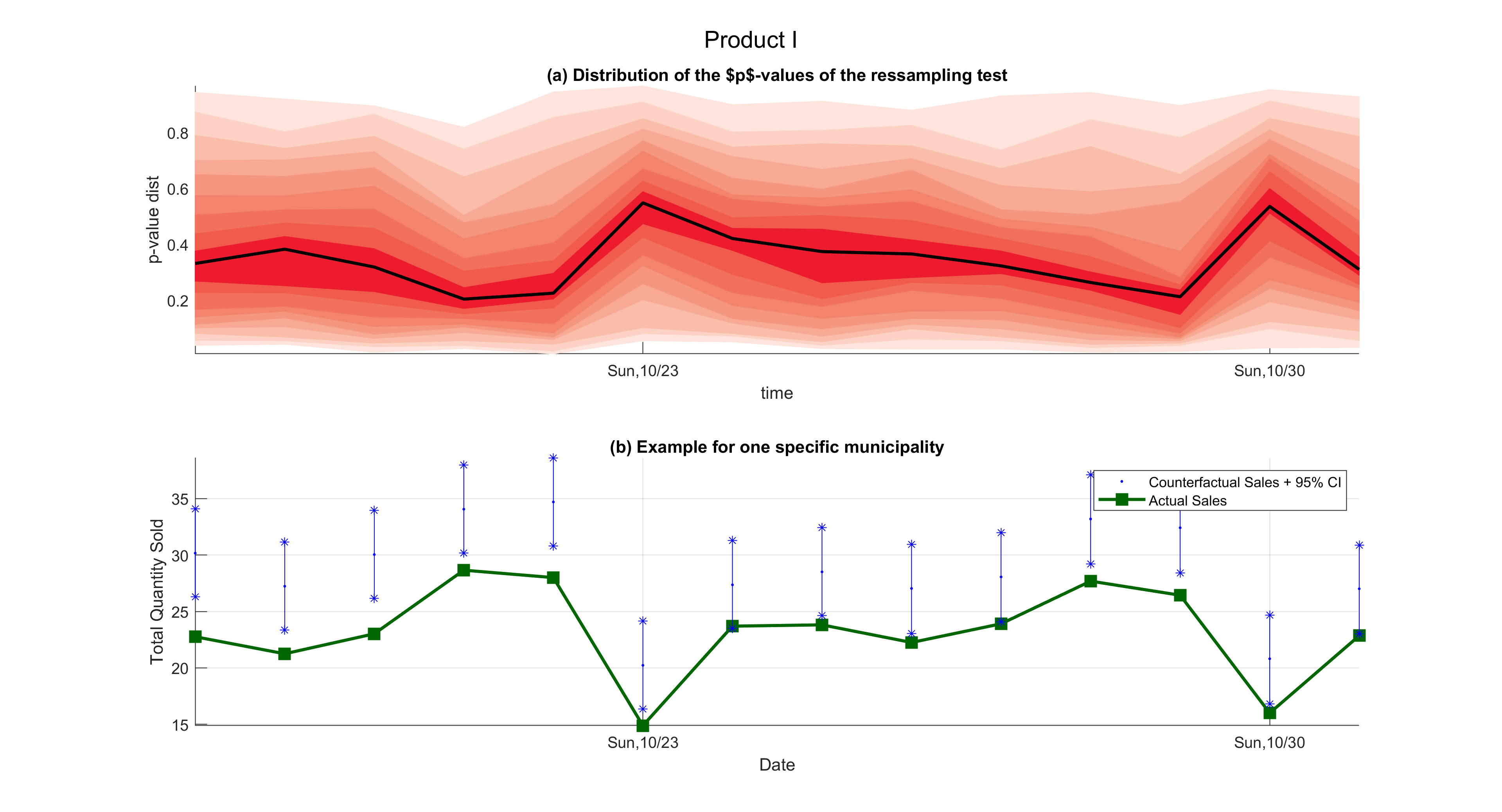}
\label{F:lasa_res1}
\end{figure}

\begin{table}[htbp]
\centering
\caption{\textbf{Average Treatment ($\Delta$) Estimation under the Null.}}
\label{T:simul_est_1}
\begin{minipage}{0.9\linewidth}
\begin{footnotesize}
The table reports descriptive statistics for the average treatment estimation under the null of no effect ($\delta_{T_0+1}=0$). The table reports the mean, median, and mean squared error (MSE) of the estimator $\widehat{\Delta}$ for one post-intervention observation. Panel (a) considers the case where the counterfactual is estimated by a LASSO regression of the treated unit on all the peers. This is the Artificial Counterfactual (ArCo) approach proposed by \citet*{cCrMmM2018}. Panel (b) presents the results when the counterfactual is estimated by principal component regression (PCR), i.e., an ordinary least squares (OLS) regression of the treated unit on factors computed from the pool of peers. This is equivalent to the method of \citet*{lGtM2016}. The number of factors is determined by the eigenvalue ratio test of \citet*{sAaH2013}. Finally, Panel (c) displays the results of the \texttt{FarmTreat} methodology. Between brackets we report the same statistics but with the model estimated using the full sample as advocated by \citet*{vCkWyZ2020a}.
\end{footnotesize}
\end{minipage}
\resizebox{0.9\linewidth}{!}{
\begin{threeparttable}
\begin{tabular}{ccccccccccccc}
\hline
    && \multicolumn{11}{c}{\textbf{\underline{Panel(a): LASSO (ArCo) - \citet*{cCrMmM2018}}}}\\
    && \multicolumn{3}{c}{\underline{Mean}} && \multicolumn{3}{c}{\underline{Median}}  && \multicolumn{3}{c}{\underline{MSE}} \\
    && $n=T$ & $n=2\times T$ & $n=3\times T$ && $n=T$ & $n=2\times T$ & $n=3\times T$ && $n=T$ & $n=2\times T$ & $n=3\times T$  \\
\cline{3-5}
\cline{7-9}
\cline{11-13}
\multirow{ 2}{*}{$T=50$}  &&  -0.036  &  0.010   &  0.023   &&  -0.073  &   0.023  &  0.074   &&  1.114  &  0.975  &  0.901  \\
                          &&  [0.017] & [0.003]  & [0.000]  &&  [0.024] & [-0.004] & [0.008]  && [0.254] & [0.178] & [0.188] \\
\multirow{ 2}{*}{$75$}    &&  -0.084  &   0.025  &  -0.011  &&  -0.080  &   0.025  &  0.015   &&  0.832  &  0.785  &  0.763  \\
                          && [-0.027] & [-0.018] & [-0.011] && [-0.035] & [-0.018] & [0.001]  && [0.383] & [0.216] & [0.206] \\
\multirow{ 2}{*}{$100$}   &&  -0.016  &   0.068  &   0.026  && -0.021   &  0.079   &  0.028   &&  0.732  &  0.674  &  0.632  \\
                          && [-0.020] & [-0.005] & [-0.022] && [0.014]  & [0.014]  & [0.009]  && [0.317] & [0.311] & [0.264] \\
\multirow{ 2}{*}{$150$}   &&  0.004   &   0.037  & -0.070   &&   0.021  &   0.038  &  -0.083  &&  0.608  &  0.655  &  0.590  \\
                          && [-0.000] & [-0.020] & [0.022]  && [-0.016] & [-0.037] & [-0.012] && [0.362] & [0.288] & [0.349] \\
\multirow{ 2}{*}{$250$}   &&  -0.013  &  -0.026  & -0.028   &&   0.021  &  -0.040  &  -0.039  &&  0.539  &  0.517  &  0.566  \\
                          && [-0.000] & [-0.020] & [0.022]  && [-0.016] & [-0.037] & [-0.012] && [0.362] & [0.288] & [0.349] \\
\multirow{ 2}{*}{$500$}   &&  0.018   & -0.028   &  0.052   &&  0.032   & -0.062   &  0.042   &&  0.419  & 0.382   &  0.424 \\
                          && [0.002]  & [0.016]  & [-0.011] && [0.009]  & [0.007]  & [0.005]  && [0.321] & [0.300] & [0.280] \\
\multirow{ 2}{*}{$1,000$} &&  0.029   &  0.033   & -0.028   &&  0.047   &  0.049   &  0.005   &&  0.323  &  0.378  &  0.350 \\
                          && [-0.026] & [-0.035] & [0.029]  && [-0.053] & [-0.026] &  [0.054] && [0.274] & [0.303] & [0.275] \\
\hline
\\
    && \multicolumn{11}{c}{\textbf{\underline{Panel(b): PCR - \citet{lGtM2016}}}}\\
    && \multicolumn{3}{c}{\underline{Mean}} && \multicolumn{3}{c}{\underline{Median}}  && \multicolumn{3}{c}{\underline{MSE}} \\
     && $n=T$ & $n=2\times T$ & $n=3\times T$ && $n=T$ & $n=2\times T$ & $n=3\times T$ && $n=T$ & $n=2\times T$ & $n=3\times T$  \\
\cline{3-5}
\cline{7-9}
\cline{11-13}
\multirow{ 2}{*}{$T=50$}  &&  -0.030  &  -0.001  &  0.038   && -0.120   &  0.008   &  0.015   &&  1.259  &  0.916  &  0.842  \\
                          &&  [0.032] & [-0.013] & [0.014]  && [0.004]  & [-0.053] & [0.042]  && [0.605] & [0.665] & [0.696] \\
\multirow{ 2}{*}{$75$}    && -0.041   &  0.011   & -0.002   &&  0.014   &  0.023   & -0.005   &&  0.957  &  0.958  &  0.893  \\
                          && [-0.037] & [-0.047] & [0.008]  && [-0.004] & [-0.057] & [0.018]  && [0.727] & [0.767] & [0.670] \\
\multirow{ 2}{*}{$100$}   &&  -0.065  &   0.083  &  0.014   &&  -0.087  &  0.080   & -0.011   &&  0.989  &  0.807  & 0.863   \\
                          && [-0.019] & [-0.012] & [-0.012] && [-0.001] & [0.024]  & [0.004]  && [0.683] & [0.719] & [0.636] \\
\multirow{ 2}{*}{$150$}   &&  -0.045  &  -0.022  & -0.093   &&  -0.005  &  -0.040  & -0.083   &&  1.071  &  0.860  & 0.914   \\
                          && [-0.017] & [-0.031] & [0.036]  && [-0.004] & [-0.036] & [0.023]  && [0.732] & [0.701] & [0.772] \\
\multirow{ 2}{*}{$250$}   &&  -0.042  &  -0.045  &  -0.038  &&  -0.041  &  -0.057  &  -0.018  &&  0.982  &  0.778  &  0.861  \\
                          && [-0.008] & [-0.047] & [-0.015] && [-0.001] & [-0.007] & [-0.025] && [0.728] & [0.762] & [0.778] \\
\multirow{ 2}{*}{$500$}   &&  0.006   &  0.001   &  0.070   &&  0.066   & -0.011   &  0.033   &&  0.765  &  0.692  &  0.758  \\
                          && [-0.018] & [0.057]  & [-0.006] && [-0.009] & [0.015]  & [-0.078] && [0.740] & [0.809] & [0.751] \\
\multirow{ 2}{*}{$1,000$} &&  0.028   &  0.050   &  -0.046  &&   0.075  &   0.049  & -0.052   &&  0.720  &  0.783  &  0.763  \\
                          && [-0.024] & [-0.051] &  [0.057] && [-0.026] & [-0.053] & [0.076]  && [0.739] & [0.801] & [0.790] \\
\hline
\\
    && \multicolumn{11}{c}{\textbf{\underline{Panel(c): \texttt{FarmTreat}}}}\\
    && \multicolumn{3}{c}{\underline{Mean}} && \multicolumn{3}{c}{\underline{Median}}  && \multicolumn{3}{c}{\underline{MSE}} \\
     && $n=T$ & $n=2\times T$ & $n=3\times T$ && $n=T$ & $n=2\times T$ & $n=3\times T$ && $n=T$ & $n=2\times T$ & $n=3\times T$  \\
\cline{3-5}
\cline{7-9}
\cline{11-13}
\multirow{ 2}{*}{$T=50$}  && -0.024   & -0.036   &  0.035   && -0.080   &  -0.026  &  0.026   &&  0.964  &  0.684  &  0.630  \\
                          &&  [0.017] & [0.003]  & [0.000]  && [0.024]  & [-0.004] & [0.008]  && [0.254] & [0.178] & [0.188] \\
\multirow{ 2}{*}{$75$}    && -0.048   &  -0.017  & -0.019   &&  -0.030  &  -0.036  & -0.015   &&  0.607  &  0.529  &  0.471  \\
                          && [-0.027] & [-0.018] & [-0.011] && [-0.035] & [-0.018] & [0.001]  && [0.383] & [0.216] & [0.206] \\
\multirow{ 2}{*}{$100$}   &&  -0.011  &   0.048  &   0.014  && -0.036   &  0.061   & 0.009    &&  0.548  &  0.377  &  0.404  \\
                          && [-0.020] & [-0.005] & [-0.022] && [0.014]  & [0.014]  & [0.009]  && [0.317] & [0.311] & [0.264] \\
\multirow{ 2}{*}{$150$}   &&  -0.063  &   0.016  & -0.055   &&  -0.035  &   0.012  &  -0.063  &&  0.585  &  0.343  &  0.343  \\
                          && [-0.000] & [-0.020] & [0.022]  && [-0.016] & [-0.037] & [-0.012] && [0.362] & [0.288] & [0.349] \\
\multirow{ 2}{*}{$250$}   &&  -0.033  &  0.003   & -0.024   &&  -0.048  &   0.007  & -0.025   &&  0.453  &  0.312  &  0.301  \\
                          &&  [0.002] & [-0.022] & [0.001]  && [-0.018] & [-0.003] & [0.025]  && [0.311] & [0.313] & [0.317] \\
\multirow{ 2}{*}{$500$}   &&   0.017  & -0.021   &  0.026   &&  0.007   & -0.010   &  0.019   &&  0.301  &  0.260  & 0.269   \\
                          &&  [0.002] & [0.016]  & [-0.011] && [0.009]  & [0.007]  & [0.005]  && [0.321] & [0.300] & [0.280] \\
\multirow{ 2}{*}{$1,000$} &&  0.031   &  0.036   & -0.034   &&  0.035   & 0.068    & -0.024   &&  0.246  &  0.291  & 0.263   \\
                          && [-0.026] & [-0.035] & [0.029]  && [-0.053] & [-0.026] & [0.054]  && [0.274] & [0.303] & [0.275] \\
\hline
\end{tabular}
\end{threeparttable}}
\end{table}

\begin{table}[htbp]
\centering
\caption{\textbf{Average Treatment ($\Delta$) Estimation under the Alternative.}}
\label{T:simul_est_2}
\begin{minipage}{0.9\linewidth}
\begin{footnotesize}
The table reports descriptive statistics for the average treatment estimation under the null of no effect ($\delta_{T_0+1}=2$). The table reports the mean, median, and mean squared error (MSE) of the estimator $\widehat{\Delta}$ for one post-intervention observation. Panel (a) considers the case where the counterfactual is estimated by a LASSO regression of the treated unit on all the peers. This is the Artificial Counterfactual (ArCo) approach proposed by \citet{cCrMmM2018}. Panel (b) presents the results when the counterfactual is estimated by principal component regression (PCR), i.e., an ordinary least squares (OLS) regression of the treated unit on factors computed from the pool of peers. This is equivalent to the method of \citet{lGtM2016}. The number of factors is determined by the eigenvalue ratio test of \citet*{sAaH2013}. Finally, Panel (c) displays the results of the \texttt{FarmTreat} methodology. Between brackets we report the same statistics but with the model estimated using the full sample as advocated by \citet*{vCkWyZ2020a}.
\end{footnotesize}
\end{minipage}
\resizebox{0.9\linewidth}{!}{
\begin{threeparttable}
\begin{tabular}{ccccccccccccc}
\hline
    && \multicolumn{11}{c}{\textbf{\underline{Panel(a): LASSO (ArCo) - \citet{cCrMmM2018}}}}\\
    && \multicolumn{3}{c}{\underline{Mean}} && \multicolumn{3}{c}{\underline{Median}}  && \multicolumn{3}{c}{\underline{MSE}} \\
    && $n=T$ & $n=2\times T$ & $n=3\times T$ && $n=T$ & $n=2\times T$ & $n=3\times T$ && $n=T$ & $n=2\times T$ & $n=3\times T$  \\
\cline{3-5}
\cline{7-9}
\cline{11-13}
\multirow{ 2}{*}{$T=50$}  &&  1.998  &  2.048  &  2.001  &&  1.938  &  2.077  &  1.973  &&  0.982  &  0.972  &  0.812  \\
                          && [0.945] & [0.830] & [0.826] && [0.811] & [0.680] & [0.694] && [1.597] & [1.763] & [1.711] \\
\multirow{ 2}{*}{$75$}    &&  2.002  &  2.025  &  1.946  &&  1.955  &  2.012  &  1.913  &&  0.871  &  0.821  &  0.828  \\
                          && [1.132] & [1.013] & [0.993] && [1.028] & [0.882] & [0.866] && [1.216] & [1.397] & [1.372] \\
\multirow{ 2}{*}{$100$}   &&  2.003  &  1.998  &  2.087  &&  2.025  &  2.024  &  2.060  &&  0.737  &  0.691  &  0.681  \\
                          && [1.316] & [1.250] & [1.177] && [1.236] & [1.169] & [1.093] && [0.899] & [1.011] & [1.098] \\
\multirow{ 2}{*}{$150$}   &&  2.014  &  2.015  &  1.967  &&  2.051  &  2.025  &  1.985  &&  0.561  &  0.617  &  0.587  \\
                          && [1.486] & [1.435] & [1.323] && [1.458] & [1.376] & [1.262] && [0.670] & [0.722] & [0.853]  \\
\multirow{ 2}{*}{$250$}   &&  2.037  &  1.989  &  2.033  &&  2.033  &  2.061  &  2.022  &&  0.497  &  0.550  &  0.491  \\
                          && [1.581] & [1.532] & [1.526] && [1.593] & [1.509] & [1.476] && [0.535] & [0.564] & [0.631] \\
\multirow{ 2}{*}{$500$}   &&  2.047  &  2.022  &  1.960  &&  2.043  &  2.036  &  1.949  &&  0.388  &  0.392  &  0.383  \\
                          && [1.717] & [1.639] & [1.696] && [1.719] & [1.638] & [1.694] && [0.392] & [0.394] & [0.399] \\
\multirow{ 2}{*}{$1,000$} &&  1.974  &  2.014  &  1.969  &&  1.957  &  2.036  &  1.987  &&  0.380  &  0.334  &  0.378  \\
                          && [1.810] & [1.785] & [1.715] && [1.815] & [1.792] & [1.725] && [0.297] & [0.341] & [0.350]  \\
\hline
\\
    && \multicolumn{11}{c}{\textbf{\underline{Panel(b): PCR - \citet{lGtM2016}}}}\\
    && \multicolumn{3}{c}{\underline{Mean}} && \multicolumn{3}{c}{\underline{Median}}  && \multicolumn{3}{c}{\underline{MSE}} \\
     && $n=T$ & $n=2\times T$ & $n=3\times T$ && $n=T$ & $n=2\times T$ & $n=3\times T$ && $n=T$ & $n=2\times T$ & $n=3\times T$  \\
\cline{3-5}
\cline{7-9}
\cline{11-13}
\multirow{ 2}{*}{$T=50$}  &&  1.941  &  2.071  &  2.025  &&   1.981  &  2.100  &  1.932  &&  1.150  &  0.944  &  0.897  \\
                          && [1.524] & [1.718] & [1.757] &&  [1.527] & [1.682] & [1.732] && [1.161] & [0.717] & [0.652] \\
\multirow{ 2}{*}{$75$}    &&  1.998  &  2.004  &  1.990  &&   1.998  &  2.020  &  1.988  &&  1.237  &  0.955  &  0.946   \\
                          && [1.637] & [1.813] & [1.789] &&  [1.716] & [1.812] & [1.786] && [0.985] & [0.793] & [0.741]\\
\multirow{ 2}{*}{$100$}   &&  2.019  &  1.962  &  2.061  &&   1.977  &  1.937  &  2.024  &&  1.050  &  0.920  &  0.810    \\
                          && [1.662] & [1.929] & [1.857] &&  [1.708] & [1.926] & [1.820] && [1.028] & [0.689] & [0.750]\\
\multirow{ 2}{*}{$150$}   &&  1.995  &  1.988  &  1.954  &&   1.997  &  1.978  &  1.937  &&  0.941  &  0.838  &  0.790    \\
                          && [1.776] & [1.867] & [1.806] &&  [1.871] & [1.859] & [1.802] && [1.000] & [0.772] & [0.838]\\
\multirow{ 2}{*}{$250$}   &&  2.032  &  1.970  &  2.016  &&   2.009  &  1.983  &  1.979  &&  0.843  &  0.802  &  0.723   \\
                          && [1.893] & [1.893] & [1.901] &&  [1.863] & [1.890] & [1.921] && [0.837] & [0.737] & [0.814]\\
\multirow{ 2}{*}{$500$}   &&  2.013  &  2.074  &  1.964  &&   2.025  &  2.047  &  1.976  &&  0.777  &  0.758  &  0.731   \\
                          && [1.969] & [1.969] & [2.037] &&  [2.029] & [1.971] & [1.997] && [0.744] & [0.669] & [0.731]\\
s\multirow{ 2}{*}{$1,000$} &&  2.026  &  2.029  &  1.936  &&   2.055  &  2.046  &  1.965  &&  0.786  &  0.765  &  0.800   \\
                          && [2.010] & [2.033] & [1.939] &&  [2.062] & [2.037] & [1.933] && [0.714] & [0.770] & [0.751] \\
\hline
\\
    && \multicolumn{11}{c}{\textbf{\underline{Panel(c): \texttt{FarmTreat}}}}\\
    && \multicolumn{3}{c}{\underline{Mean}} && \multicolumn{3}{c}{\underline{Median}}  && \multicolumn{3}{c}{\underline{MSE}} \\
     && $n=T$ & $n=2\times T$ & $n=3\times T$ && $n=T$ & $n=2\times T$ & $n=3\times T$ && $n=T$ & $n=2\times T$ & $n=3\times T$  \\
\cline{3-5}
\cline{7-9}
\cline{11-13}
\multirow{ 2}{*}{$T=50$}  &&  1.933  &  2.052  &  2.044  &&  1.938  &  2.062  &  2.031  &&  0.800  &  0.712  &  0.633   \\
                          && [0.986] & [0.865] & [0.796] && [0.982] & [0.603] & [0.509] && [1.595] & [1.713] & [1.874] \\
\multirow{ 2}{*}{$75$}    &&  1.995  &  2.010  &  1.985  &&  1.990  &  2.007  &  1.996  &&  0.913  &  0.503  &  0.510   \\
                          && [1.423] & [1.508] & [1.439] && [1.515] & [1.533] & [1.459] && [0.828] & [0.689] & [0.745] \\
\multirow{ 2}{*}{$100$}   &&  2.007  &  1.982  &  2.080  &&  2.000  &  1.965  &  2.094  &&  0.565  &  0.418  &  0.395   \\
                          && [1.597] & [1.765] & [1.752] && [1.639] & [1.772] & [1.769] && [0.660] & [0.366] & [0.363] \\
\multirow{ 2}{*}{$150$}   &&  1.997  &  2.003  &  1.986  &&  2.050  &  2.029  &  1.962  &&  0.509  &  0.320  &  0.311   \\
                          && [1.701] & [1.843] & [1.789] && [1.770] & [1.833] & [1.785] && [0.567] & [0.303] & [0.322] \\
\multirow{ 2}{*}{$250$}   &&  2.019  &  2.000  &  2.033  &&  2.017  &  2.000  &  2.014  &&  0.363  &  0.322  &  0.294   \\
                          && [1.875] & [1.908] & [1.887] && [1.865] & [1.919] & [1.892] && [0.332] & [0.288] & [0.279] \\
\multirow{ 2}{*}{$500$}   &&  2.037  &  2.005  &  1.964  &&  2.049  &  2.023  &  1.981  &&  0.262  &  0.247  &  0.262   \\
                          && [1.967] & [1.933] & [1.980] && [1.976] & [1.938] & [1.978] && [0.290] & [0.251] & [0.256] \\
\multirow{ 2}{*}{$1,000$} &&  1.993  &  2.008  &  1.985  &&  2.005  &  2.003  &  1.968  &&  0.278  &  0.264  &  0.271   \\
                          && [2.007] & [2.013] & [1.945] && [2.008] & [2.042] & [1.957] && [0.245] & [0.283] & [0.259] \\
\hline
\end{tabular}
\end{threeparttable}}
\end{table}

\begin{table}[htbp]
\centering
\caption{\textbf{Rejection Rates under the Null (empirical size)}}
\label{T:simul_size}
\begin{minipage}{0.9\linewidth}
\begin{footnotesize}
The table reports the rejection rates of the ressampling test under the null. Panel (a) considers the case where the counterfactual is estimated by a LASSO regression of the treated unit on all the peers. This is the Artificial Counterfactual (ArCo) approach proposed by \citet{cCrMmM2018}. Panel (b) presents the results when the counterfactual is estimated by principal component regression (PCR), i.e., an ordinary least squares (OLS) regression of the treated unit on factors computed from the pool of peers. This is equivalent to the method of \citet{lGtM2016}. The number of factors is determined by the eigenvalue ratio test of \citet*{sAaH2013}. Finally, Panel (c) displays the results of the \texttt{FarmTreat} methodology. Between brackets we report the rejection rates but with the model estimated using the full sample as advocated by \citet*{vCkWyZ2020a}.
\end{footnotesize}
\end{minipage}
\resizebox{0.9\linewidth}{!}{
\begin{threeparttable}
\begin{tabular}{ccccccccccccc}
\hline
    && \multicolumn{11}{c}{\textbf{\underline{Panel(a): LASSO (ArCo) - \citet{cCrMmM2018}}}}\\
    && \multicolumn{3}{c}{\underline{$\alpha=0.01$}} && \multicolumn{3}{c}{\underline{$\alpha=0.05$}}  && \multicolumn{3}{c}{\underline{$\alpha=0.10$}} \\
    && $n=T$ & $n=2\times T$ & $n=3\times T$ && $n=T$ & $n=2\times T$ & $n=3\times T$ && $n=T$ & $n=2\times T$ & $n=3\times T$  \\
\cline{3-5}
\cline{7-9}
\cline{11-13}
\multirow{ 2}{*}{$T=50$}  &&  0.294  &  0.398  &  0.326  &&  0.398  &  0.490  &  0.430  &&  0.452  &  0.562  &  0.492 \\
                          && [0.028] & [0.028] & [0.026] && [0.052] & [0.070] & [0.072] && [0.096] & [0.102] & [0.110]\\
\multirow{ 2}{*}{$75$}    &&  0.156  &  0.244  &  0.254  &&  0.260  &  0.408  &  0.372  &&  0.354  &  0.476  &  0.452 \\
                          && [0.022] & [0.026] & [0.018] && [0.054] & [0.064] & [0.062] && [0.134] & [0.114] & [0.110]\\
\multirow{ 2}{*}{$100$}   &&  0.096  &  0.160  &  0.220  &&  0.210  &  0.282  &  0.316  &&  0.288  &  0.366  &  0.394 \\
                          && [0.016] & [0.024] & [0.010] && [0.050] & [0.078] & [0.056] && [0.092] & [0.124] & [0.086]\\
\multirow{ 2}{*}{$150$}   &&  0.090  &  0.114  &  0.118  &&  0.166  &  0.228  &  0.220  &&  0.252  &  0.304  &  0.290 \\
                          && [0.010] & [0.012] & [0.014] && [0.046] & [0.044] & [0.052] && [0.104] & [0.086] & [0.118]\\
\multirow{ 2}{*}{$250$}   &&  0.064  &  0.050  &  0.060  &&  0.146  &  0.146  &  0.142  &&  0.198  &  0.218  &  0.230 \\
                          && [0.010] & [0.014] & [0.014] && [0.044] & [0.052] & [0.060] && [0.092] & [0.116] & [0.116]\\
\multirow{ 2}{*}{$500$}   &&  0.032  &  0.024  &  0.040  &&  0.110  &  0.102  &  0.108  &&  0.172  &  0.150  &  0.176 \\
                          && [0.016] & [0.014] & [0.004] && [0.062] & [0.052] & [0.046] && [0.112] & [0.106] & [0.102]\\
\multirow{ 2}{*}{$1,000$} &&  0.012  &  0.024  &  0.026  &&  0.068  &  0.096  &  0.082  &&  0.122  &  0.166  &  0.160 \\
                          && [0.010] & [0.022] & [0.014] && [0.048] & [0.048] & [0.054] && [0.088] & [0.110] & [0.106]\\
\hline
\\
    && \multicolumn{11}{c}{\textbf{\underline{Panel(b): PCR - \citet{lGtM2016}}}}\\
    && \multicolumn{3}{c}{\underline{$\alpha=0.01$}} && \multicolumn{3}{c}{\underline{$\alpha=0.05$}}  && \multicolumn{3}{c}{\underline{$\alpha=0.10$}} \\
    && $n=T$ & $n=2\times T$ & $n=3\times T$ && $n=T$ & $n=2\times T$ & $n=3\times T$ && $n=T$ & $n=2\times T$ & $n=3\times T$  \\
\cline{3-5}
\cline{7-9}
\cline{11-13}
\multirow{ 2}{*}{$T=50$}  && 0.152  &  0.042  &  0.040  &&  0.216  &  0.104  &  0.080  &&  0.242  &  0.162  &  0.126 \\
                          &&[0.018] & [0.020] & [0.022] && [0.040] & [0.064] & [0.052] && [0.084] & [0.108] & [0.102]\\
\multirow{ 2}{*}{$75$}    && 0.100  &  0.032  &  0.022  &&  0.134  &  0.122  &  0.068  &&  0.194  &  0.184  &  0.150 \\
                          &&[0.018] & [0.008] & [0.012] && [0.064] & [0.064] & [0.066] && [0.130] & [0.130] & [0.100]\\
\multirow{ 2}{*}{$100$}   && 0.086  &  0.012  &  0.010  &&  0.138  &  0.060  &  0.066  &&  0.194  &  0.100  &  0.132 \\
                          &&[0.006] & [0.008] & [0.008] && [0.056] & [0.058] & [0.044] && [0.110] & [0.110] & [0.082]\\
\multirow{ 2}{*}{$150$}   && 0.084  &  0.020  &  0.024  &&  0.128  &  0.078  &  0.088  &&  0.176  &  0.118  &  0.144 \\
                          &&[0.012] & [0.010] & [0.016] && [0.046] & [0.040] & [0.070] && [0.116] & [0.102] & [0.120]\\
\multirow{ 2}{*}{$250$}   && 0.026  &  0.014  &  0.026  &&  0.080  &  0.052  &  0.078  &&  0.128  &  0.112  &  0.130 \\
                          &&[0.010] & [0.016] & [0.014] && [0.038] & [0.052] & [0.060] && [0.102] & [0.106] & [0.104]\\
\multirow{ 2}{*}{$500$}   && 0.018  &  0.010  &  0.010  &&  0.060  &  0.046  &  0.048  &&  0.110  &  0.084  &  0.122 \\
                          &&[0.014] & [0.016] & [0.006] && [0.036] & [0.062] & [0.060] && [0.090] & [0.100] & [0.118]\\
\multirow{ 2}{*}{$1,000$} && 0.008  &  0.002  &  0.010  &&  0.050  &  0.056  &  0.052  &&  0.096  &  0.102  &  0.104 \\
                          &&[0.012] & [0.018] & [0.012] && [0.058] & [0.064] & [0.054] && [0.084] & [0.114] & [0.110]\\
\hline
\\
    && \multicolumn{11}{c}{\textbf{\underline{Panel(c): \texttt{FarmTreat}}}}\\
    && \multicolumn{3}{c}{\underline{$\alpha=0.01$}} && \multicolumn{3}{c}{\underline{$\alpha=0.05$}}  && \multicolumn{3}{c}{\underline{$\alpha=0.10$}} \\
    && $n=T$ & $n=2\times T$ & $n=3\times T$ && $n=T$ & $n=2\times T$ & $n=3\times T$ && $n=T$ & $n=2\times T$ & $n=3\times T$  \\
\cline{3-5}
\cline{7-9}
\cline{11-13}
\multirow{ 2}{*}{$T=50$}  && 0.332  &  0.400  &  0.362  &&  0.388  &  0.468  &  0.460  &&   0.434  &  0.532  &  0.496  \\
                          &&[0.018] & [0.020] & [0.028] && [0.052] & [0.056] & [0.058] &&  [0.080] & [0.084] & [0.092] \\
\multirow{ 2}{*}{$75$}    && 0.120  &  0.084  &  0.110  &&  0.186  &  0.182  &  0.190  &&   0.248  &  0.292  &  0.272  \\
                          &&[0.020] & [0.010] & [0.018] && [0.066] & [0.064] & [0.048] &&  [0.146] & [0.112] & [0.088] \\
\multirow{ 2}{*}{$100$}   && 0.096  &  0.028  &  0.028  &&  0.158  &  0.096  &  0.098  &&   0.208  &  0.168  &  0.160  \\
                          &&[0.004] & [0.024] & [0.016] && [0.046] & [0.058] & [0.052] &&  [0.120] & [0.118] & [0.078] \\
\multirow{ 2}{*}{$150$}   && 0.078  &  0.022  &  0.026  &&  0.156  &  0.080  &  0.084  &&   0.206  &  0.134  &  0.140  \\
                          &&[0.014] & [0.012] & [0.008] && [0.054] & [0.052] & [0.070] &&  [0.094] & [0.104] & [0.118] \\
\multirow{ 2}{*}{$250$}   && 0.028  &  0.006  &  0.022  &&  0.096  &  0.066  &  0.070  &&   0.134  &  0.136  &  0.132  \\
                          &&[0.014] & [0.012] & [0.014] && [0.046] & [0.044] & [0.056] &&  [0.102] & [0.098] & [0.094] \\
\multirow{ 2}{*}{$500$}   && 0.014  &  0.010  &  0.022  &&  0.052  &  0.058  &  0.044  &&   0.124  &  0.090  &  0.092  \\
                          &&[0.010] & [0.012] & [0.004] && [0.058] & [0.052] & [0.044] &&  [0.124] & [0.118] & [0.094] \\
\multirow{ 2}{*}{$1,000$} && 0.008  &  0.016  &  0.012  &&  0.058  &  0.054  &  0.060  &&   0.092  &  0.124  &  0.114  \\
                          &&[0.012] & [0.022] & [0.008] && [0.052] & [0.060] & [0.050] &&  [0.110] & [0.114] & [0.098] \\
\hline
\end{tabular}
\end{threeparttable}}
\end{table}

\begin{table}[htbp]
\centering
\caption{\textbf{Rejection Rates under the Alternative (empirical power)}}
\label{T:simul_power}
\begin{minipage}{0.9\linewidth}
\begin{footnotesize}
The table reports the rejection rates of the ressampling test under the alternative. Panel (a) considers the case where the counterfactual is estimated by a LASSO regression of the treated unit on all the peers. This is the Artificial Counterfactual (ArCo) approach proposed by \citet{cCrMmM2018}. Panel (b) presents the results when the counterfactual is estimated by principal component regression (PCR), i.e., an ordinary least squares (OLS) regression of the treated unit on factors computed from the pool of peers. This is equivalent to the method of \citet{lGtM2016}. The number of factors is determined by the eigenvalue ratio test of \citet*{sAaH2013}. Finally, Panel (c) displays the results of the \texttt{FarmTreat} methodology. Between brackets we report the rejection rates but with the model estimated using the full sample as advocated by \citet*{vCkWyZ2020a}.
\end{footnotesize}
\end{minipage}
\resizebox{0.9\linewidth}{!}{
\begin{threeparttable}
\begin{tabular}{ccccccccccccc}
\hline
    && \multicolumn{11}{c}{\textbf{\underline{Panel(a): LASSO (ArCo) - \citet{cCrMmM2018}}}}\\
    && \multicolumn{3}{c}{\underline{$\alpha=0.01$}} && \multicolumn{3}{c}{\underline{$\alpha=0.05$}}  && \multicolumn{3}{c}{\underline{$\alpha=0.10$}} \\
    && $n=T$ & $n=2\times T$ & $n=3\times T$ && $n=T$ & $n=2\times T$ & $n=3\times T$ && $n=T$ & $n=2\times T$ & $n=3\times T$  \\
\cline{3-5}
\cline{7-9}
\cline{11-13}
\multirow{ 2}{*}{$T=50$}  && 0.764 &  0.826 &  0.856 &&  0.850 &  0.894 &  0.906 &&  0.894 &  0.920 &  0.926 \\
                          &&[0.394]& [0.448]& [0.448]&& [0.514]& [0.562]& [0.596]&& [0.616]& [0.626]& [0.678]\\
\multirow{ 2}{*}{$75$}    && 0.728 &  0.812 &  0.806 &&  0.830 &  0.886 &  0.882 &&  0.878 &  0.912 &  0.918 \\
                          &&[0.412]& [0.462]& [0.494]&& [0.606]& [0.642]& [0.660]&& [0.702]& [0.754]& [0.750]\\
\multirow{ 2}{*}{$100$}   && 0.744 &  0.800 &  0.816 &&  0.858 &  0.886 &  0.888 &&  0.906 &  0.914 &  0.916 \\
                          &&[0.464]& [0.514]& [0.540]&& [0.646]& [0.722]& [0.728]&& [0.736]& [0.804]& [0.820]\\
\multirow{ 2}{*}{$150$}   && 0.778 &  0.766 &  0.766 &&  0.892 &  0.870 &  0.858 &&  0.908 &  0.910 &  0.900 \\
                          &&[0.596]& [0.586]& [0.554]&& [0.756]& [0.736]& [0.728]&& [0.824]& [0.830]& [0.812]\\
\multirow{ 2}{*}{$250$}   && 0.812 &  0.780 &  0.808 &&  0.912 &  0.878 &  0.892 &&  0.946 &  0.920 &  0.922 \\
                          &&[0.674]& [0.634]& [0.602]&& [0.824]& [0.794]& [0.780]&& [0.872]& [0.864]& [0.856]\\
\multirow{ 2}{*}{$500$}   && 0.856 &  0.854 &  0.836 &&  0.944 &  0.938 &  0.932 &&  0.960 &  0.958 &  0.964 \\
                          &&[0.744]& [0.700]& [0.756]&& [0.880]& [0.888]& [0.882]&& [0.930]& [0.932]& [0.948]\\
\multirow{ 2}{*}{$1,000$} && 0.860 &  0.878 &  0.838 &&  0.948 &  0.944 &  0.942 &&  0.962 &  0.974 &  0.972 \\
                          &&[0.808]& [0.774]& [0.772]&& [0.922]& [0.916]& [0.896]&& [0.966]& [0.954]& [0.954] \\
\hline
\\
    && \multicolumn{11}{c}{\textbf{\underline{Panel(b): PCR - \citet{lGtM2016}}}}\\
    && \multicolumn{3}{c}{\underline{$\alpha=0.01$}} && \multicolumn{3}{c}{\underline{$\alpha=0.05$}}  && \multicolumn{3}{c}{\underline{$\alpha=0.10$}} \\
    && $n=T$ & $n=2\times T$ & $n=3\times T$ && $n=T$ & $n=2\times T$ & $n=3\times T$ && $n=T$ & $n=2\times T$ & $n=3\times T$  \\
\cline{3-5}
\cline{7-9}
\cline{11-13}
\multirow{ 2}{*}{$T=50$}  && 0.488 &  0.510 &  0.448 &&  0.654 &  0.676 &  0.660 &&  0.728 &  0.744 &  0.740 \\
                          &&[0.266]& [0.308]& [0.344]&& [0.450]& [0.542]& [0.562]&& [0.560]& [0.644]& [0.668]\\
\multirow{ 2}{*}{$75$}    && 0.448 &  0.414 &  0.426 &&  0.658 &  0.642 &  0.622 &&  0.754 &  0.750 &  0.722 \\
                          &&[0.282]& [0.322]& [0.328]&& [0.488]& [0.576]& [0.570]&& [0.624]& [0.706]& [0.700]\\
\multirow{ 2}{*}{$100$}   && 0.400 &  0.338 &  0.390 &&  0.624 &  0.610 &  0.646 &&  0.752 &  0.710 &  0.744 \\
                          &&[0.264]& [0.320]& [0.290]&& [0.504]& [0.604]& [0.570]&& [0.626]& [0.720]& [0.706]\\
\multirow{ 2}{*}{$150$}   && 0.464 &  0.418 &  0.398 &&  0.672 &  0.632 &  0.630 &&  0.764 &  0.740 &  0.738 \\
                          &&[0.372]& [0.362]& [0.346]&& [0.574]& [0.574]& [0.560]&& [0.676]& [0.684]& [0.680]\\
\multirow{ 2}{*}{$250$}   && 0.412 &  0.414 &  0.400 &&  0.654 &  0.642 &  0.650 &&  0.752 &  0.744 &  0.754 \\
                          &&[0.354]& [0.368]& [0.378]&& [0.594]& [0.604]& [0.612]&& [0.704]& [0.710]& [0.708]\\
\multirow{ 2}{*}{$500$}   && 0.392 &  0.434 &  0.376 &&  0.650 &  0.666 &  0.628 &&  0.766 &  0.788 &  0.750 \\
                          &&[0.374]& [0.360]& [0.386]&& [0.644]& [0.628]& [0.664]&& [0.760]& [0.752]& [0.760]\\
\multirow{ 2}{*}{$1,000$} && 0.412 &  0.434 &  0.362 &&  0.668 &  0.640 &  0.604 &&  0.760 &  0.754 &  0.702 \\
                          &&[0.418]& [0.436]& [0.344]&& [0.640]& [0.630]& [0.628]&& [0.734]& [0.746]& [0.744]\\
\hline
\\
    && \multicolumn{11}{c}{\textbf{\underline{Panel(c): \texttt{FarmTreat}}}}\\
    && \multicolumn{3}{c}{\underline{$\alpha=0.01$}} && \multicolumn{3}{c}{\underline{$\alpha=0.05$}}  && \multicolumn{3}{c}{\underline{$\alpha=0.10$}} \\
    && $n=T$ & $n=2\times T$ & $n=3\times T$ && $n=T$ & $n=2\times T$ & $n=3\times T$ && $n=T$ & $n=2\times T$ & $n=3\times T$  \\
\cline{3-5}
\cline{7-9}
\cline{11-13}
\multirow{ 2}{*}{$T=50$}  && 0.826 &  0.854 &  0.884 &&  0.892 &  0.920 &  0.938 &&  0.912 &  0.940 &  0.946 \\
                          &&[0.458]& [0.586]& [0.626]&& [0.602]& [0.712]& [0.768]&& [0.680]& [0.776]& [0.838]\\
\multirow{ 2}{*}{$75$}    && 0.774 &  0.812 &  0.800 &&  0.894 &  0.906 &  0.912 &&  0.940 &  0.954 &  0.950 \\
                          &&[0.618]& [0.680]& [0.676]&& [0.752]& [0.834]& [0.858]&& [0.852]& [0.902]& [0.920]\\
\multirow{ 2}{*}{$100$}   && 0.758 &  0.800 &  0.818 &&  0.886 &  0.916 &  0.958 &&  0.934 &  0.950 &  0.980 \\
                          &&[0.662]& [0.752]& [0.762]&& [0.816]& [0.910]& [0.894]&& [0.868]& [0.956]& [0.940]\\
\multirow{ 2}{*}{$150$}   && 0.852 &  0.862 &  0.878 &&  0.950 &  0.960 &  0.964 &&  0.978 &  0.972 &  0.976 \\
                          &&[0.764]& [0.832]& [0.834]&& [0.860]& [0.946]& [0.928]&& [0.902]& [0.972]& [0.952]\\
\multirow{ 2}{*}{$250$}   && 0.884 &  0.872 &  0.908 &&  0.952 &  0.962 &  0.968 &&  0.972 &  0.974 &  0.994 \\
                          &&[0.872]& [0.872]& [0.874]&& [0.940]& [0.956]& [0.948]&& [0.968]& [0.980]& [0.976]\\
\multirow{ 2}{*}{$500$}   && 0.918 &  0.910 &  0.878 &&  0.978 &  0.976 &  0.974 &&  0.990 &  0.986 &  0.988 \\
                          &&[0.878]& [0.912]& [0.908]&& [0.964]& [0.968]& [0.976]&& [0.984]& [0.986]& [0.984]\\
\multirow{ 2}{*}{$1,000$} && 0.898 &  0.916 &  0.924 &&  0.974 &  0.978 &  0.970 &&  0.988 &  0.988 &  0.990 \\
                          &&[0.918]& [0.890]& [0.888]&& [0.974]& [0.974]& [0.968]&& [0.990]& [0.984]& [0.988]\\
\hline
\end{tabular}
\end{threeparttable}}
\end{table}

\begin{table}[htbp]
\caption{\textbf{Experiments.}}\label{T:experiments}
\begin{minipage}{0.9\linewidth}
\begin{footnotesize}
The table shows, for each product considered in the paper, the sample, the period when the experiment was carried out,
the type of the experiment (price increase or decrease), the magnitude of the price change, and the number of cities in the control
and treatment groups.
\end{footnotesize}
\end{minipage}
\resizebox{0.9\linewidth}{!}{
\begin{threeparttable}
\begin{tabular}{ccccccc}
\hline
\textbf{Product} & \textbf{Sample}            & \textbf{Experiment Period} & \textbf{Experiment Type} & \textbf{Magnitude} & \textbf{Control Group} & \textbf{Treatment Group}\\
\hline
I       & Feb-13-2016 -- Oct-31-2016 & Oct-18-2016 -- Oct-31-2016 & Price increase  & 10.58\%  & 318 & 97  \\
II      & May-14-2016 -- Jan-23-2017 & Jan-17-2017 -- Jan-23-2017 & Price increase  & 5.01\%   & 321 & 102 \\
III     & Feb-13-2016 -- Oct-31-2016 & Oct-18-2016 -- Oct-31-2016 & Price increase  & 20.13\%  & 309 & 106 \\
IV      & May-14-2016 -- Jan-23-2017 & Jan-17-2017 -- Jan-23-2017 & Price reduction & 18.20\%  & 321 & 100 \\
V       & Aug-14-2016 -- May-02-2017 & Apr-19-2017 -- May-02-2017 & Price reduction & 10.05\%  & 328 & 110 \\
\hline
\end{tabular}
\end{threeparttable}}
\end{table}

\begin{table}[htbp]
\caption{\textbf{Results: Estimation and Inference.}}\label{T:estimation}
\begin{minipage}{0.9\linewidth}
\begin{footnotesize}
The table reports estimation results.
In each panel we report, for each product, the minimum, the 5\%-, 25\%-, 50\%-, 75\%-, and 95\%-quantiles, maximum, average, and standard deviation
for a variety of different statistics. We consider the distribution over the treated municipalities.
In Panel (a) we report the results for the R-squared of the pre-intervention model.
Panel (b) displays the results for the average intervention effect over the experiment period ($\Delta$).
Panels (c) and (d) depict the results for the $p$-values of the ressampling test for the null hypothesis
$\mathcal{H}_0:\delta_t = 0, \forall t\in \{T_{0}+1,\dots, T\}$ using respectively the test statistic
$\phi(\widehat \delta_{T_0+1},\ldots, \widehat  \delta_T)=\sum_{t=T_0+1}^T \widehat  \delta_t^2$
or $\phi(\widehat \delta_{T_0+1},\ldots, \widehat \delta_T)=\sum_{t=T_0+1}^T|\widehat  \delta_t|$. 
Finally, Panel (e) reports the results for the $p$-values for the test for the idiosyncratic contribution. 
\end{footnotesize}
\end{minipage}
\resizebox{0.9\linewidth}{!}{
\begin{threeparttable}
\begin{tabular}{ccccccccccc}
\hline
\multicolumn{11}{c}{\underline{Panel (a): \textbf{R-squared}}} \\
Product && Min & 5\%-quantile & 25\%-quantile & Median & 75\%-quantile & 95\% quantile & Max & Average & Std. Dev \\
\hline
I       && 0.1134 & 0.1949 & 0.3605 & 0.4913 & 0.6215 & 0.7542 & 0.9077 & 0.4865 & 0.1749  \\
II      && 0.4669 & 0.7237 & 0.8744 & 0.9298 & 0.9552 & 0.9863 & 0.9959 & 0.8984 & 0.0926  \\
III     && 0.1190 & 0.3094 & 0.5236 & 0.7101 & 0.8341 & 0.9278 & 0.9540 & 0.6750 & 0.1981  \\
IV      && 0.4028 & 0.6744 & 0.8822 & 0.9327 & 0.9633 & 0.9865 & 0.9988 & 0.8972 & 0.1073  \\
V       && 0.0366 & 0.0527 & 0.1040 & 0.1670 & 0.2795 & 0.4258 & 0.6517 & 0.1995 & 0.1260  \\
\\
\multicolumn{11}{c}{\underline{Panel (b): \textbf{Average Treatment Effect (over time): $\Delta$}}} \\
Product && Min & 5\%-quantile & 25\%-quantile & Median & 75\%-quantile & 95\% quantile & Max & Average & Std. Dev \\
\hline
I       && -18.9451 & -16.3384 &  -7.7380 & -3.2545 & -1.1891 & 1.4580 &  3.8921 & -4.8996 &  5.3115 \\
II      && -44.4114 & -26.8298 & -14.6979 & -7.2601 & -3.6173 & 3.7216 & 42.2832 & -9.0446 & 11.1052 \\
III     && -48.6364 & -17.2678 &  -6.3314 & -2.5950 & -0.6001 & 0.7309 &  9.0886 & -4.8875 &  7.9211 \\
IV      &&  -2.9729 &  -1.7697 &  -0.4535 &  0.2842 &  1.3057 & 3.7759 &  6.5858 &  0.5660 &  1.7131 \\
V       &&  -1.5141 &  -0.9080 &  -0.4308 & -0.1544 &  0.1455 & 0.6984 &  1.6005 & -0.1440 &  0.4949 \\
\\
\multicolumn{11}{c}{\underline{Panel (c): \textbf{$p$-value of the test on squared values}}} \\
Product && Min & 5\%-quantile & 25\%-quantile & Median & 75\%-quantile & 95\% quantile & Max & Average & Std. Dev \\
\hline
I       && 0 &      0 & 0.0755 & 0.2468 & 0.5660 & 0.8623 & 0.9617 & 0.3338 & 0.2961 \\
II      && 0 & 0.0091 & 0.0826 & 0.3409 & 0.6116 & 0.9331 & 1.0000 & 0.3774 & 0.2992 \\
III     && 0 &      0 & 0.0809 & 0.2872 & 0.5702 & 0.9191 & 0.9702 & 0.3488 & 0.2913\\
IV      && 0 &      0 & 0.1178 & 0.3781 & 0.7149 & 0.9711 & 1.0000 & 0.4151 & 0.3292 \\
V       && 0 & 0.0596 & 0.3021 & 0.6234 & 0.9234 & 1.0000 & 1.0000 & 0.5998 & 0.3285 \\
\\
\multicolumn{11}{c}{\underline{Panel (d): \textbf{$p$-value of the test on absolute values}}} \\
Product && Min & 5\%-quantile & 25\%-quantile & Median & 75\%-quantile & 95\% quantile & Max & Average & Std. Dev \\
\hline
I       && 0 &      0 & 0.0457 & 0.1660 & 0.6053 & 0.8894 & 0.9957 & 0.3093 & 0.3128 \\
II      && 0 &      0 & 0.0620 & 0.2831 & 0.6074 & 0.9190 & 0.9917 & 0.3500 & 0.3059 \\
III     && 0 &      0 & 0.0511 & 0.2383 & 0.5957 & 0.9174 & 0.9787 & 0.3333 & 0.3119 \\
IV      && 0 &      0 & 0.0992 & 0.3988 & 0.6860 & 0.9463 & 1.0000 & 0.4113 & 0.3311 \\
V       && 0 & 0.0553 & 0.2511 & 0.6553 & 0.9234 & 1.0000 & 1.0000 & 0.6006 & 0.3346 \\
\\
\multicolumn{11}{c}{\underline{Panel (e): \textbf{$p$-value of the test for idiosyncratic contribution}}} \\
Product && Min & 5\%-quantile & 25\%-quantile & Median & 75\%-quantile & 95\% quantile & Max & Average & Std. Dev \\
\hline
I       &&     0 &      0 & 0.0120 & 0.0700 & 0.2415 & 0.6651 & 0.7380 & 0.1570 & 0.1975 \\
II      &&0.0120 & 0.0232 & 0.0560 & 0.1400 & 0.2540 & 0.5012 & 0.7000 & 0.1819 & 0.1531 \\
III     &&     0 &      0 & 0.0140 & 0.0720 & 0.1480 & 0.3296 & 0.5100 & 0.1041 & 0.1155 \\
IV      &&0.0140 & 0.0430 & 0.0970 & 0.1770 & 0.3050 & 0.4400 & 0.6900 & 0.2082 & 0.1365 \\
V       &&0.0040 & 0.0120 & 0.1860 & 0.3340 & 0.5120 & 0.7680 & 0.8540 & 0.3524 & 0.2208 \\
\hline
\end{tabular}
\end{threeparttable}}
\end{table}

\begin{table}[htbp]
\caption{\textbf{Results: Elasticities and Optimal Prices.}}\label{T:price}
\begin{minipage}{0.9\linewidth}
\begin{footnotesize}
The table reports elasticities estimates as well the percentage difference between the current prices and the optimal price maximizing profit.
In each panel we report, for each product, the minimum, the 5\%-, 25\%-, 50\%-, 75\%-, and 95\%-quantiles, maximum, average, and standard deviation
for a given statistic. We consider the distribution over the selected treated municipalities. \textbf{We only report results concerning the cities where the estimated $\Delta$ has the correct sign and the effects are
statistical significance at the 10\% level}. The last column indicates the fraction of cities that satisfy the criterium described above.
In Panel (a) we report the results for the estimated elasticities. In Panel (b) we show the results for the difference between the current
price and the optimal price.
\end{footnotesize}
\end{minipage}
\resizebox{0.9\linewidth}{!}{
\begin{threeparttable}
\begin{tabular}{cccccccccccc}
\hline
\multicolumn{12}{c}{\underline{Panel (a): \textbf{Elasticities}}} \\
Product && Min & 5\%-quantile & 25\%-quantile & Median & 75\%-quantile & 95\% quantile & Max & Average & Std. Dev & Fraction \\
\hline
I       &&    -6.5063 &  -4.9033 &  -3.8701 &  -3.2394 & -2.8788 & -1.6952 & -0.9111 &  -3.3737 &  1.0605 & 0.2887 \\
II      &&   -17.6172 & -15.9339 & -12.4079 &  -9.2212 & -5.3922 & -2.1923 & -1.7181 &  -9.1675 &  4.2633 & 0.2549 \\
III     &&    -3.3421 &  -3.2692 &  -2.8714 &  -2.3356 & -1.7182 & -1.2854 & -1.1852 &  -2.3187 &  0.6597 & 0.2642 \\
IV      &&   -23.7060 & -20.9949 &  -9.8284 &  -7.2089 & -2.9208 & -1.8260 & -1.7398 &  -8.4089 &  6.2820 & 0.2100 \\
V       &&   -43.7705 & -43.7705 & -24.8448 & -19.7129 & -9.8875 & -3.6541 & -3.6541 & -20.3380 & 13.2366 & 0.0636 \\
\\
\multicolumn{12}{c}{\underline{Panel (b): \textbf{Price Discrepancies (\% Difference)}}} \\
Product && Min & 5\%-quantile & 25\%-quantile & Median & 75\%-quantile & 95\% quantile & Max & Average & Std. Dev & Fraction \\
\hline
I       &&    -15.9781 & -13.3710 & -10.7336 &  -8.2254 &  -6.2937 &  7.0732 & 31.2139 &  -6.8061 & 8.4973 & 0.2887 \\
II      &&    -21.4575 & -21.1495 & -20.2659 & -18.8623 & -15.0229 & -1.1660 &  4.8060 & -16.4801 & 6.3444 & 0.2549 \\
III     &&    -10.3946 & -10.0601 &  -7.9305 &  -3.9473 &   3.7551 & 13.5708 & 16.8302 &  -1.7678 & 7.6748 & 0.2642 \\
IV      &&    -17.6040 & -17.2995 & -14.5927 & -12.7773 &  -2.5345 &  7.7192 &  9.0263 &  -9.1644 & 8.3262 & 0.2100 \\
V       &&    -18.5970 & -18.5970 & -17.7256 & -17.2029 & -13.4599 & -6.0562 & -6.0562 & -15.2362 & 4.5547 & 0.0636 \\
\hline
\end{tabular}
\end{threeparttable}}
\end{table}

\appendix

\onehalfspacing

\setlength{\abovedisplayskip}{2.5pt}
\setlength{\belowdisplayskip}{2.5pt}

\section{Supplement Introduction}

This is a supplement to the paper ``Do We Exploit all Information for Counterfactual Analysis? Benefits of Factor Models and Idiosyncratic Correction''. The document is organized as follows. In Section \ref{S:random} we describe the algorithm used to split the cities into the treatment and control groups. Section \ref{S:empirical} contains additional empirical results. More specifically, in Section \ref{S:information} we compare the empirical results when the ArCo methodology of \citet*{cCrMmM2018} and the Principal Component Regression (PCR) as in \citet{lGtM2016} are used to estimate the counterfactuals. In Section \ref{S:trends} we evaluate different approaches to model trending behavior in the data, while in Section \ref{S:aggregation} we present the results at a state-level aggregation. The proof of the main result in the paper is presented in Section \ref{App: Main Results}.

\section{Randomization Algorithm}\label{S:random}

In this section we describe the algorithm used to split the municipalities into two different groups according to a set of characteristics. Once the groups are formed we randomly label them as treatment and control groups.

Let $\b{Z}$ be a $n\times J$ matrix of municipalities' variables, where each column $j$ is a different characteristic (covariate) and each row  $i$ is a municipality, $n$ is the number of municipalities and $J$ is the number of covariates. We consider the following variables: human development index, employment, GDP per capita, population,  female population, literate population, average household income (total), household income (urban areas), number of stores, and number of convenience stores.

The goal is to match the average of each characteristic of the treatment group with the control group. Once each group of municipalities is created, each group is further divided into two other groups, resulting in four different sets of municipalities. The experiments were carried on different combinations of the groups. In the paper, we report only one set of the experiments.

The optimization problem is defined as
\[
\begin{split}
&\widehat{\b\alpha}=\arg\min_{\b\alpha}\frac{1}{J}\sum_{j=1}^{J}\left|\frac{1}{\sum_{i=1}^n\alpha_i}\sum_{i=1}^{n} \alpha_i Z_{i,j} - \frac{1}{\sum_{i=1}^n(1-\alpha_i)}\sum_{i=1}^{n}(1-\alpha_i) Z_{i,j} \right|\\
&\textnormal{subject to: } \sum_{i=1}^n \alpha_i = K \,\textnormal{and}\,\alpha_i \in \{0,1\}\,\forall\,i,
\end{split}
\]
where $\b\alpha=(\alpha_1,\ldots,\alpha_n)$', $\alpha_i=1$ if municipality $i$ belongs to the first group and $\alpha_i=0$ otherwise; $K$ is the number of municipalities in the first group. The optimization problem above can be transformed into a mixed integer program.

\section{Additional Empirical Results}\label{S:empirical}

In this section we report a number of additional empirical results with the aim of showing the robustness and advantages of the proposed methodology.

\subsection{Additional Plots}

\begin{figure}
\caption{Data for Product II.}
\begin{spacing}{}
\scriptsize
Panel (a) reports the daily sales divided by the number of stores aggregated for all cities as well as for the treatment and control groups.
The plot also indicates the date of the intervention.
Panels (b) and (c) display the distribution of the average sales per store over
time across municipalities in the treatment and control groups, respectively.
Panels (d) and (e) present fan plots of sales across municipalities in the treatment and control groups for each given time point.
The black curves represent the cross-sectional mean over time and the vertical green line indicates the date of intervention.
\vspace{0.2cm}
\end{spacing}
\centering
\includegraphics[width=\linewidth]{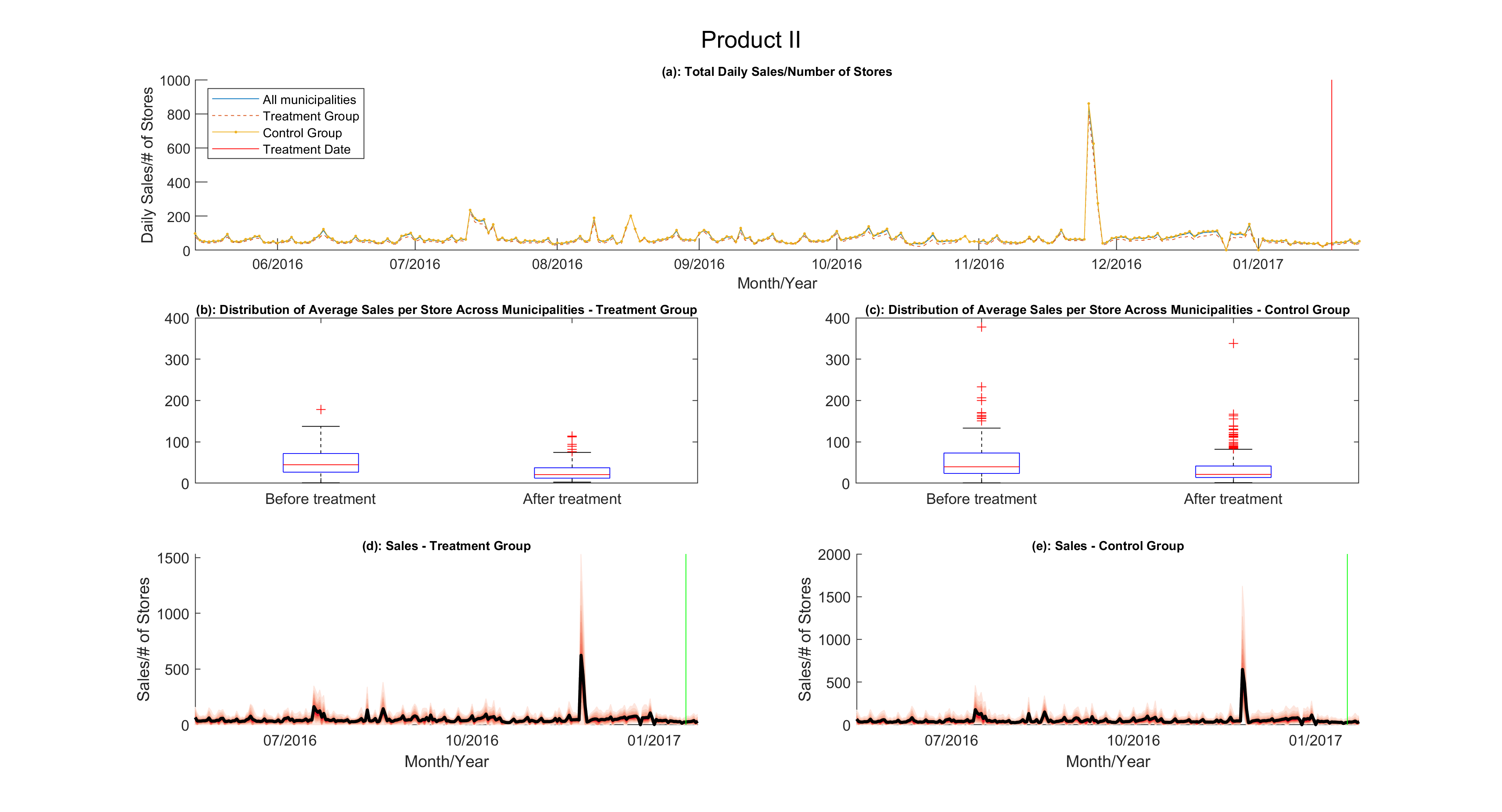}
\label{F:lasa2}
\end{figure}

\begin{figure}
\caption{Data for Product III.}
\begin{spacing}{}
\scriptsize
Panel (a) reports the daily sales divided by the number of stores aggregated for all cities as well as for the treatment and control groups.
The plot also indicates the date of the intervention.
Panels (b) and (c) display the distribution of the average sales per store over
time across municipalities in the treatment and control groups, respectively.
Panels (d) and (e) present fan plots of sales across municipalities in the treatment and control groups for each given time point.
The black curves represent the cross-sectional mean over time and the vertical green line indicates the date of intervention.
\vspace{0.2cm}
\end{spacing}
\centering
\includegraphics[width=\linewidth]{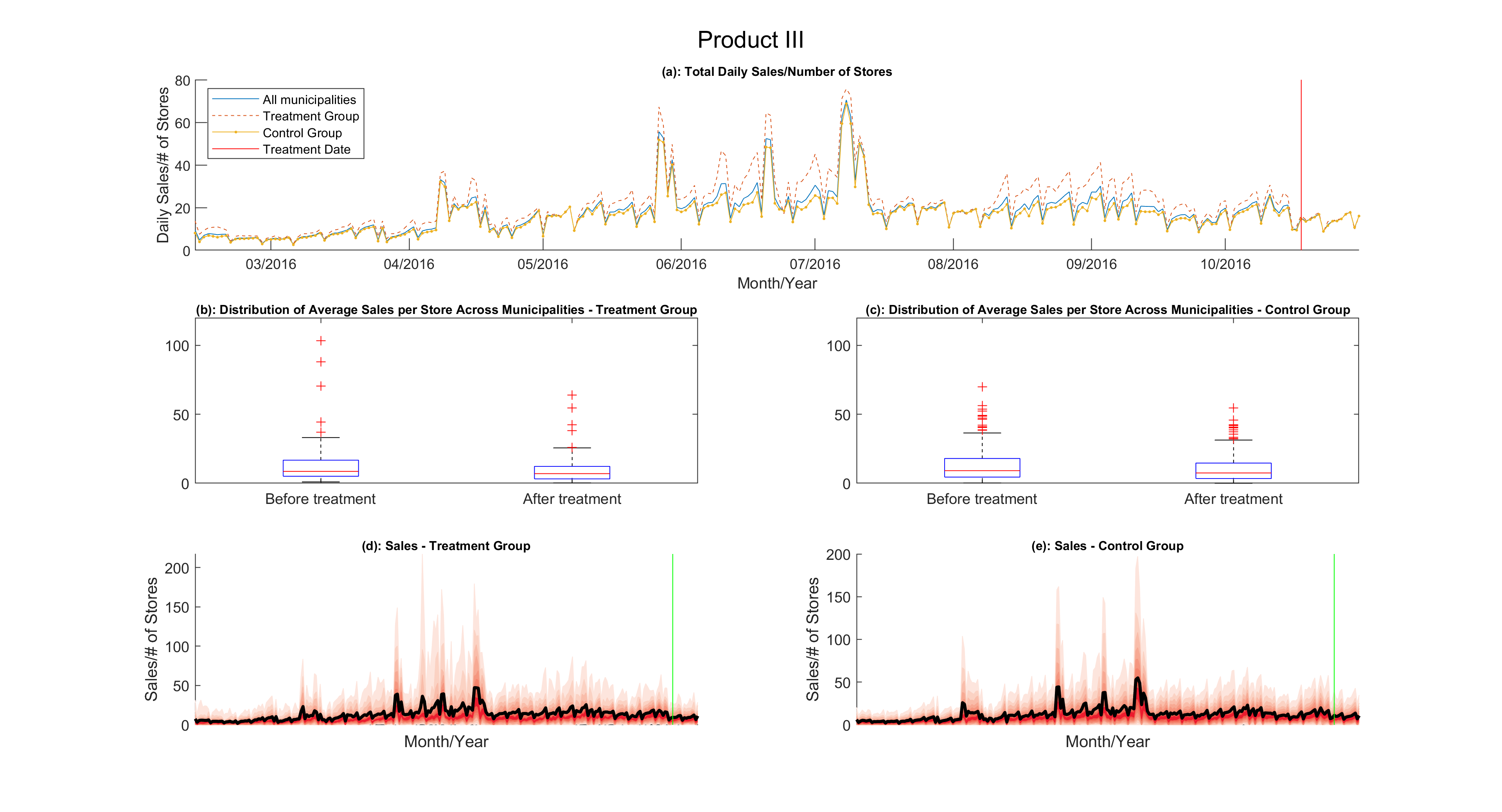}
\label{F:lasa3}
\end{figure}

\begin{figure}
\caption{Data for Product IV.}
\begin{spacing}{}
\scriptsize
Panel (a) reports the daily sales divided by the number of stores aggregated for all cities as well as for the treatment and control groups.
The plot also indicates the date of the intervention.
Panels (b) and (c) display the distribution of the average sales per store over
time across municipalities in the treatment and control groups, respectively.
Panels (d) and (e) present fan plots of sales across municipalities in the treatment and control groups for each given time point.
The black curves represent the cross-sectional mean over time and the vertical green line indicates the date of intervention.
\vspace{0.2cm}
\end{spacing}
\centering
\includegraphics[width=\linewidth]{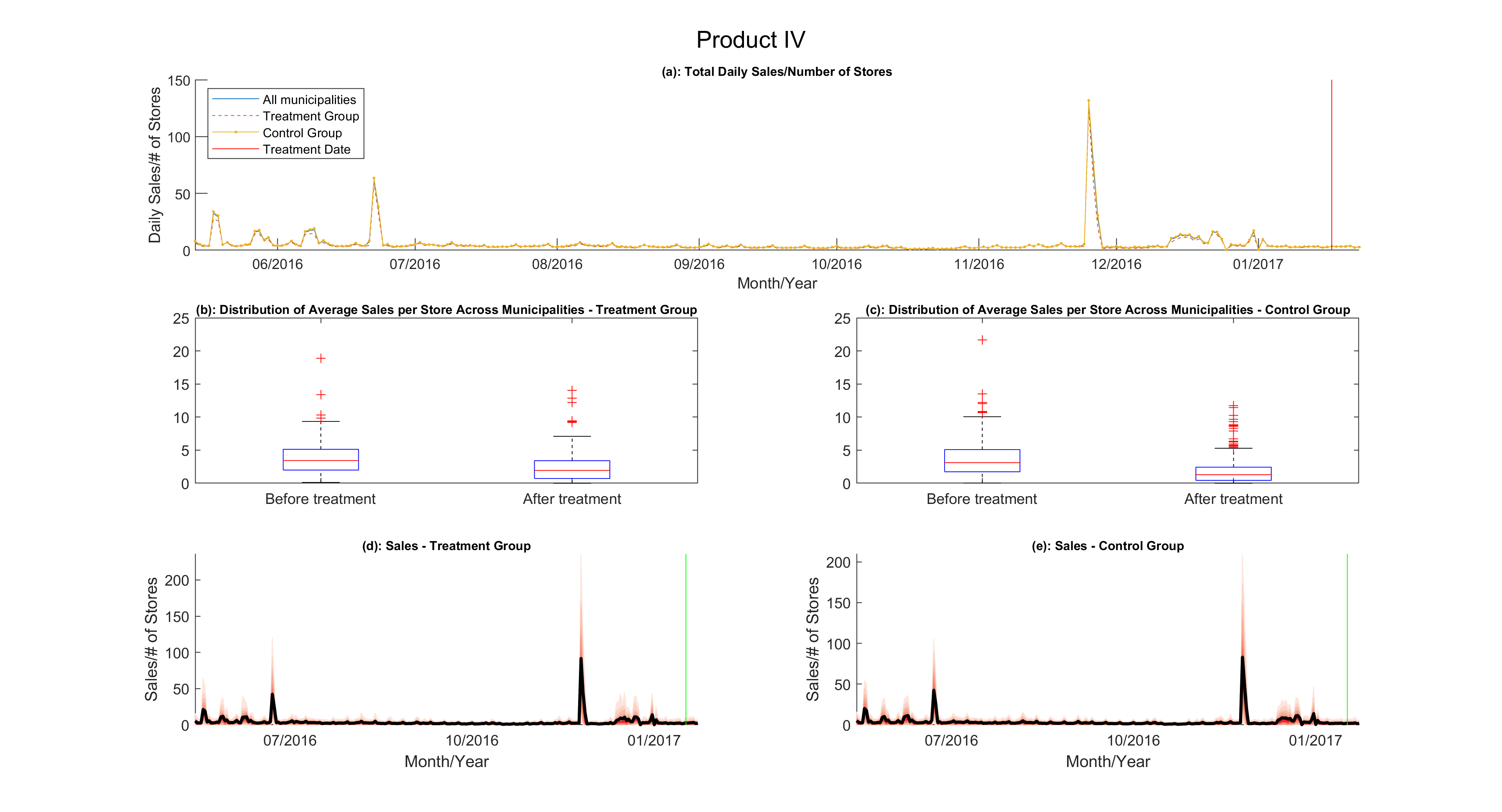}
\label{F:lasa4}
\end{figure}

\begin{figure}
\caption{Data for Product V.}
\begin{spacing}{}
\scriptsize
Panel (a) reports the daily sales divided by the number of stores aggregated for all cities as well as for the treatment and control groups.
The plot also indicates the date of the intervention.
Panels (b) and (c) display the distribution of the average sales per store over
time across municipalities in the treatment and control groups, respectively.
Panels (d) and (e) present fan plots of sales across municipalities in the treatment and control groups for each given time point.
The black curves represent the cross-sectional mean over time and the vertical green line indicates the date of intervention.
\vspace{0.2cm}
\end{spacing}
\centering
\includegraphics[width=\linewidth]{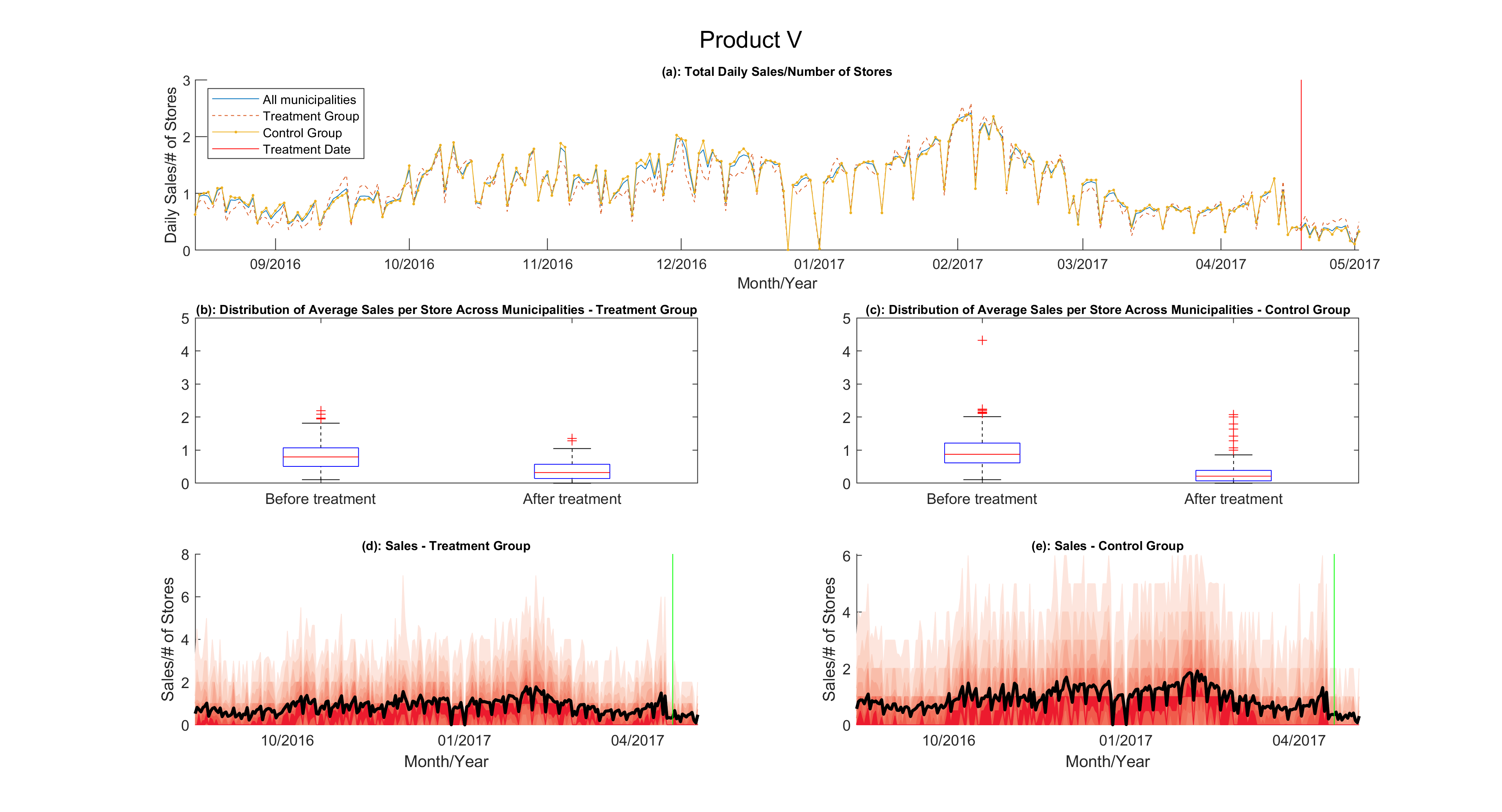}
\label{F:lasa5}
\end{figure}

Figures \ref{F:lasa2}--\ref{F:lasa5} display relevant data for Products II-V. Panel (a) in the figures reports the daily sales at each group of municipalities (all, treatment, and control) divided by the number of stores in each group. More specifically, the plot shows the daily evolution of $q_{\textnormal{all},t\cdot}^{(j)}=\frac{1}{s}\sum_{i=1}^nq_{it}^{(j)}$, $q_{\textnormal{control},t\cdot}^{(j)}=\frac{1}{s_0}\sum_{i=1}^{n_0}q_{it}^{(j)}$, and $q_{\textnormal{treatment},t\cdot}^{(j)}=\frac{1}{s_1}\sum_{i=n_0+1}^nq_{it}^{(j)}$. The plot shows the data before and after price changes and the intervention date is represented by the horizontal line. Panels (b) and (c) display the distribution across municipalities of the time averages of $\widetilde{q}_{it}^{(j)}$, before and after the intervention and for the treatment and control groups, respectively.
Panels (d) and (e) present fan plots for the evolution of $\widetilde{q}_{it}^{(j)}$. The black curves there represent the cross-sectional means over time.

\begin{figure}
\caption{Results for Product II}
\begin{spacing}{}
\scriptsize
Panel (a) displays a fan plot, across $n_1$ municipalities in the treatment group, of the $p$-values of the re-sampling test
for the null $\mathscr{H}_{0}: \delta_{t} = 0$ at each time $t$ after the treatment. The black curve represents the median $p$-value
across municipalities over $t$. Panel (b) shows an example for one municipality.
The panel depicts the actual and counterfactual sales per store for the post-treatment period.
95\% confidence intervals for the counterfactual path is also displayed.
\vspace{0.3cm}
\end{spacing}
\centering
\includegraphics[width=\linewidth]{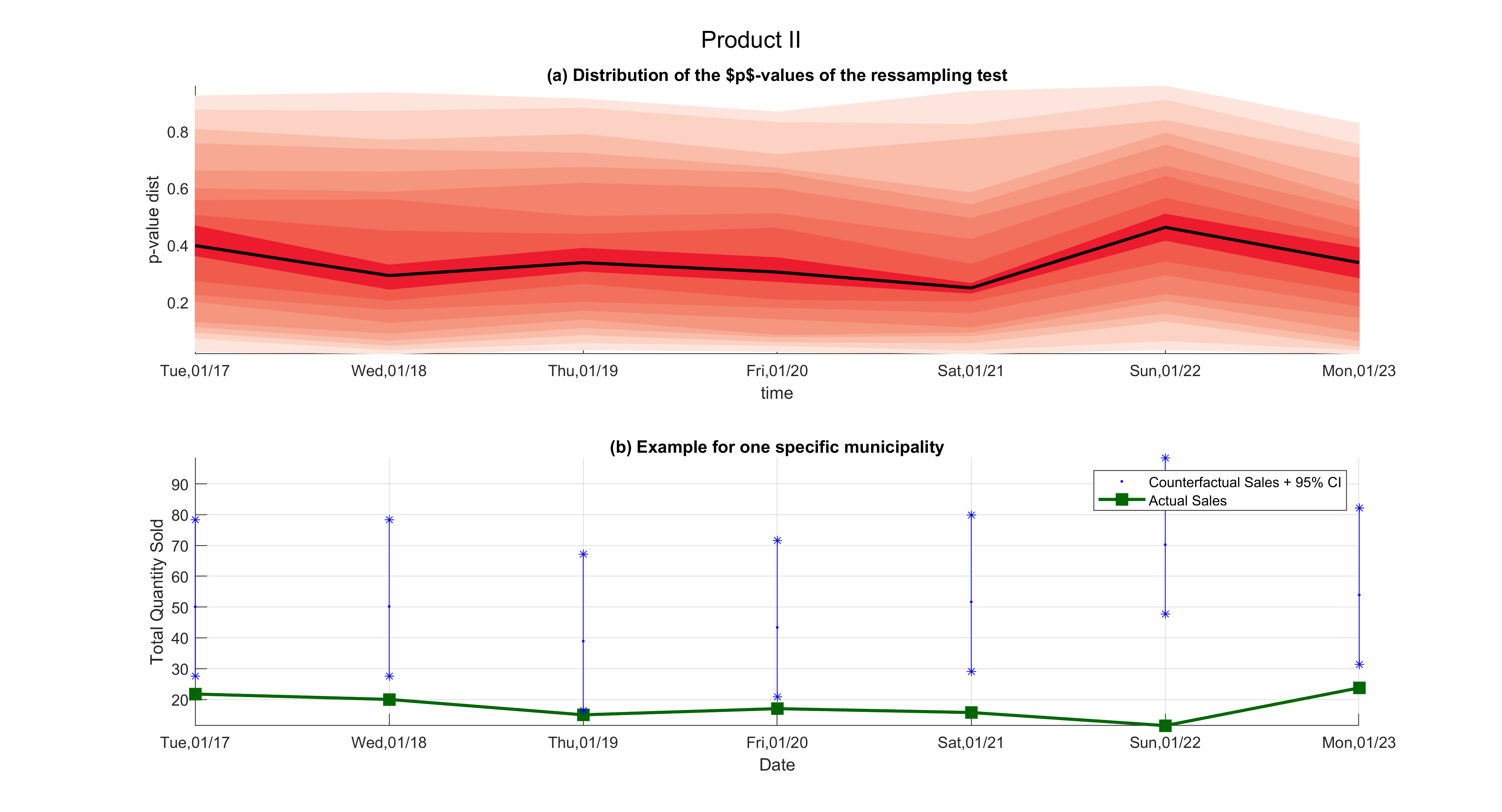}
\label{F:lasa_res2}
\end{figure}

\begin{figure}
\caption{Results for Product III}
\begin{spacing}{}
\scriptsize
Panel (a) displays a fan plot, across $n_1$ municipalities in the treatment group, of the $p$-values of the re-sampling test
for the null $\mathscr{H}_{0}: \delta_{t} = 0$ at each time $t$ after the treatment. The black curve represents the median $p$-value
across municipalities over $t$. Panel (b) shows an example for one municipality.
The panel depicts the actual and counterfactual sales per store for the post-treatment period.
95\% confidence intervals for the counterfactual path is also displayed.
\vspace{0.3cm}
\end{spacing}
\centering
\includegraphics[width=\linewidth]{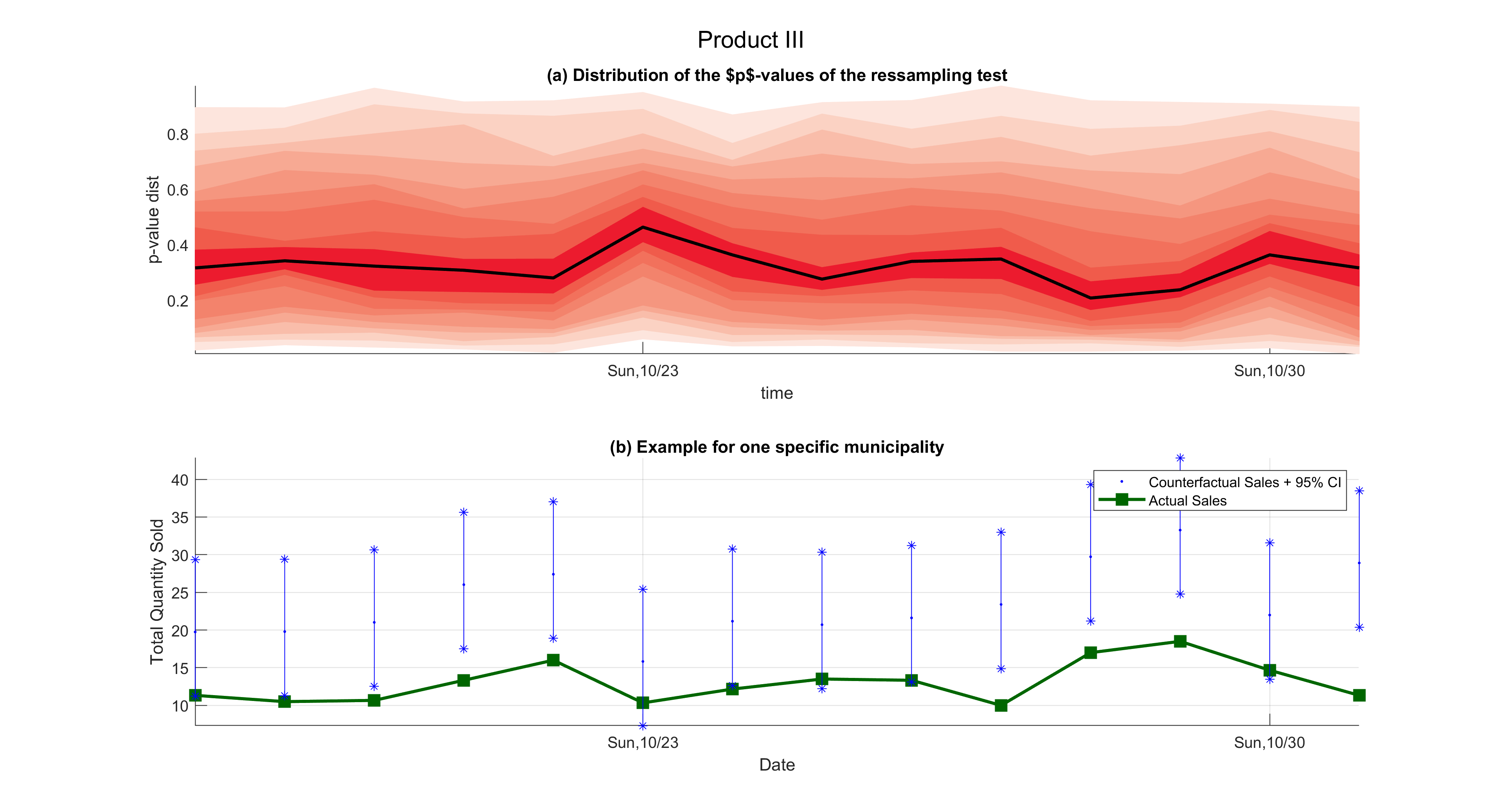}
\label{F:lasa_res3}
\end{figure}

\begin{figure}
\caption{Results for Product IV}
\begin{spacing}{}
\scriptsize
Panel (a) displays a fan plot, across $n_1$ municipalities in the treatment group, of the $p$-values of the re-sampling test
for the null $\mathscr{H}_{0}: \delta_{t} = 0$ at each time $t$ after the treatment. The black curve represents the median $p$-value
across municipalities over $t$. Panel (b) shows an example for one municipality.
The panel depicts the actual and counterfactual sales per store for the post-treatment period.
95\% confidence intervals for the counterfactual path is also displayed.
\vspace{0.3cm}
\end{spacing}
\centering
\includegraphics[width=\linewidth]{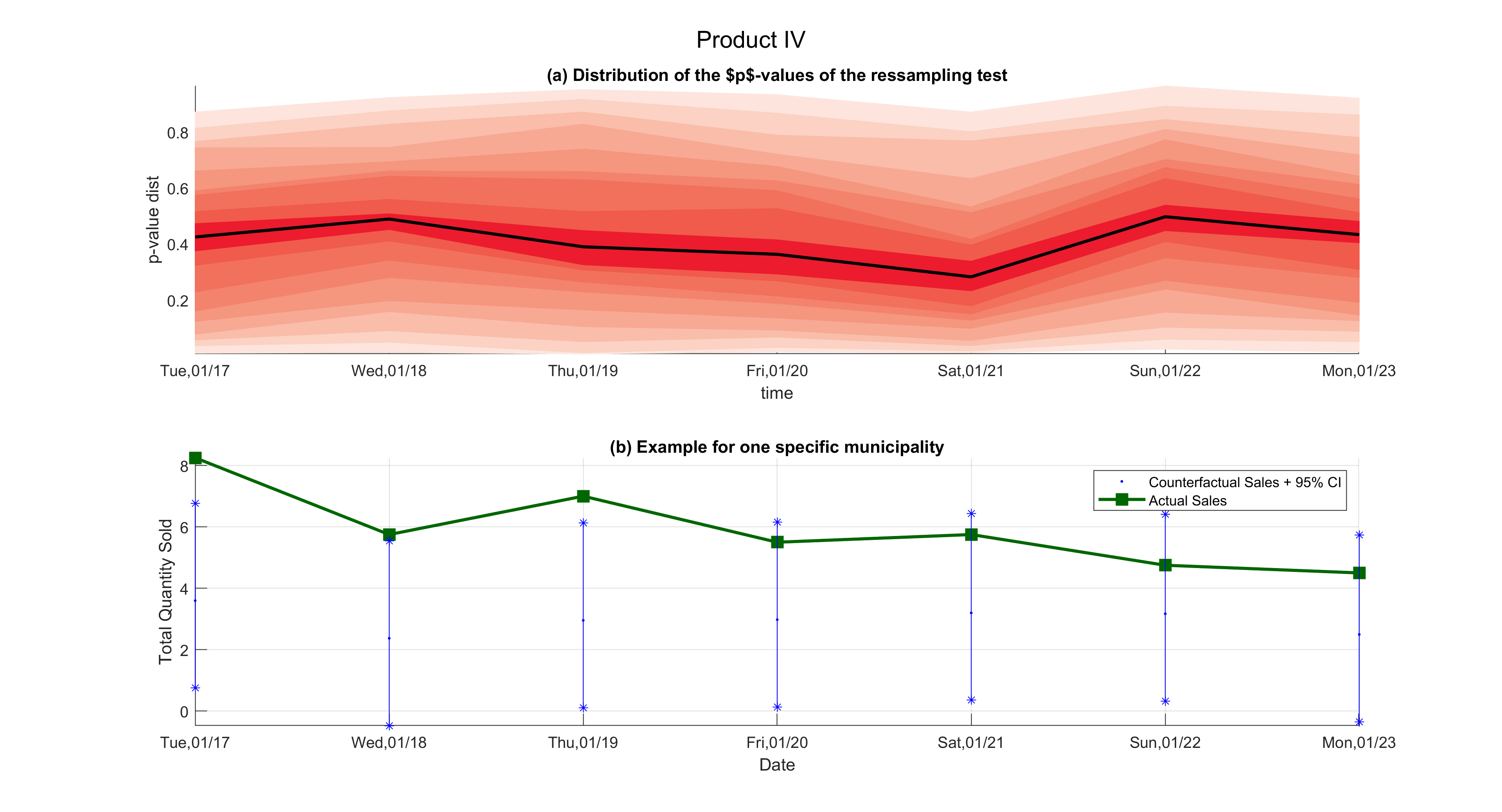}
\label{F:lasa_res4}
\end{figure}

\begin{figure}
\caption{Results for Product V}
\begin{spacing}{}
\scriptsize
Panel (a) displays a fan plot, across $n_1$ municipalities in the treatment group, of the $p$-values of the re-sampling test
for the null $\mathscr{H}_{0}: \delta_{t} = 0$ at each time $t$ after the treatment. The black curve represents the median $p$-value
across municipalities over $t$. Panel (b) shows an example for one municipality.
The panel depicts the actual and counterfactual sales per store for the post-treatment period.
95\% confidence intervals for the counterfactual path is also displayed.
\vspace{0.3cm}
\end{spacing}
\centering
\includegraphics[width=\linewidth]{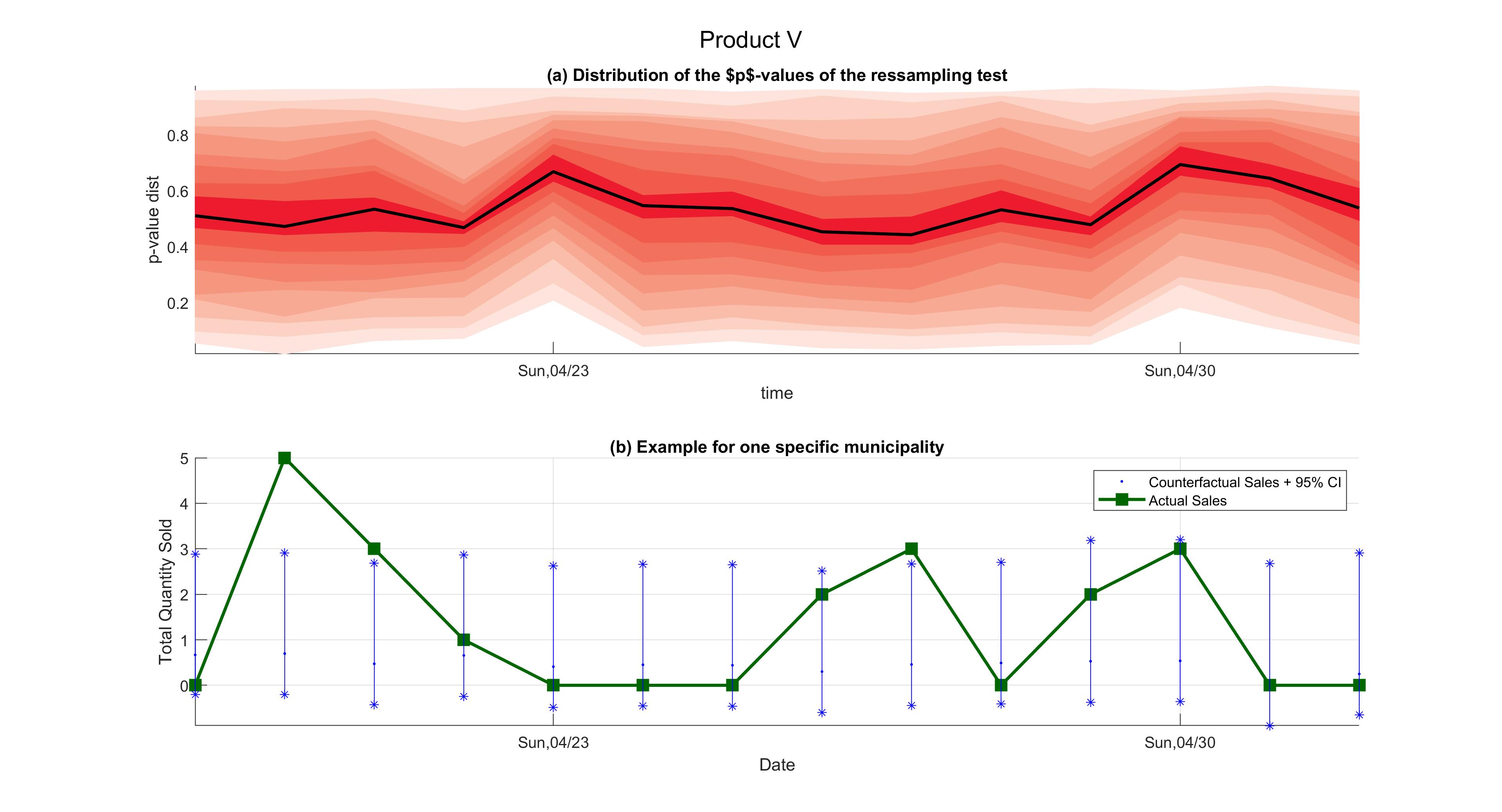}
\label{F:lasa_res5}
\end{figure}

Figures \ref{F:lasa_res2}--\ref{F:lasa_res5} display some estimation results. For each product, Panel (a) in the figures displays a fan plot of the $p$-values of the ressampling test for the null hypothesis $\mathcal{H}_{0}: \delta_t = 0$ for each given $t$ after the treatment, using the test statistic $\phi(\widehat \delta_t) = |\widehat \delta_t|$, which is the same as using the test statistic $\widehat \delta_t^2$. The black curve represents the cross-sectional median across time $t$. Panel (b) shows an example for one municipality. The panel shows the actual and counterfactual sales per store for the post-treatment period. 95\% confidence intervals for the counterfactual path are also displayed.

\begin{figure}[htbp]
\caption{Daily Inventory Distribution.}
\begin{spacing}{}
\scriptsize
\vspace{0.2cm}
\end{spacing}
\centering
{\stackunder[4pt]{\includegraphics[width=0.48\linewidth]{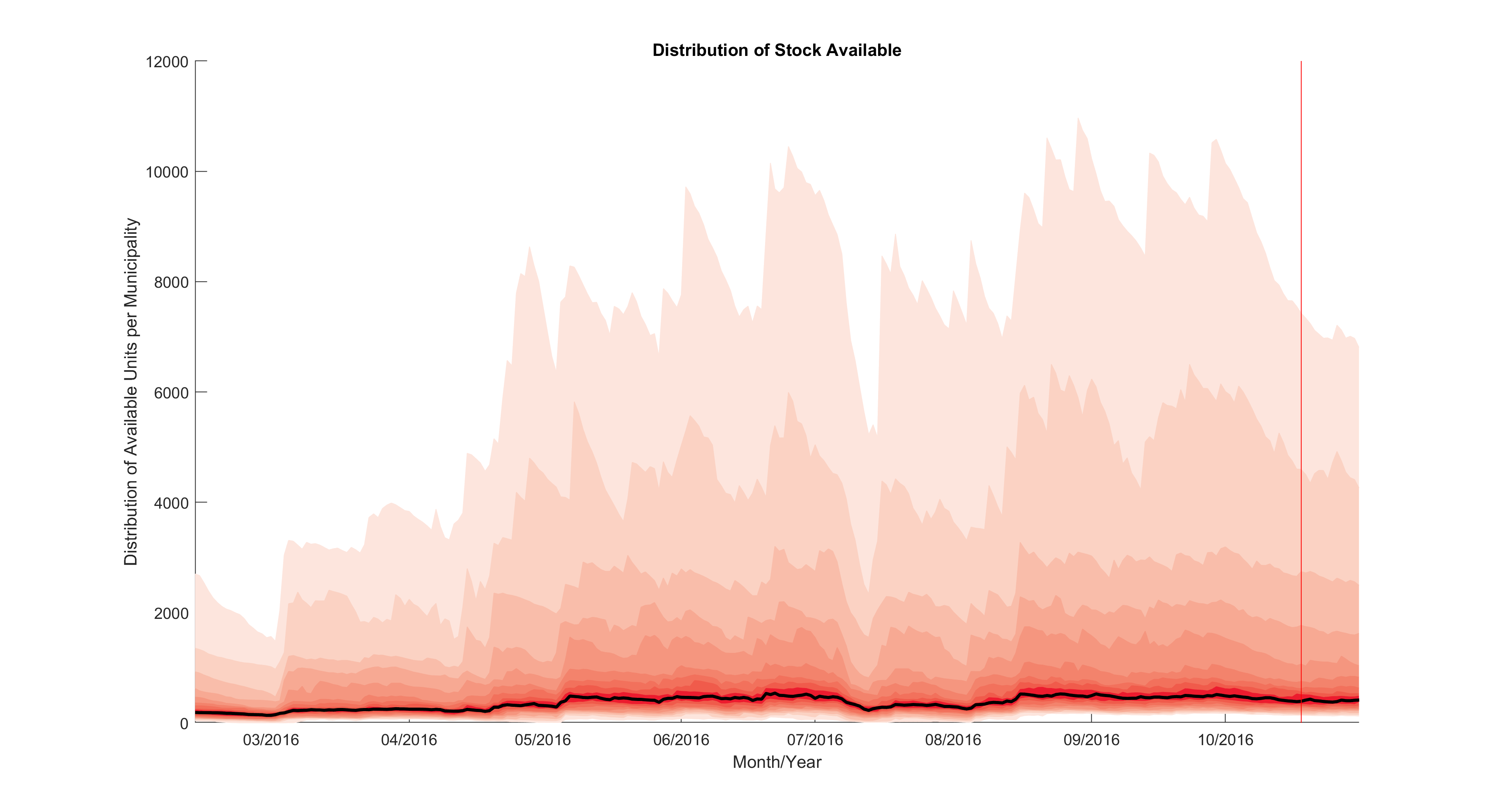}}{{\scriptsize{Product I}}}}
{\stackunder[4pt]{\includegraphics[width=0.48\linewidth]{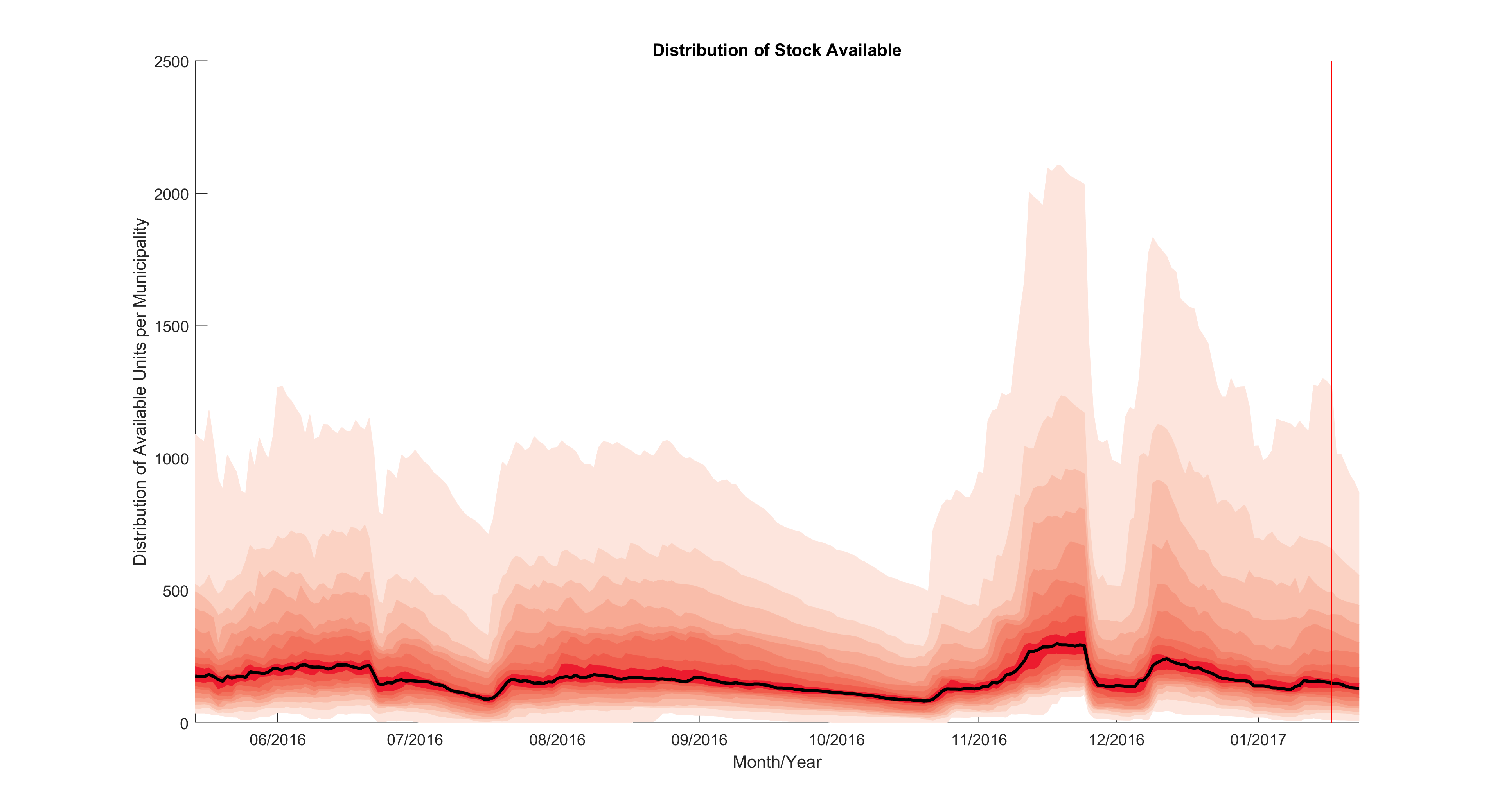}}{{\scriptsize{Product II}}}}\\
{\stackunder[4pt]{\includegraphics[width=0.48\linewidth]{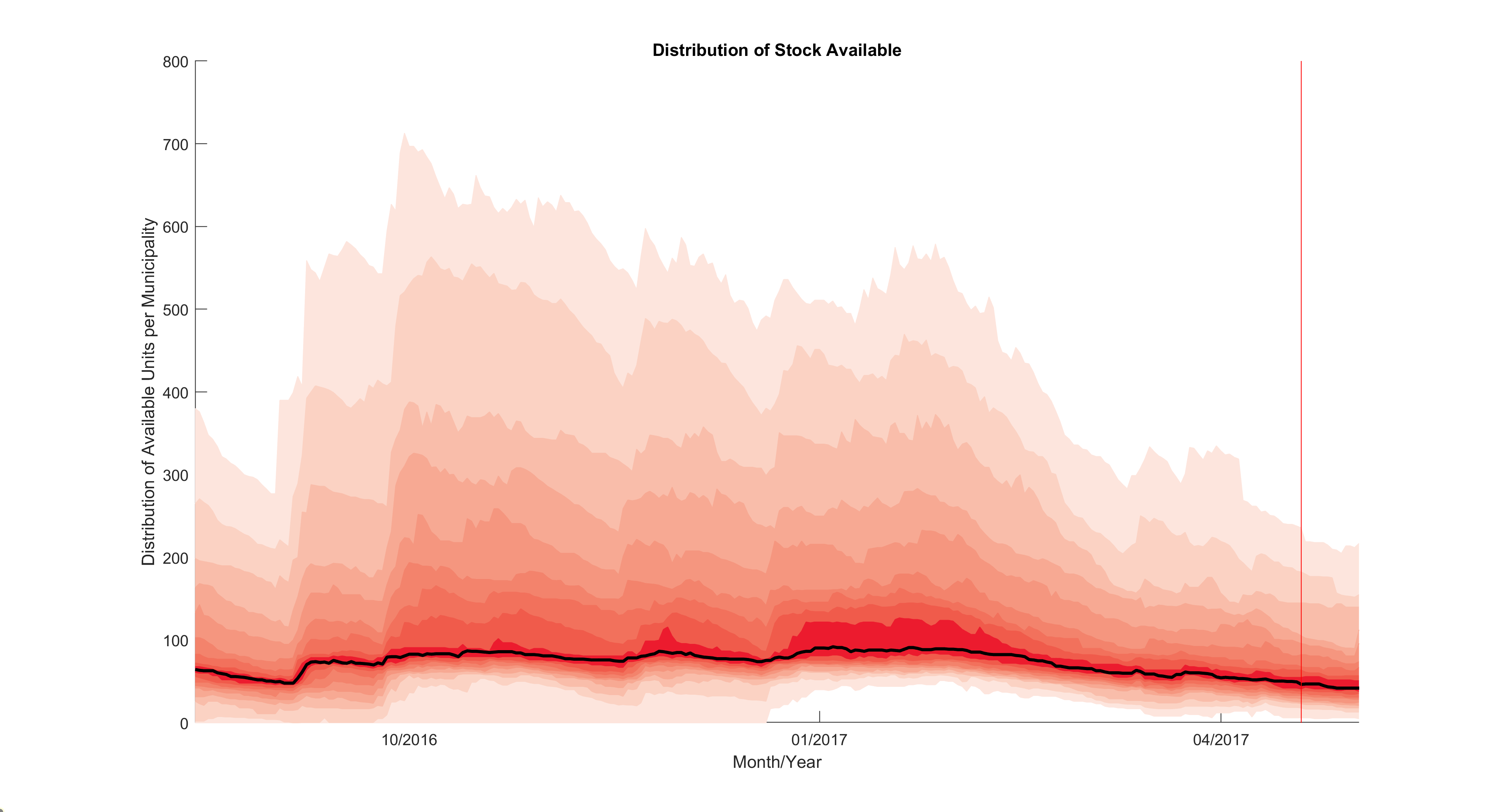}}{{\scriptsize{Product III}}}}
{\stackunder[4pt]{\includegraphics[width=0.48\linewidth]{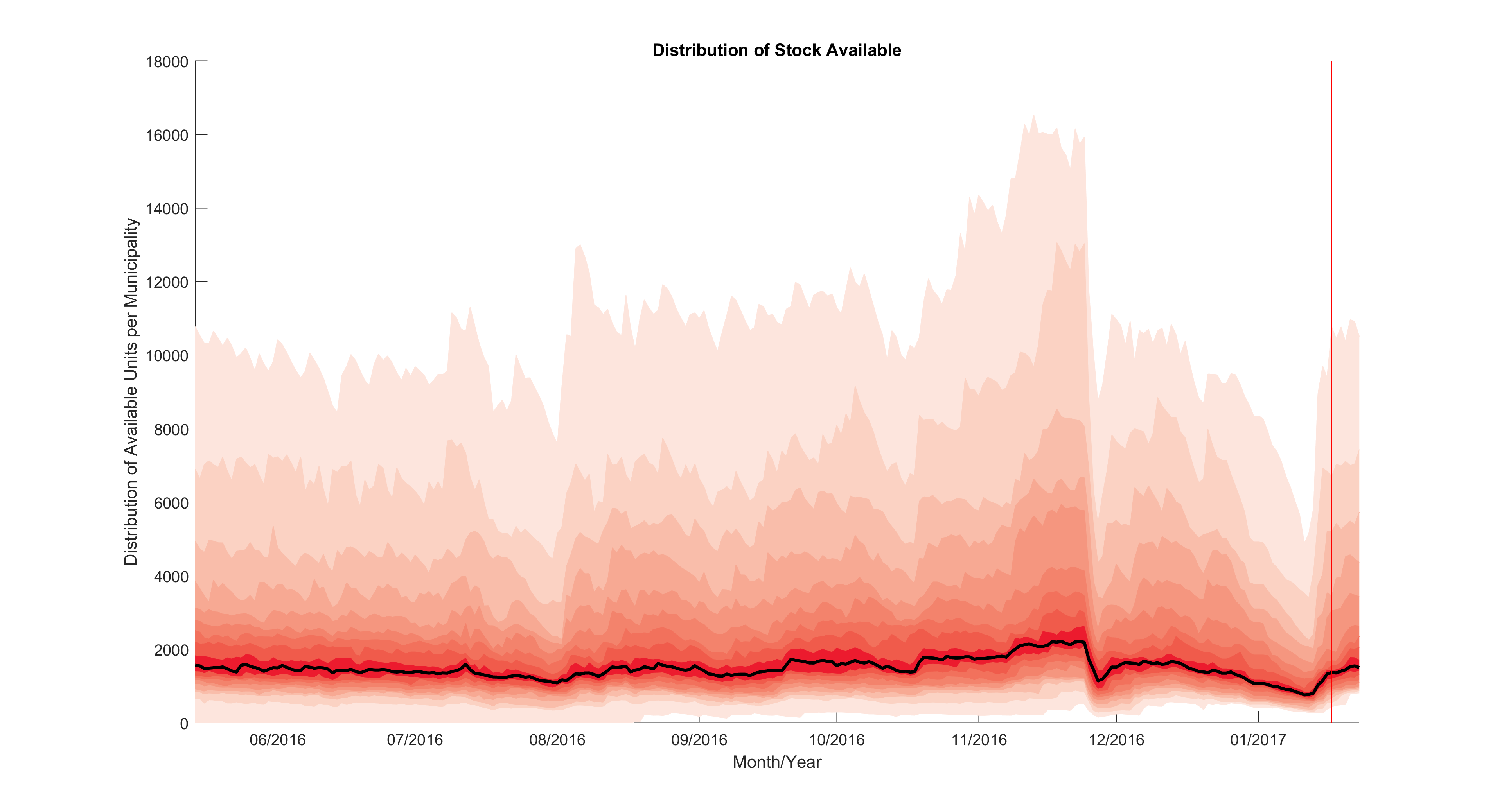}}{{\scriptsize{Product IV}}}}\\
{\stackunder[4pt]{\includegraphics[width=0.48\linewidth]{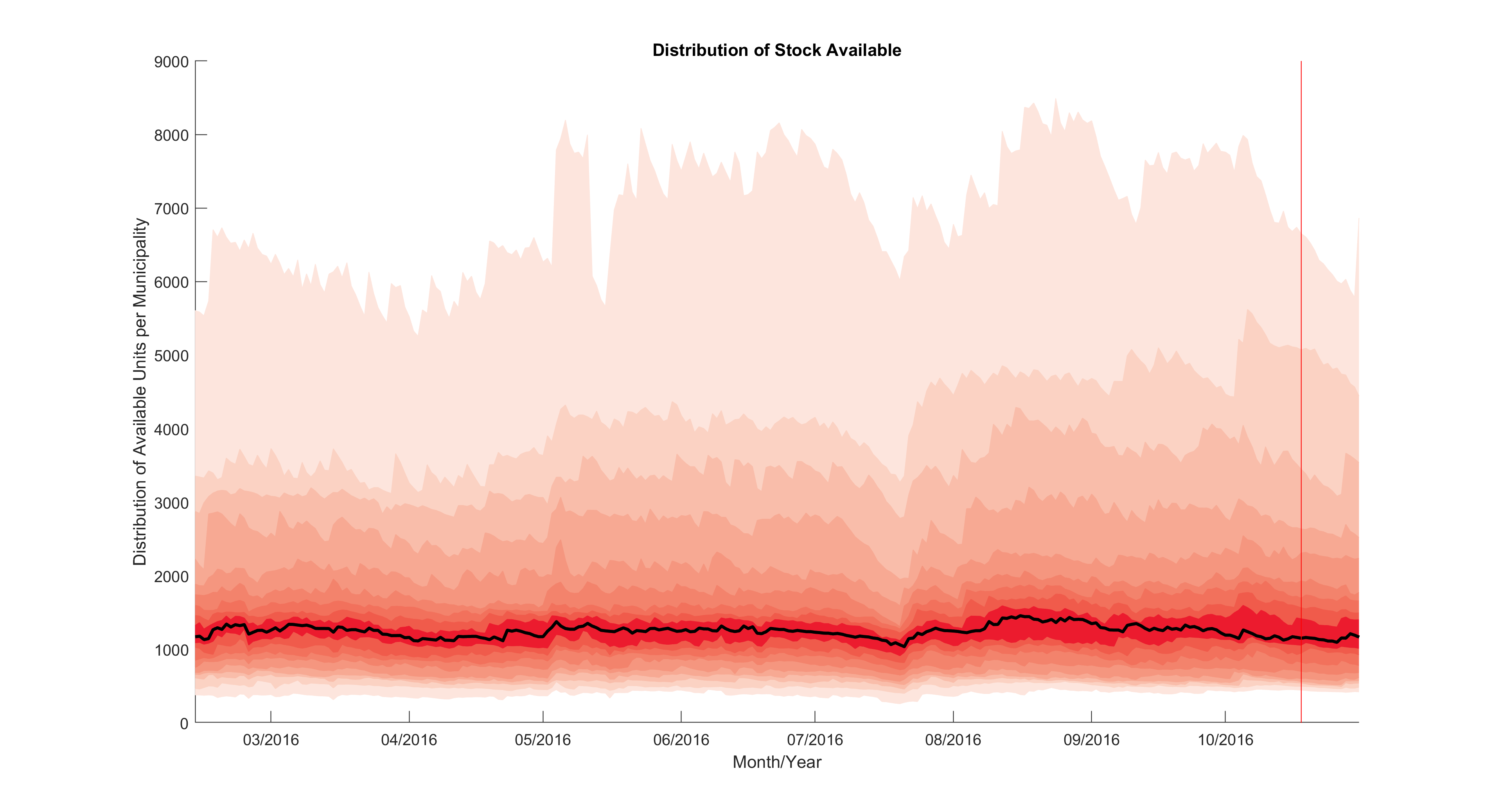}}{{\scriptsize{Product V}}}}
\label{F:inventory}
\end{figure}

Figure \ref{F:inventory} displays the distribution of the daily evolution of the inventory of each product across different municipalities.

\subsection{Effects of Additional Information: \texttt{ArCo}, \texttt{PCR}, and \texttt{FarmTreat}}\label{S:information}

We report the estimation results when either the \texttt{ArCo} methodology of \citet{cCrMmM2018} or principal component regression (PCR) in the spirit of \citet{lGtM2016} are used. For the ArCo methodology we construct counterfactuals by estimating a LASSO regression of $\widetilde{q}_{it}^{(j)}$ on the values of $\widetilde{q}_{kt}^{(j)}$, where $k\in\{1,\ldots,n\}/i$. Note that we do not include any other regressor. For PCR, we consider the first two stages of the \texttt{FarmTreat} methodology.

The \texttt{ArCo} results are reported in Tables \ref{T:estimation_arco} and \ref{T:price_arco} while the results for the PCR method are shown in Tables \ref{T:estimation_pcr} and \ref{T:price_pcr}. Some interesting facts emerge from the tables. First, the \texttt{ArCo} and \texttt{FarmTreat} show similar results, with the later having a slightly better pre-intervention fit. One key difference, however, is the substantially smaller number of municipalities with significant intervention effects when the ArCo methodology is considered. Comparing the PCR approach and the  \texttt{FarmTreat}, we can clearly see an improvement in the pre-intervention fit, as expected. As in the \texttt{ArCo} method, the \texttt{PCR} approach yields a smaller fraction of cities with significant effects of the price changes. Finally, one important point to highlight is that all three methods suggests that on average the current prices must be decreased.

\subsection{Effects of Trends}\label{S:trends}

Tables \ref{T:estimation_trend} and \ref{T:price_trend} report the results of
the \texttt{FarmTreat} methodology is used without detrending the data in the
first step. Compared to the baseline results presented in Tables
\ref{T:estimation} and \ref{T:price} in the main text we highlight the
following facts. First, the counterfactual model adjustment is similar with
only marginal differences concerning the pre-intervention R-squared. Second,
without detrending, the average treatment effects are smaller but the
rejection rates are higher. Third, the number of municipalities where the
estimated $\Delta$ has the correct sign and is statistically significant at
the 10\% level is much smaller when we do not include a linear trend in the
first step of the methodology, specially in the case of Product V. We note
that for this last product, the recommendation is a price increase and not
decrease. For the other four products, the conclusions are similar as the
baseline case.

\subsection{State-Level Aggregation}\label{S:aggregation}

Tables \ref{T:estimation_state} and \ref{T:price_state} report the results of
the \texttt{FarmTreat} methodology applied to data aggregated at the state
level. The control and treatment groups at the state-level are constructed by aggregating the untreated and treated municipalities in each state, respectively.
From the tables we see that for Products IV and V we do not find
significant effects at the state level. This is mainly due to heterogeneity
across municipalities within each state. On the other hand, for Products I,
II and III we find significant effects of price changes on sales. On the
average, the optimal price for Products III and V are higher than the actual
ones, whereas for Product IV the \texttt{FarmTreat} method indicates that on
average the prices should be reduced. However, even for this products the
effects are significant in only a fraction of states. These results,
corroborates the huge municipality heterogeneity.

\subsection{Before-and-After Estimation}\label{S:ba}

Table \ref{T:estimation_ba} reports estimation the average treatment effect using the before-and-after estimator.
In each panel we report, for each product, the minimum, the 5\%-, 25\%-, 50\%-, 75\%-, and 95\%-quantiles, maximum, average, and standard deviation
for a variety of different statistics. We consider the distribution over the treated municipalities.

\section{Proof of the Main Result} \label{App: Main Results}

Before proving our main result, we define below the compatibility constant for convenience.
\begin{Def}\label{D:GIF}
For a $(n\times n)$ matrix $\b{M}$, a set $\S\subseteq[n]$  and a scalar $\zeta \geq 0$, the compatibility constant is given by
\begin{equation}\label{E:GIF}
\kappa(\b{M},\S,\zeta):=\inf\left\{\frac{\|\b{x}^T \b{M} \b{ x} \|}{\sqrt{|\S|}}{\|\b{x}_\S\|_1}:\b{x}\in\R^n:\|\b{x}_{\S^c}\|_1\leq\xi\|\b{x}_{\S}\|_1 \right\}.
\end{equation}
Moreover, we say that $(\b{M},\S,\zeta)$ satisfies the compatibility condition if $\kappa(\b M,\S,\zeta)>0$.
\end{Def}
The compatibility constant is related to $\ell_1$-eigenvalue of $\b M$ restricted to a cone in $\R^n$.

\subsection{Proof of Proposition \ref{C:Main}}

The fact that $\|\widehat{\b\theta}_1-\b\theta_1\|_1 = O_P(\xi|\m{S}_0|)$ follows from Theorem 3 in \citet*{jFrMmM2021}. We are left to show the  second part. By the triangle inequality, for $t>T_0$:
\begin{align*}
|\widehat{\alpha}_t-\alpha_t - V_t| &=|(\widehat{\b\gamma}_1 -\b\gamma_{1})' \b W_{1t} + \widehat{\b\lambda}_1'\widehat{\b F}_t -\b\lambda_1'\b F_t +\widehat{\b \theta}_1' \widehat{\b U}_{-1t} - \b \theta_1'\b U_{-1t} |\\
&\leq |(\widehat{\b\gamma}_1 -\b\gamma_{1})' \b W_{1t}| + |\widehat{U}_{1t} -U_{1t}| +|\widehat{\b \theta}_1' \widehat{\b U}_{-1t} - \b \theta_1'\b U_{-1t} |.
\end{align*}
Using Hölder's inequality, the third term can be further bounded as
\begin{align*}
|\widehat{\b \theta}_1' \widehat{\b U}_{-1t} - \b \theta_1'\b U_{-1t} | &\leq |\widehat{\b \theta}_1'(\widehat{\b U}_{-1t} - \b U_{-1t} )| + |( \widehat{\b \theta}_1 -\b \theta_1)'\b U_{-1t}|\\
&\leq \|\widehat{\b \theta}_1\|_1\|\widehat{\b U}_{-1t} - \b U_{-1t}\|_\infty + \| \widehat{\b \theta}_1 -\b \theta_1\|_1\|\b U_{-1t}\|_\infty\\
&\leq (\|\b \theta_1\|_1 +\| \widehat{\b \theta}_1 -\b \theta_1\|_1 )\|\widehat{\b U}_{-1t} - \b U_{-1t}\|_\infty + \| \widehat{\b \theta}_1 -\b \theta_1\|_1\|\b U_{-1t}\|_\infty\\
&=O_P[(\|\b \theta_1\|_1 + \upsilon|\S_0|\psi^{-1}(T))\upsilon+ \upsilon|\S_0|\psi^{-1}(T)\psi^{-1}(n) ].
\end{align*}
Combining the last two expressions we are left with
\begin{align*}
|\widehat{\alpha}_t-\alpha_t - V_t|
&\leq |(\widehat{\b\gamma}_1 -\b\gamma_{1})' \b W_{1t}| +( 1+ \|\b \theta_1\|_1 +\| \widehat{\b \theta}_1 -\b \theta_1\|_1 )\|\widehat{\b U}_{t} - \b U_{t}\|_\infty + \| \widehat{\b \theta}_1 -\b \theta_1\|_1\|\b U_{t}\|_\infty.
\end{align*}

The first term is $O_P(1/\sqrt{T})$ by Assumption \ref{Ass:Moments}(a). The second is $O_P(|\m{S}_0|\eta)$ because by Assumption \ref{Ass:Moments}(d) we have that $\|\b\theta_1\|_1\leq|\S_0|\|\b\theta_1\|_\infty\leq C|\S_0|$ and $\| \widehat{\b \theta}_1 -\b \theta_1\|_1 = O_P(1)$ under the assumptions of the Proposition.  Finally, the third term is $O_P(\xi|\m{S}_0| n^{1/p})$ by Assumption \ref{Ass:Moments}(b) and the maximum inequality. Therefore we conclude that
\[\widehat{\alpha}_t-\alpha_t - V_t = O_P\left(T^{-1/2} + |\m{S}_0|\eta + \xi|\m{S}_0| n^{1/p}\right) = O_P\left[|\m{S}_0|(\eta+\xi n^{1/p} )\right].\]

\begin{table}[htbp]
\caption{\textbf{Results: Estimation and Inference (ArCo).}}\label{T:estimation_arco}
\begin{minipage}{0.9\linewidth}
\begin{footnotesize}
The table reports estimation results using the ArCo methodology of \citet{cCrMmM2018}.
In each panel we report, for each product, the minimum, the 5\%-, 25\%-, 50\%-, 75\%-, and 95\%-quantiles, maximum, average, and standard deviation
for a variety of different statistics. We consider the distribution over the treated municipalities.
In Panel (a) we report the results for the R-squared of the pre-intervention model.
Panel (b) displays the results for the average intervention effect over the experiment period ($\Delta$).
Panels (c) and (d) depict the results for the $p$-values of the ressampling test for the null hypothesis
$\mathcal{H}_0:\delta_t = 0, \forall t\in \{T_{0}+1,\dots, T\}$ using respectively the test statistic
$\phi(\widehat \delta_{T_0+1},\ldots, \widehat  \delta_T)=\sum_{t=T_0+1}^T \widehat  \delta_t^2$
or $\phi(\widehat \delta_{T_0+1},\ldots, \widehat \delta_T)=\sum_{t=T_0+1}^T|\widehat  \delta_t|$.
Finally, Panel (e) reports the results for the $p$-values for the test for the idiosyncratic contribution.
\end{footnotesize}
\end{minipage}
\resizebox{0.9\linewidth}{!}{
\begin{threeparttable}
\begin{tabular}{ccccccccccc}
\hline
\multicolumn{11}{c}{\underline{Panel (a): \textbf{R-squared}}} \\
Product && Min & 5\%-quantile & 25\%-quantile & Median & 75\%-quantile & 95\% quantile & Max & Average & Std. Dev \\
\hline
I       &&          0 & 0.1421 & 0.3367 & 0.4389 & 0.6276 & 0.7821 & 0.8958 & 0.4641 & 0.1981  \\
II      &&     0.4448 & 0.6555 & 0.8691 & 0.9218 & 0.9575 & 0.9851 & 0.9958 & 0.8899 & 0.1073  \\
III     &&     0.0639 & 0.3119 & 0.4957 & 0.6937 & 0.8181 & 0.9115 & 0.9679 & 0.6554 & 0.2018  \\
IV      &&     0.3688 & 0.6902 & 0.8823 & 0.9262 & 0.9635 & 0.9888 & 0.9987 & 0.8984 & 0.1056  \\
V       &&     0      & 0      & 0      & 0.0966 & 0.2210 & 0.4319 & 0.6975 & 0.1452 & 0.1545  \\
\\
\multicolumn{11}{c}{\underline{Panel (b): \textbf{Average Treatment Effect (over time): $\Delta$}}} \\
Product && Min & 5\%-quantile & 25\%-quantile & Median & 75\%-quantile & 95\% quantile & Max & Average & Std. Dev \\
\hline
I       &&  -20.1194 & -12.0679 & -6.0420 & -2.9966 & -0.6335 & 1.7254 &  7.2911 & -3.6588 &  4.3075 \\
II      &&  -40.6070 & -25.4886 & -9.9769 & -3.1266 &  0.2057 & 9.9614 & 59.7638 & -4.2132 & 11.6643 \\
III     &&  -37.8542 &  -8.5142 & -3.3295 & -1.0079 &  0.2364 & 3.7909 &  9.6714 & -2.0799 &  5.8070 \\
IV      &&   -2.5440 &  -1.6212 & -0.5723 &  0.1673 &  1.4634 & 3.8332 &  6.4165 &  0.4945 &  1.6339 \\
V       &&   -1.2218 &  -0.8548 & -0.4922 & -0.2797 &  0.0044 & 0.4945 &  1.1978 & -0.2476 &  0.4234 \\
\\
\multicolumn{11}{c}{\underline{Panel (c): \textbf{$p$-value of the test on squared values}}} \\
Product && Min & 5\%-quantile & 25\%-quantile & Median & 75\%-quantile & 95\% quantile & Max & Average & Std. Dev \\
\hline
I       &&          0 & 0.0085 & 0.1830 & 0.3702 & 0.6351 & 0.9147 & 0.9787 & 0.4077 & 0.2838 \\
II      &&          0 & 0.0388 & 0.2273 & 0.4876 & 0.7521 & 0.9521 & 1.0000 & 0.4905 & 0.2967 \\
III     &&          0 & 0.0306 & 0.2638 & 0.4894 & 0.6638 & 0.8928 & 0.9915 & 0.4735 & 0.2658 \\
IV      &&          0 &      0 & 0.0888 & 0.3802 & 0.7004 & 0.9029 & 0.9793 & 0.4092 & 0.3162 \\
V       &&          0 & 0.0894 & 0.3574 & 0.6936 & 0.9149 & 1.0000 & 1.0000 & 0.6452 & 0.3015 \\
\\
\multicolumn{11}{c}{\underline{Panel (d): \textbf{$p$-value of the test on absolute values}}} \\
Product && Min & 5\%-quantile & 25\%-quantile & Median & 75\%-quantile & 95\% quantile & Max & Average & Std. Dev \\
\hline
I       &&          0 &      0 & 0.0787 & 0.3149 & 0.5691 & 0.8960 & 0.9872 & 0.3593 & 0.2995 \\
II      &&          0 & 0.0223 & 0.1818 & 0.5021 & 0.7273 & 0.9504 & 1.0000 & 0.4753 & 0.3095\\
III     &&          0 &      0 & 0.2681 & 0.4532 & 0.6766 & 0.8655 & 1.0000 & 0.4624 & 0.2671 \\
IV      &&          0 &      0 & 0.1033 & 0.3946 & 0.7066 & 0.9318 & 0.9876 & 0.4124 & 0.3220\\
V       &&          0 & 0.1234 & 0.3787 & 0.6745 & 0.8809 & 0.9957 & 1.0000 & 0.6284 & 0.2886 \\
\hline
\end{tabular}
\end{threeparttable}}
\end{table}

\newpage

\begin{table}[htbp]
\caption{\textbf{Results: Elasticities and Optimal Prices (ArCo).}}\label{T:price_arco}
\begin{minipage}{0.9\linewidth}
\begin{footnotesize}
The table reports elasticities estimates as well the percentage difference between the current prices and the optimal price maximizing profit when the ArCo methodology by \citet{cCrMmM2018} is used.
In each panel we report, for each product, the minimum, the 5\%-, 25\%-, 50\%-, 75\%-, and 95\%-quantiles, maximum, average, and standard deviation
for a given statistic. We consider the distribution over the selected treated municipalities. \textbf{We only report results concerning the cities where the estimated $\Delta$ has the correct sign and the effects are
statistical significance at the 10\% level}. The last column indicates the fraction of cities that satisfy the criterium described above.
In Panel (a) we report the results for the estimated elasticities. In Panel (b) we show the results for the difference between the current
price and the optimal price.
\end{footnotesize}
\end{minipage}
\resizebox{0.9\linewidth}{!}{
\begin{threeparttable}
\begin{tabular}{cccccccccccc}
\hline
\multicolumn{12}{c}{\underline{Panel (a): \textbf{Elasticities}}} \\
Product && Min & 5\%-quantile & 25\%-quantile & Median & 75\%-quantile & 95\% quantile & Max & Average & Std. Dev & Fraction \\
\hline
I       &&    -6.1256 &   -6.0847 &  -3.4902 &  -2.7145 & -2.1592 & -1.3585 & -1.2700 &  -3.1159 &  1.4785 & 0.1443 \\
II      &&   -16.8229 &  -16.8229 & -12.6336 & -11.4970 & -7.5925 & -4.0746 & -4.0746 & -10.5119 &  3.9061 & 0.0882 \\
III     &&    -3.0759 &   -3.0759 &  -2.8387 &  -2.0602 & -1.8896 & -1.6480 & -1.6480 &  -2.2876 &  0.5477 & 0.0755 \\
IV      &&   -44.2416 &  -34.5020 & -11.9050 &  -6.5419 & -4.5606 & -2.4450 & -1.9634 & -10.6109 & 10.1396 & 0.2400 \\
V       &&  -135.5289 & -135.5289 & -19.1707 & -10.1509 & -5.4511 & -4.3575 & -4.3575 & -30.8017 & 51.5716 & 0.0545 \\
\\
\multicolumn{12}{c}{\underline{Panel (b): \textbf{Price Discrepancies (\% Difference)}}} \\
Product && Min & 5\%-quantile & 25\%-quantile & Median & 75\%-quantile & 95\% quantile & Max & Average & Std. Dev & Fraction \\
\hline
I       &&   -15.5005 & -15.4441 &  -9.3372 &  -5.2233 &  -0.5061 &  13.6720 &  15.7075 &  -4.2981 & 8.5470 & 0.1443 \\
II      &&   -21.3235 & -21.3235 & -20.3202 & -19.9466 & -17.6746 & -12.0246 & -12.0246 & -18.6809 & 2.8620 & 0.0882 \\
III     &&    -9.0999 &  -9.0999 &  -7.7113 &  -1.0822 &   1.1540 &   4.9837 &   4.9837 &  -2.4244 & 5.1882 & 0.0755 \\
IV      &&   -18.5830 & -18.2201 & -15.5075 & -12.0222 &  -8.7234 &   1.1269 &   5.7524 & -11.2130 & 5.9686 & 0.2400 \\
V       &&   -19.3704 & -19.3704 & -17.1312 & -14.8087 & -10.5669 &  -8.2649 &  -8.2649 & -14.1585 & 4.1117 & 0.0545 \\
\hline
\end{tabular}
\end{threeparttable}}
\end{table}

\newpage

\begin{table}[htbp]
\caption{\textbf{Results: Estimation and Inference (PCR).}}\label{T:estimation_pcr}
\begin{minipage}{0.9\linewidth}
\begin{footnotesize}
The table reports estimation results using principal component regressions.
In each panel we report, for each product, the minimum, the 5\%-, 25\%-, 50\%-, 75\%-, and 95\%-quantiles, maximum, average, and standard deviation
for a variety of different statistics. We consider the distribution over the treated municipalities.
In Panel (a) we report the results for the R-squared of the pre-intervention model.
Panel (b) displays the results for the average intervention effect over the experiment period ($\Delta$).
Panels (c) and (d) depict the results for the $p$-values of the ressampling test for the null hypothesis
$\mathcal{H}_0:\delta_t = 0, \forall t\in \{T_{0}+1,\dots, T\}$ using respectively the test statistic
$\phi(\widehat \delta_{T_0+1},\ldots, \widehat  \delta_T)=\sum_{t=T_0+1}^T \widehat  \delta_t^2$
or $\phi(\widehat \delta_{T_0+1},\ldots, \widehat \delta_T)=\sum_{t=T_0+1}^T|\widehat  \delta_t|$.
\end{footnotesize}
\end{minipage}
\resizebox{0.9\linewidth}{!}{
\begin{threeparttable}
\begin{tabular}{ccccccccccc}
\hline
\multicolumn{11}{c}{\underline{Panel (a): \textbf{R-squared}}} \\
Product && Min & 5\%-quantile & 25\%-quantile & Median & 75\%-quantile & 95\% quantile & Max & Average & Std. Dev \\
\hline
I       && 0.1115  &  0.1727  &  0.3307  &  0.4491  &  0.6011  &  0.7294  &  0.7892  &  0.4517  &  0.1707  \\
II      && 0.2549  &  0.4633  &  0.7428  &  0.8345  &  0.8815  &  0.9456  &  0.9759  &  0.7898  &  0.1445  \\
III     && 0.1026  &  0.1588  &  0.2489  &  0.3466  &  0.5095  &  0.6296  &  0.6944  &  0.3751  &  0.1545  \\
IV      && 0.1300  &  0.2384  &  0.5805  &  0.7173  &  0.8236  &  0.8941  &  0.9627  &  0.6723  &  0.1996  \\
V       && 0.0255  &  0.0366  &  0.0739  &  0.1236  &  0.2068  &  0.3815  &  0.5079  &  0.1545  &  0.1033  \\
\\
\multicolumn{11}{c}{\underline{Panel (b): \textbf{Average Treatment Effect (over time): $\Delta$}}} \\
Product && Min & 5\%-quantile & 25\%-quantile & Median & 75\%-quantile & 95\% quantile & Max & Average & Std. Dev \\
\hline
I       &&-21.9722  & -17.1898 &  -7.6521 &  -3.4870 &  -1.0735  &  1.6398  &  3.6122  & -5.0798  &  5.6688 \\
II      &&-47.0186  & -32.5355 & -15.2901 &  -7.5150 &  -2.8772  &  9.9514  & 40.2040  & -9.2082  & 12.9511 \\
III     &&-55.4751  & -17.1204 &  -7.2165 &  -3.4482 &  -0.6900  &  1.8316  &  8.8381  & -5.6288  &  9.8650 \\
IV      && -4.3269  & -1.9948  &  -0.7039 &   0.2394 &   1.5064  &  4.1167  &  7.3901  &  0.5691  &  1.9752 \\
V       && -2.0826  & -0.9796  &  -0.5058 &  -0.1766 &   0.1292  &  0.6744  &  1.6705  & -0.1831  &  0.5190 \\
\\
\multicolumn{11}{c}{\underline{Panel (c): \textbf{$p$-value of the test on squared values}}} \\
Product && Min & 5\%-quantile & 25\%-quantile & Median & 75\%-quantile & 95\% quantile & Max & Average & Std. Dev \\
\hline
I       && 0      & 0       &  0.0723  &  0.2553  &  0.6170  &  0.8985  &  0.9915  &  0.3445  &  0.3063 \\
II      && 0.0289 & 0.0421  &  0.2355  &  0.4566  &  0.6901  &  0.8983  &  0.9752  &  0.4697  &  0.2844 \\
III     && 0      & 0.0664  &  0.2809  &  0.4511  &  0.6340  &  0.9336  &  1.0000  &  0.4624  &  0.2478 \\
IV      && 0      & 0.0723  &  0.2459  &  0.4153  &  0.7169  &  0.9917  &  1.0000  &  0.4794  &  0.2878 \\
V       && 0      & 0.0596  &  0.3277  &  0.6511  &  0.9319  &  1.0000  &  1.0000  &  0.6050  &  0.3302 \\
\\
\multicolumn{11}{c}{\underline{Panel (d): \textbf{$p$-value of the test on absolute values}}} \\
Product && Min & 5\%-quantile & 25\%-quantile & Median & 75\%-quantile & 95\% quantile & Max & Average & Std. Dev \\
\hline
I       && 0  &       0  &  0.0511  &  0.2128  &  0.6053  &  0.9019  &  0.9957  &  0.3199  &  0.3199 \\
II      && 0  &  0.0207  &  0.1570  &  0.4256  &  0.6942  &  0.9298  &  0.9628  &  0.4481  &  0.3038 \\
III     && 0  &  0.0102  &  0.2128  &  0.4128  &  0.5957  &  0.9319  &  1.0000  &  0.4248  &  0.2689 \\
IV      && 0  &  0.0517  &  0.2149  &  0.4070  &  0.7521  &  0.9690  &  1.0000  &  0.4710  &  0.3001 \\
V       && 0  &  0.0511  &  0.2681  &  0.6638  &  0.9234  &  1.0000  &  1.0000  &  0.6084  &  0.3362 \\
\hline
\end{tabular}
\end{threeparttable}}
\end{table}

\newpage

\begin{table}[htbp]
\caption{\textbf{Results: Elasticities and Optimal Prices (PCR).}}\label{T:price_pcr}
\begin{minipage}{0.9\linewidth}
\begin{footnotesize}
The table reports elasticities estimates as well the percentage difference between the current prices and the optimal price maximizing profit when the counterfactuals are estimated by principal component regression.
In each panel we report, for each product, the minimum, the 5\%-, 25\%-, 50\%-, 75\%-, and 95\%-quantiles, maximum, average, and standard deviation
for a given statistic. We consider the distribution over the selected treated municipalities. \textbf{We only report results concerning the cities where the estimated $\Delta$ has the correct sign and the effects are
statistical significance at the 10\% level}. The last column indicates the fraction of cities that satisfy the criterium described above.
In Panel (a) we report the results for the estimated elasticities. In Panel (b) we show the results for the difference between the current
price and the optimal price.
\end{footnotesize}
\end{minipage}
\resizebox{0.9\linewidth}{!}{
\begin{threeparttable}
\begin{tabular}{cccccccccccc}
\hline
\multicolumn{12}{c}{\underline{Panel (a): \textbf{Elasticities}}} \\
Product && Min & 5\%-quantile & 25\%-quantile & Median & 75\%-quantile & 95\% quantile & Max & Average & Std. Dev & Fraction \\
\hline
I       &&    -6.5287 &  -5.5838 &  -4.3723 &  -3.5023 & -2.9266 & -1.7697 & -0.9696 &  -3.5689 &  1.1443 & 0.2784 \\
II      &&   -17.7671 & -17.5199 & -14.5000 & -13.1484 & -8.6987 & -2.8126 & -1.9998 & -11.8098 &  4.2565 & 0.1275 \\
III     &&    -3.3805 &  -3.3805 &  -3.2669 &  -2.9047 & -2.7249 & -2.3503 & -2.3503 &  -2.9405 &  0.3565 & 0.0755 \\
IV      &&   -15.8735 & -15.8735 & -12.4477 & -11.0990 & -9.2416 & -1.0297 & -1.0297 & -10.3376 &  4.6432 & 0.0700\\
V       &&   -36.2752 & -36.2752 & -25.0377 & -15.3515 & -5.8105 & -3.3284 & -3.3284 & -16.8591 & 12.4347 & 0.0545 \\
\\
\multicolumn{12}{c}{\underline{Panel (b): \textbf{Price Discrepancies (\% Difference)}}} \\
Product && Min & 5\%-quantile & 25\%-quantile & Median & 75\%-quantile & 95\% quantile & Max & Average & Std. Dev & Fraction \\
\hline
I       &&   -16.0044 & -14.6686 & -12.1884 &  -9.3865 &  -6.5784 &  6.3136 & 27.9067 &  -7.6159 &  8.1823 & 0.2784 \\
II      &&   -21.4814 & -21.4382 & -20.8472 & -20.4928 & -18.5425 & -2.0321 &  0.7073 & -18.4545 &  5.8816 & 0.1275 \\
III     &&   -10.5643 & -10.5643 & -10.0456 &  -8.1187 &  -6.9926 & -4.0815 & -4.0815 &  -8.1200 &  2.1928 & 0.0755 \\
IV      &&   -16.5633 & -16.5633 & -15.6963 & -15.2083 & -14.2557 & 28.8451 & 28.8451 &  -9.0521 & 16.7294 & 0.0700 \\
V       &&   -18.3610 & -18.3610 & -17.7424 & -16.3414 & -11.1343 & -4.7169 & -4.7169 & -14.1062 &  5.2776 & 0.0545 \\
\hline
\end{tabular}
\end{threeparttable}}
\end{table}

\newpage


\begin{table}[htbp]
\caption{\textbf{Results: Estimation and Inference (no trend).}}\label{T:estimation_trend}
\begin{minipage}{0.9\linewidth}
\begin{footnotesize}
The table reports estimation results without the trend component in the counterfactual model.
In each panel we report, for each product, the minimum, the 5\%-, 25\%-, 50\%-, 75\%-, and 95\%-quantiles, maximum, average, and standard deviation
for a variety of different statistics. We consider the distribution over the treated municipalities aggregated at the state level.
In Panel (a) we report the results for the R-squared of the pre-intervention model.
Panel (b) displays the results for the average intervention effect over the experiment period ($\Delta$).
Panels (c) and (d) depict the results for the $p$-values of the ressampling test for the null hypothesis
$\mathcal{H}_0:\delta_t = 0, \forall t\in \{T_{0}+1,\dots, T\}$ using respectively the test statistic
$\phi(\widehat \delta_{T_0+1},\ldots, \widehat  \delta_T)=\sum_{t=T_0+1}^T \widehat  \delta_t^2$
or $\phi(\widehat \delta_{T_0+1},\ldots, \widehat \delta_T)=\sum_{t=T_0+1}^T|\widehat  \delta_t|$.
Finally, Panel (e) reports the results for the $p$-values for the test for the idiosyncratic contribution.
\end{footnotesize}
\end{minipage}
\resizebox{0.9\linewidth}{!}{
\begin{threeparttable}
\begin{tabular}{ccccccccccc}
\hline
\multicolumn{11}{c}{\underline{Panel (a): \textbf{R-squared}}} \\
Product && Min & 5\%-quantile & 25\%-quantile & Median & 75\%-quantile & 95\% quantile & Max & Average & Std. Dev \\
\hline
I       && 0.1112  &  0.1983  &  0.3463  &  0.4910 &   0.6302  &  0.7556  &  0.9029 &   0.4869 &   0.1777\\
II      && 0.4876  &  0.6913  &  0.8721  &  0.9280 &   0.9563  &  0.9850  &  0.9945 &   0.9007 &   0.0905\\
III     && 0.1141  &  0.2904  &  0.5243  &  0.7085 &   0.8324  &  0.9336  &  0.9600 &   0.6736 &   0.2041\\
IV      && 0.3824  &  0.6693  &  0.8802  &  0.9344 &   0.9632  &  0.9869  &  0.9986 &   0.8969 &   0.1101\\
V       && 0.0243  &  0.0378  &  0.0895  &  0.1461 &   0.2706  &  0.4143  &  0.6396 &   0.1876 &   0.1321\\
\\
\multicolumn{11}{c}{\underline{Panel (b): \textbf{Average Treatment Effect (over time): $\Delta$}}} \\
Product && Min & 5\%-quantile & 25\%-quantile & Median & 75\%-quantile & 95\% quantile & Max & Average & Std. Dev \\
\hline
I       && -16.1305 & -11.3082 &  -5.0625  & -2.6195 &  -0.8542 &   1.4314 &   9.9071&   -3.3187  &  4.1600\\
II      && -46.3695 & -27.3151 & -10.4665  & -4.1799 &  -0.6947 &   6.8649 &  58.5092&   -5.9179  & 11.9431\\
III     && -26.5438 &  -9.0437 &  -3.0657  & -0.9038 &   0.6108 &   4.8286 &  16.0986&   -1.5804  &  5.2233\\
IV      && -3.9357  & -1.6404  & -0.5186   & 0.2410  &  1.2506  &  4.0381  &  6.3938 &   0.5208   & 1.7143\\
V       && -1.0360  & -0.5738  & -0.2827   &-0.1076  &  0.1859  &  0.7279  &  1.0478 &  -0.0468   & 0.3770\\
\\
\multicolumn{11}{c}{\underline{Panel (c): \textbf{$p$-value of the test on squared values}}} \\
Product && Min & 5\%-quantile & 25\%-quantile & Median & 75\%-quantile & 95\% quantile & Max & Average & Std. Dev \\
\hline
I       && 0 &   0.0143 &   0.1564  &  0.4213 &   0.6298  &  0.8764 &   0.9872 &   0.4112  &  0.2786\\
II      && 0 &   0.0198 &   0.1818  &  0.4628 &   0.7273  &  0.9793 &   1.0000 &   0.4626  &  0.3138\\
III     && 0 &   0.0170 &   0.2638  &  0.4745 &   0.7064  &  0.9583 &   1.0000 &   0.4864  &  0.2839\\
IV      && 0 &        0 &   0.1302  &  0.3802 &   0.7025  &  0.9545 &   0.9876 &   0.4074  &  0.3110\\
V       && 0 &   0.0766 &   0.3447  &  0.8170 &   0.9872  &  1.0000 &   1.0000 &   0.6779  &  0.3238\\
\\
\multicolumn{11}{c}{\underline{Panel (d): \textbf{$p$-value of the test on absolute values}}} \\
Product && Min & 5\%-quantile & 25\%-quantile & Median & 75\%-quantile & 95\% quantile & Max & Average & Std. Dev \\
\hline
I       &&  0 &        0  &  0.1000 &   0.4000  &  0.5936 &   0.8979  &  0.9957  &  0.3785 &   0.2885\\
II      &&  0 &   0.0025  &  0.1446 &   0.4029  &  0.7355 &   0.9694  &  1.0000  &  0.4471 &   0.3213\\
III     &&  0 &        0  &  0.2170 &   0.4787  &  0.7234 &   0.9149  &  1.0000  &  0.4757 &   0.2918\\
IV      &&  0 &        0  &  0.0992 &   0.3616  &  0.7066 &   0.9360  &  0.9917  &  0.4000 &   0.3137\\
V       &&  0 &   0.1064  &  0.4340 &   0.8021  &  0.9915 &   1.0000  &  1.0000  &  0.6974 &   0.3104\\
\\
\multicolumn{11}{c}{\underline{Panel (e): \textbf{$p$-value of the test for idiosyncratic contribution}}} \\
Product && Min & 5\%-quantile & 25\%-quantile & Median & 75\%-quantile & 95\% quantile & Max & Average & Std. Dev \\
\hline
I       &&         0 &      0 & 0.0300 & 0.0820 & 0.2525 & 0.6636 & 0.9500 & 0.1824 & 0.2145 \\
II      &&    0.0080 & 0.0224 & 0.0620 & 0.1250 & 0.2520 & 0.4940 & 0.6760 & 0.1771 & 0.1488 \\
III     &&         0 &      0 & 0.0080 & 0.0590 & 0.1540 & 0.3468 & 0.5460 & 0.0995 & 0.1180 \\
IV      &&    0.0300 & 0.0470 & 0.0990 & 0.1920 & 0.2800 & 0.4410 & 0.6400 & 0.2093 & 0.1317 \\
V       &&         0 & 0.0240 & 0.1160 & 0.2890 & 0.4180 & 0.7080 & 0.8400 & 0.2977 & 0.2083 \\
\hline
\end{tabular}
\end{threeparttable}}
\end{table}

\newpage

\begin{table}[htbp]
\caption{\textbf{Results: Elasticities and Optimal Prices (no trend).}}\label{T:price_trend}
\begin{minipage}{0.9\linewidth}
\begin{footnotesize}
The table reports elasticities estimates as well the percentage difference between the current prices and the optimal price maximizing profit.
In each panel we report, for each product, the minimum, the 5\%-, 25\%-, 50\%-, 75\%-, and 95\%-quantiles, maximum, average, and standard deviation
for a given statistic. We consider the distribution over the selected treated municipalities. \textbf{We only report results concerning the cities where the estimated $\Delta$ has the correct sign and the effects are
statistical significance at the 10\% level}. The last column indicates the fraction of cities that satisfy the criterium described above.
In Panel (a) we report the results for the estimated elasticities. In Panel (b) we show the results for the difference between the current
price and the optimal price.
\end{footnotesize}
\end{minipage}
\resizebox{0.9\linewidth}{!}{
\begin{threeparttable}
\begin{tabular}{cccccccccccc}
\hline
\multicolumn{12}{c}{\underline{Panel (a): \textbf{Elasticities}}} \\
Product && Min & 5\%-quantile & 25\%-quantile & Median & 75\%-quantile & 95\% quantile & Max & Average & Std. Dev & Fraction \\
\hline
I       &&     -6.1709 &  -6.1363 &  -4.3408 &  -2.9859 &  -2.2372 &  -1.5650 &  -1.4268 &  -3.4141 & 1.4124 & 0.1753 \\
II      &&    -17.2147 & -16.8507 & -12.9427 & -11.9334 &  -8.7978 &  -3.4640 &  -2.8945 & -10.9383 & 3.8642 & 0.1569 \\
III     &&     -2.8147 &  -2.8147 &  -2.4532 &  -1.8840 &  -1.6626 &  -1.6254 &  -1.6254 &  -2.0550 & 0.4905 & 0.0755 \\
IV      &&    -32.7958 & -24.6827 & -11.4079 &  -6.6815 &  -3.9395 &  -2.6159 &  -2.4158 &  -8.5285 & 7.1821 & 0.2000 \\
V       &&    -30.6356 & -30.6356 & -28.5022 & -25.0506 & -23.6188 & -15.6706 & -15.6706 & -24.7547 & 5.2214 & 0.0545 \\
\\
\multicolumn{12}{c}{\underline{Panel (b): \textbf{Price Discrepancies (\% Difference)}}} \\
Product && Min & 5\%-quantile & 25\%-quantile & Median & 75\%-quantile & 95\% quantile & Max & Average & Std. Dev & Fraction \\
\hline
I       &&   -15.5604 & -15.5143 & -12.1436 &  -6.9179 &  -1.3130 &   8.7227 &  11.3806 &  -6.3656 & 7.4661 & 0.1753 \\
II      &&   -21.3911 & -21.3250 & -20.4324 & -20.1047 & -18.5991 &  -9.0741 &  -7.0214 & -18.6833 & 3.6201 & 0.1569 \\
III     &&    -7.5915 &  -7.5915 &  -4.5816 &   1.2230 &   4.7269 &   5.4073 &   5.4073 &   0.0690 & 5.2832 & 0.0755 \\
IV      &&   -18.1886 & -17.4421 & -15.3303 & -12.2240 &  -7.0189 &  -0.4867 &   0.9835 & -10.4346 & 5.6478 & 0.2000 \\
V       &&   -18.1073 & -18.1073 & -17.9851 & -17.7375 & -17.6224 & -16.5487 & -16.5487 & -17.6231 & 0.5603 & 0.0545 \\
\hline
\end{tabular}
\end{threeparttable}}
\end{table}

\newpage


\begin{table}[htbp]
\caption{\textbf{Results: Estimation and Inference (state level).}}\label{T:estimation_state}
\begin{minipage}{0.9\linewidth}
\begin{footnotesize}
The table reports estimation results at the state level.
In each panel we report, for each product, the minimum, the 5\%-, 25\%-, 50\%-, 75\%-, and 95\%-quantiles, maximum, average, and standard deviation
for a variety of different statistics. We consider the distribution over the treated municipalities aggregated at the state level.
In Panel (a) we report the results for the R-squared of the pre-intervention model.
Panel (b) displays the results for the average intervention effect over the experiment period ($\Delta$).
Panels (c) and (d) depict the results for the $p$-values of the ressampling test for the null hypothesis
$\mathcal{H}_0:\delta_t = 0, \forall t\in \{T_{0}+1,\dots, T\}$ using respectively the test statistic
$\phi(\widehat \delta_{T_0+1},\ldots, \widehat  \delta_T)=\sum_{t=T_0+1}^T \widehat  \delta_t^2$
or $\phi(\widehat \delta_{T_0+1},\ldots, \widehat \delta_T)=\sum_{t=T_0+1}^T|\widehat  \delta_t|$.
Finally, Panel (e) reports the results for the $p$-values for the test for the idiosyncratic contribution.
\end{footnotesize}
\end{minipage}
\resizebox{0.9\linewidth}{!}{
\begin{threeparttable}
\begin{tabular}{ccccccccccc}
\hline
\multicolumn{11}{c}{\underline{Panel (a): \textbf{R-squared}}} \\
Product && Min & 5\%-quantile & 25\%-quantile & Median & 75\%-quantile & 95\% quantile & Max & Average & Std. Dev \\
\hline
I       &&  0.3553 & 0.3804 & 0.6776 & 0.8027 & 0.8969 & 0.9573 & 0.9603 & 0.7593 & 0.1814  \\
II      &&  0.8830 & 0.8895 & 0.9410 & 0.9812 & 0.9934 & 0.9962 & 0.9962 & 0.9659 & 0.0351  \\
III     &&  0.2983 & 0.3604 & 0.6552 & 0.7726 & 0.8651 & 0.9422 & 0.9642 & 0.7329 & 0.1763  \\
IV      &&  0.7566 & 0.8014 & 0.9377 & 0.9684 & 0.9874 & 0.9952 & 0.9962 & 0.9480 & 0.0587  \\
V       &&  0.0996 & 0.1249 & 0.1795 & 0.3048 & 0.5028 & 0.8898 & 0.9024 & 0.3687 & 0.2298  \\
\\
\multicolumn{11}{c}{\underline{Panel (b): \textbf{Average Treatment Effect (over time): $\Delta$}}} \\
Product && Min & 5\%-quantile & 25\%-quantile & Median & 75\%-quantile & 95\% quantile & Max & Average & Std. Dev \\
\hline
I       &&   -4.9172 &  -4.6035 & -3.3475 & -1.8782 &  0.1261 &  1.8318 &  1.8886 & -1.6650 & 2.1306 \\
II      &&  -17.5812 & -16.8032 & -9.1995 & -2.6898 &  2.0167 & 13.5124 & 14.5276 & -2.9574 & 9.0027 \\
III     &&   -7.5112 &  -6.7144 & -3.5990 & -0.9333 &  0.3148 & 14.7754 & 32.3728 & -0.2965 & 7.6855 \\
IV      &&   -2.0756 &  -1.7365 & -0.7512 & -0.3154 &  0.3816 &  0.8821 &  1.0061 & -0.2839 & 0.7757 \\
V       &&   -0.7695 &  -0.6821 & -0.3421 & -0.1571 & -0.0078 &  0.3294 &  0.4216 & -0.1837 & 0.2904 \\
\\
\multicolumn{11}{c}{\underline{Panel (c): \textbf{$p$-value of the test on squared values}}} \\
Product && Min & 5\%-quantile & 25\%-quantile & Median & 75\%-quantile & 95\% quantile & Max & Average & Std. Dev \\
\hline
I       &&      0 &      0 & 0.0511 & 0.2511 & 0.4883 & 0.8736 & 0.9957 & 0.3041 & 0.2823 \\
II      &&      0 &      0 & 0.0548 & 0.3017 & 0.5610 & 0.9731 & 0.9876 & 0.3717 & 0.3427 \\
III     &&      0 & 0.0194 & 0.2043 & 0.4128 & 0.7511 & 0.9387 & 0.9830 & 0.4503 & 0.3201\\
IV      && 0.0331 & 0.0793 & 0.2500 & 0.4215 & 0.5723 & 0.7901 & 0.8017 & 0.4166 & 0.2206 \\
V       &&      0 & 0.0070 & 0.3170 & 0.8426 & 0.9543 & 0.9930 & 1.0000 & 0.6470 & 0.3741 \\
\\
\multicolumn{11}{c}{\underline{Panel (d): \textbf{$p$-value of the test on absolute values}}} \\
Product && Min & 5\%-quantile & 25\%-quantile & Median & 75\%-quantile & 95\% quantile & Max & Average & Std. Dev \\
\hline
I       &&         0 &      0 & 0.0032 & 0.3404 & 0.4553 & 0.9109 & 0.9957 & 0.3189 & 0.2930 \\
II      &&         0 &      0 & 0.0207 & 0.3430 & 0.4824 & 0.9599 & 0.9628 & 0.3493 & 0.3421 \\
III     &&         0 &      0 & 0.1372 & 0.3830 & 0.7660 & 0.9257 & 0.9617 & 0.4213 & 0.3406 \\
IV      &&    0.0331 & 0.0605 & 0.2397 & 0.4793 & 0.5981 & 0.7837 & 0.8140 & 0.4256 & 0.2377 \\
V       &&         0 &      0 & 0.3319 & 0.7830 & 0.9340 & 0.9930 & 1.0000 & 0.6237 & 0.3796 \\
\\
\multicolumn{11}{c}{\underline{Panel (e): \textbf{$p$-value of the test for idiosyncratic contribution}}} \\
Product && Min & 5\%-quantile & 25\%-quantile & Median & 75\%-quantile & 95\% quantile & Max & Average & Std. Dev \\
\hline
I       &&         0 &      0 &      0 & 0.0020 & 0.0210 & 0.1724 & 0.1920 & 0.0314 & 0.0593 \\
II      &&         0 &      0 & 0.0110 & 0.0380 & 0.0545 & 0.2964 & 0.3440 & 0.0616 & 0.0893 \\
III     &&         0 &      0 & 0.0105 & 0.0420 & 0.0990 & 0.5157 & 0.6080 & 0.1017 & 0.1640 \\
IV      &&    0.0940 & 0.1164 & 0.1780 & 0.2040 & 0.2310 & 0.5657 & 0.6420 & 0.2408 & 0.1292 \\
V       &&         0 &      0 & 0.0350 & 0.0820 & 0.1350 & 0.5735 & 0.6780 & 0.1370 & 0.1767 \\
\hline
\end{tabular}
\end{threeparttable}}
\end{table}

\newpage

\begin{table}[htbp]
\caption{\textbf{Results: Elasticities and Optimal Prices (state level).}}\label{T:price_state}
\begin{minipage}{0.9\linewidth}
\begin{footnotesize}
The table reports elasticities estimates as well the percentage difference between the current prices and the optimal price maximizing profit.
In each panel we report, for each product, the minimum, the 5\%-, 25\%-, 50\%-, 75\%-, and 95\%-quantiles, maximum, average, and standard deviation
for a given statistic. We consider the distribution over the selected treated municipalities. \textbf{We only report results concerning the cities where the estimated $\Delta$ has the correct sign and the effects are
statistical significance at the 10\% level}. The last column indicates the fraction of cities that satisfy the criterium described above.
In Panel (a) we report the results for the estimated elasticities. In Panel (b) we show the results for the difference between the current
price and the optimal price.
\end{footnotesize}
\end{minipage}
\resizebox{0.9\linewidth}{!}{
\begin{threeparttable}
\begin{tabular}{cccccccccccc}
\hline
\multicolumn{12}{c}{\underline{Panel (a): \textbf{Elasticities}}} \\
Product && Min & 5\%-quantile & 25\%-quantile & Median & 75\%-quantile & 95\% quantile & Max & Average & Std. Dev & Fraction \\
\hline
I       &&      -1.8282 &  -1.8282 & -1.7647 & -1.5235 & -1.1270 & -0.8921 & -0.8921 & -1.4431 & 0.3720 & 0.2222 \\
II      &&     -11.5235 & -11.5235 & -7.7493 & -5.8728 & -4.4528 & -4.1237 & -4.1237 & -6.5147 & 2.9592 & 0.1852 \\
III     &&      -3.2089 &  -3.2089 & -2.9427 & -1.9333 & -1.0804 & -0.9708 & -0.9708 & -2.0116 & 1.1006 & 0.1481 \\
IV      &&   -- & -- & -- & -- & -- & -- & -- & -- & -- & --  \\
V       &&   -- & -- & -- & -- & -- & -- & -- & -- & -- & --  \\
\\
\multicolumn{12}{c}{\underline{Panel (b): \textbf{Price Discrepancies (\% Difference)}}} \\
Product && Min & 5\%-quantile & 25\%-quantile & Median & 75\%-quantile & 95\% quantile & Max & Average & Std. Dev & Fraction \\
\hline
I       &&     3.6865 &   3.6865 &   4.6703 &   9.3255 &  20.7040 &  32.3850 &  32.3850 &  13.3495 & 11.2129 & 0.2222 \\
II      &&   -19.9566 & -19.9566 & -17.4338 & -15.7817 & -13.0450 & -12.1705 & -12.1705 & -15.5676 &  3.0364 & 0.1852 \\
III     &&    -9.7734 &  -9.7734 &  -8.2241 &   4.9918 &  21.4037 &  26.1490 &  26.1490 &   6.5898 & 17.5845 & 0.1481 \\
IV      &&  -- & -- & -- & -- & -- & -- & -- & -- & -- & -- \\
V       &&  -- & -- & -- & -- & -- & -- & -- & -- & -- & -- \\
\hline
\end{tabular}
\end{threeparttable}}
\end{table}

\begin{table}[htbp]
\caption{\textbf{Results: Estimation and Inference (Before-and-After).}}\label{T:estimation_ba}
\begin{minipage}{0.9\linewidth}
\begin{footnotesize}
The table reports estimation the average treatment effect using the before-and-after estimator.
In each panel we report, for each product, the minimum, the 5\%-, 25\%-, 50\%-, 75\%-, and 95\%-quantiles, maximum, average, and standard deviation
for a variety of different statistics. We consider the distribution over the treated municipalities.
\end{footnotesize}
\end{minipage}
\resizebox{0.9\linewidth}{!}{
\begin{threeparttable}
\begin{tabular}{ccccccccccc}
\hline
Product && Min & 5\%-quantile & 25\%-quantile & Median & 75\%-quantile & 95\% quantile & Max & Average & Std. Dev \\
\hline
I       &&   -23.8652 & -17.2270 &  -8.1333 &  -4.1126 &  -1.1093 & 2.1760 & 11.5150 &  -5.2622 &   6.0399  \\
II      &&   -74.8229 & -53.2274 & -30.7149 & -18.4681 & -10.3370 & 1.8621 & 13.1138 & -22.0736 &  16.9785  \\
III     &&   -48.8512 & -15.3860 &  -5.6494 &  -2.1679 &  -0.5397 & 2.4336 & 11.1025 &  -3.9888 &   7.0311  \\
IV      &&    -5.5069 &  -4.7638 &  -2.1703 &  -1.2016 &  -0.1093 & 1.7353 &  3.5901 &  -1.2483 &   1.9274  \\
V       &&    -2.0595 &  -1.3942 &  -0.8139 &  -0.4505 &  -0.1244 & 0.4032 &  1.1809 &  -0.4682 &   0.5426  \\
\hline
\end{tabular}
\end{threeparttable}}
\end{table}

\newpage

\newpage

\newpage

\end{document}